\documentclass{article}

\usepackage{PRIMEarxiv}
\usepackage{amsmath}
\usepackage{multirow}
\usepackage{siunitx}
\usepackage[utf8]{inputenc} % allow utf-8 input
\usepackage[T1]{fontenc}    % use 8-bit T1 fonts
\usepackage{hyperref}       % hyperlinks
\usepackage{url}            % simple URL typesetting
\usepackage{booktabs}       % professional-quality tables
\usepackage{amsfonts}       % blackboard math symbols
\usepackage{nicefrac}       % compact symbols for 1/2, etc.
\usepackage{microtype}      % microtypography
\usepackage{graphicx}       % graphics
\graphicspath{{media/}}     % organize images and other figures under media/ folder

\usepackage{makecell}
\usepackage{caption}
\usepackage{authblk}  % for automatic line breaks in author block

\newcommand{\am}[1]{\textcolor{black}{{ #1}}}
\newcommand{\vs}[1]{\textcolor{black}{{ #1}}}

\title{Thermal and dimensional stability of photocatalytic material ZnPS$_3$ under extreme environmental conditions
}

\author[1,$\S$]{Abhishek Mukherjee}
\author[1,2,$\S$]{Vivian J. Santamar\'{i}a-Garc\'{i}a}
\author[3]{Damian Wlodarczyk}
\author[3]{Ajeesh K. Somakumar}
\author[3]{Piotr Sybilski}
\author[4,5]{Ryan Siebenaller}
\author[6]{Emmanuel Rowe}
\author[3]{Saranya Narayanan}
\author[4]{Michael A. Susner}
\author[2]{L. Marcelo Lozano-Sanchez}
\author[3]{Andrzej Suchocki}
\author[7]{Julio L. Palma}
\author[1,*]{Svetlana V. Boriskina}

\affil[1]{Massachusetts Institute of Technology, Cambridge, MA 02139, USA.}
\affil[2]{Tecnologico de Monterrey, Escuela de Ingeniería y Ciencias, Monterrey 64849, Mexico.}
\affil[3]{Institute of Physics, Polish Academy of Sciences, Warsaw 02-668, Poland.}
\affil[4]{Air Force Research Laboratory, Wright-Patterson AFB, OH 45433, USA.}
\affil[5]{Department of Materials Science and Engineering, The Ohio State University, Columbus, OH 43210, USA}
\affil[6]{National Research Council, Washington, D. C. 20001, USA.}
\affil[7]{Department of Chemistry, Penn State University, Lemont Furnace, PA 15456, USA.}
\affil[$\S$]{Co-first authors}
\affil[*]{sborisk@mit.edu}

\begin{document}

\maketitle

\begin{abstract}
Zinc phosphorus trisulfide (ZnPS$_3$), a promising material for photocatalysis and energy storage, is shown in this study to exhibit remarkable stability under extreme conditions. We explore its optical and structural properties under high pressure and cryogenic temperatures using photoluminescence (PL) spectroscopy, Raman scattering, and density functional theory (DFT). Our results identify a pressure-induced phase transition starting at 6.75 GPa and stabilizing by 12.5 GPa, after which ZnPS$_3$ demonstrates robust stability across a broad pressure range of 15 to 100 GPa. DFT calculations predict a semiconductor-to-semimetal transition at 100 GPa, while PL measurements reveal defect-assisted emissions that quench under pressure due to enhanced non-radiative recombination. At cryogenic temperatures, PL quenching intensifies as non-radiative processes dominate, driven by a rising Grüneisen parameter and reduced phonon population. Cryogenic X-ray diffraction (XRD) also reveals a high mean thermal expansion coefficient (TEC) of (4.369 $\pm$ 0.393) $\times$ 10$^{-5}$ K$^{-1}$, among the highest reported for 2D materials. This unique combination of tunable electronic properties under low pressure and high thermal sensitivity makes ZnPS$_3$ a strong candidate for sensing applications in extreme environments.

\end{abstract}

\keywords{Photoluminescence \and Defects \and Thermal Expansion Coefficient \and Grüneisen Parameter}

\section{Introduction}
The discovery of graphene has sparked significant interest in two-dimensional (2D) van der Waals (vdW) materials, which are characterized by strong in-plane covalent bonds and weak interlayer vdW forces. These materials exhibit unique properties, such as charge density waves \cite{hossain2017recent,chargeden_2}, anisotropic magnetic behavior \cite{Gibertini2019,Magnetic_anisotropy}, and tunable optical and electronic characteristics \cite{Optprop2D,Opt_tun,mukherjee2024synergistic}, due to their layered structure and the associated electron confinement effects. Graphene boasts an exceptionally high charge carrier mobility but lacks an electronic bandgap, which limits its applications in semiconducting and optoelectronic devices \cite{Graph}. This limitation has fueled the exploration of other 2D materials, such as transition metal dichalcogenides (TMDs), black phosphorus (BP), and boron nitride (BN), which possess intrinsic bandgaps. Both TMDs and BP feature thickness-dependent bandgaps: TMDs usually transition from indirect bandgaps (1.2–1.4 eV) in bulk to direct bandgaps (1.7–2.2 eV) in monolayers, while bandgap of BP varies from 0.3 eV in bulk to 2 eV in monolayers \cite{TMDs_Eg,TMDs_Eg_2,BP}. BP is especially promising for infrared applications due to its high anisotropy, though its bulk form is less suitable for devices operating at shorter wavelengths \cite{BP_2}. In contrast, hexagonal BN possesses a wide bandgap ($\sim$ 6.0 eV), beyond the visible wavelength range. Materials with bandgaps between 2 and 4 eV are sought after for applications in optoelectronics, photovoltaics, and UV detection.
Recently, metal phosphorus trichalcogenides (MPTs) have attracted attention owing to their intermediate bandgaps (1.3–3.5 eV), diverse electronic and magnetic properties, and nonlinear optical effects. These materials show enhanced visible light absorptance compared to TMDs and are effectively tunable, capable of forming intercalation compounds and undergoing Mott transitions under pressure \cite{Mott_evidence, Mott_2}. These properties are strongly influenced by the choice of a specific transition metal element within the MPTs.

In this study, we focus on ZnPS$_3$, a member of the metal phosphorus thiophosphate (MPS$_3$) family, which exhibits a monoclinic ${C2/m}$ lattice symmetry at ambient conditions. These materials consist of layered structures with slightly distorted octahedral sites surrounded by sulfur atoms, forming a honeycomb lattice separated by van der Waals gaps. In ZnPS$_3$, the Zn$^{2+}$ cations are positioned within ZnS$_6$ cages, while P-P dimers create a [P$_2$S$_6$]$^{4-}$ anionic sublattice that balances the charge of the cations. This anionic framework is consistent across the MPS$_3$ family, with different metal cations contributing to various material functionalities, such as magnetism or lack thereof. ZnPS$_3$ stands out for its relatively large bandgap of approximately 3.4-3.8 eV \cite{Water_splitting_ZHANG,ZnPS3_3}, its diamagnetic nature, and a second-order Jahn-Teller effect leading to subtle lattice distortions \cite{Jahn_Teller_BOUCHER,ZnPS3_3}. In prior studies, Raman spectroscopy has shown a significant broadening of low-frequency phonon modes in ZnPS$_3$, revealing the presence of these distortions,  even though they do not induce a structural phase transition in the material \cite{Lat_dyn_2,ZnPS3_3,Vib_properties}.

Additionally, ZnPS$_3$ exhibits higher specific heat at low temperatures, suggesting that lattice vibrations, rather than magnetic interactions, dominate its thermal behavior \cite{S_heat_TAKANO2004E593}. Unlike other MPS$_3$ materials, the electronic band structure of ZnPS$_3$ is stable under external perturbations such as biaxial strain or electric fields, with limited tunability of its bandgap \cite{Water_splitting_ZHANG}.  These characteristics highlight the unique electronic, structural, and thermal properties of ZnPS$_3$ within the MPS$_3$ family, largely influenced by its lattice dynamics and electronic configuration.

The tuning of physical properties in MPS$_3$ compounds is an active  area of research, encompassing techniques such as chemical substitution, doping, intercalation, and the application of external stimuli like electric fields, light, magnetic fields, and strain \cite{MPX3_review,MPX3_Emerging,MPX3_Strain_Doping}. However, ZnPS$_3$, due to its non-magnetic nature and the lack of Mott insulating behavior, has attracted less attention than other MPS$_3$ compounds with partially-filled d-orbitals, leaving its response to external stimuli underexplored. However, ZnPS$_3$ has shown considerable potential across a wide range of applications, including photocatalytic water splitting, energy storage, and humidity sensing. Its wide bandgap makes it well-suited for ultraviolet light absorption. However, ZnPS$_3$ stability under external stimuli like biaxial strain and electric fields limits opportunities to engineer its broadband light absorption, particularly in the visible spectrum \cite{Water_splitting_ZHANG}. Despite these challenges, the favorable band edge positions of ZnPS$_3$ suggest some potential for water splitting. In energy storage, ZnPS$_3$ stands out for its ability to conduct divalent Zn$^{+2}$ ions with low activation energy, making it a promising candidate for solid-state electrolytes in next-generation batteries, especially those utilizing multivalent ions \cite{ZnPS3_2}. Furthermore, its superionic conductivity in humid environments, along with its structural stability, suggests suitability of this material as a potential humidity sensor \cite{WaterVapor_ZnPS3}. These diverse applications highlight the versatility of ZnPS$_3$, yet also point to the need for further exploration of its behavior under various external perturbations to fully unlock its capabilities. Here, we present, to the best of our knowledge, the first report on room temperature, low temperature, and high pressure photoluminescence (PL) studies of ZnPS$_3$. To correlate the impacts of pressure and temperature stimuli on its electronic band structure, we performed first principle calculations within the DFT framework, complemented by Raman and X-ray spectroscopy and light absorption measurements to support our findings.

\section{Experimental Section}
\label{sec:headings}

\subsection{Material Synthesis}
We synthesized ZnPS$_3$ via conventional vapor transport methods \cite{susner2017metal}. Briefly, we combined Zn foil (Alfa Aesar Puratronic, 99.999\%), P chunks (Alfa Aesar Puratronic, 99.999+\%), and S pieces (Alfa Aesar Puratronic, 99.999\%) together in a near-stoichiometric ratio (with 10\% excess P) in an evacuated quartz ampoule (2mm wall thickness, 22 mm outside diameter, 10 cm in length) together with $\sim$100 mg I$_2$ crystals (Alfa Aesar, 99.8\%). We placed the sealed ampoule in a single-zone tube furnace, heated to 750$^\circ$C over a period of 20 hours, held at that temperature for 100 hrs, and cooled over a period of 20 hrs. The resulting crystals measured maximum 5 mm $\times$ 5 mm in area and $\sim$200 $\mu$m in thickness.

\subsection{Instruments and Methods}
Sample compositions were determined by subjecting single-crystal specimens to multiple-spot scanning electron microscopy/energy-dispersive X-ray spectroscopy (SEM/EDS) analysis using an UltiMax 100 Oxford Instruments EDS spectrometer joined with a Zeiss Sigma 300 VP SEM. Material characterization was done via the X-ray diffraction (XRD), using the PANalytical X'pert Pro equipped with a copper target (sealed tube) as the X-ray source and an X’Celerator Scientific 1D position-sensitive detector. Cryogenic XRD measurements were performed with a liquid helium cooled cryostat supplied by Oxford Instruments.

To observe Photoluminescence (PL), the samples were excited with a near-UV 325 nm He-Cd laser (Kimmon Koha IK3201R-F model). PL data were collected using a Triax 320 monochromator from ISA Yobin Yvon-Spex equipped with a Spectrum One CCD detector. A Cary 5000 spectrophotometer was used to obtain absorption spectra in the UV-Visible range. Raman spectral data were acquired using a Monovista CRS+ system from S\&I Ltd, equipped with a 0.75m long monochromator. The resolution was approximately  0.5 cm$^{-1}$. The setup was operated on Trivista Software. A semiconducting 532 nm laser (Cobolt Samba 05-01 model) was used, set at a power density of 7.5 mW/$\mu$m$^2$.

Both Raman and PL data were collected in two configurations: under high pressure at room temperature and at low temperature under ambient pressure. Absorption measurements were carried out only in the latter configuration. For the Raman Spectroscopy and PL measurements, liquid nitrogen was utilized to achieve the desired temperatures. To achieve high pressures, the samples were exfoliated using a blue PVC tape and then transferred in a Diacell CryoDAC-LT Almax easyLab Diamond Anvil Cell (DAC) with two 16-side standard-type (IIas) 2.5 mm anvils having 450 $\mu$m culets. A 4:1 methanol-ethanol mixture was used as the pressure-transmitting medium. Ruby was placed in the DAC to act as the pressure gauge. \\

\section{Computational Methods}
First principle calculations were performed using density functional theory (DFT) as implemented in the Vienna Ab initio simulation package (VASP) \cite{KRESSE1995222}. The generalized gradient approximation (GGA) with the Perdew$-$Burke$-$Ernzerhof (PBE) \cite{PBE} functional with the DFT$-$D3 \cite{DFT-D} correction was used for modeling the electron-exchange and correlation energies. Core electrons were treated using the projector augmented$-$wave (PAW) \cite{PAWmethod} pseudopotentials method with Zn 3d 4s, P 3s 3p and S 3s 3p electrons included in the valence bands. The valence Kohn$-$Sham orbitals were represented in a plane$-$wave basis with a kinetic energy cut-off of 500 eV. \vs{Spin-orbit coupling (SOC) was included in all electronic structure calculations to account for relativistic effects. However, no significant changes were observed with and without SOC, as can be seen in the band structure shown in Fig. S9 (Supplementary Information).} Tight convergence criteria were applied during the structure relaxations to ensure high accuracy in the optimized atomic positions. The convergence criterion for the electronic total energy was set to $10^{-8}$ eV. A primitive unit cell composed of 10 atoms was used for the simulations. To adequately sample the Brillouin zone, a  $12\times12\times12$  mesh was utilized based on the Monkhorst$-$Pack \cite{MonkhorstPack} sampling scheme. 

Quasi-harmonic lattice dynamics calculations were performed on the optimized structures at different pressures (0, 5, 10, 15, 20, and 25 GPa, as well as -0.2, -0.2, -0.4, -0.6, -0.8 and -1 GPa). At each of the corresponding volumes, the phonon dispersion was calculated using the Phonopy \cite{Phonopy} package combined with the density functional perturbation theory (DFPT) \cite{DFPT} as implemented in VASP \cite{KRESSE1995222}. The dynamical matrix was constructed using a $2 \times 2 \times 2$ supercell, and a $5 \times 5 \times 5$ k-point mesh was employed to sample the Brillouin zone for the DFPT calculations, ensuring accurate computation of interatomic force constants. Phonopy \cite{Phonopy} utilized a fine $20 \times 20 \times 20$ mesh to achieve high-resolution phonon dispersion curves, providing detailed insights into the vibrational properties and stability of the material. Thermal expansion and Gr\"uneisen parameters were then computed within the quasi-harmonic approximation (QHA). 

\color{black}

\section{Results and Discussion}

\begin{figure}[h]
    \centering
    \includegraphics[width=0.75\linewidth]{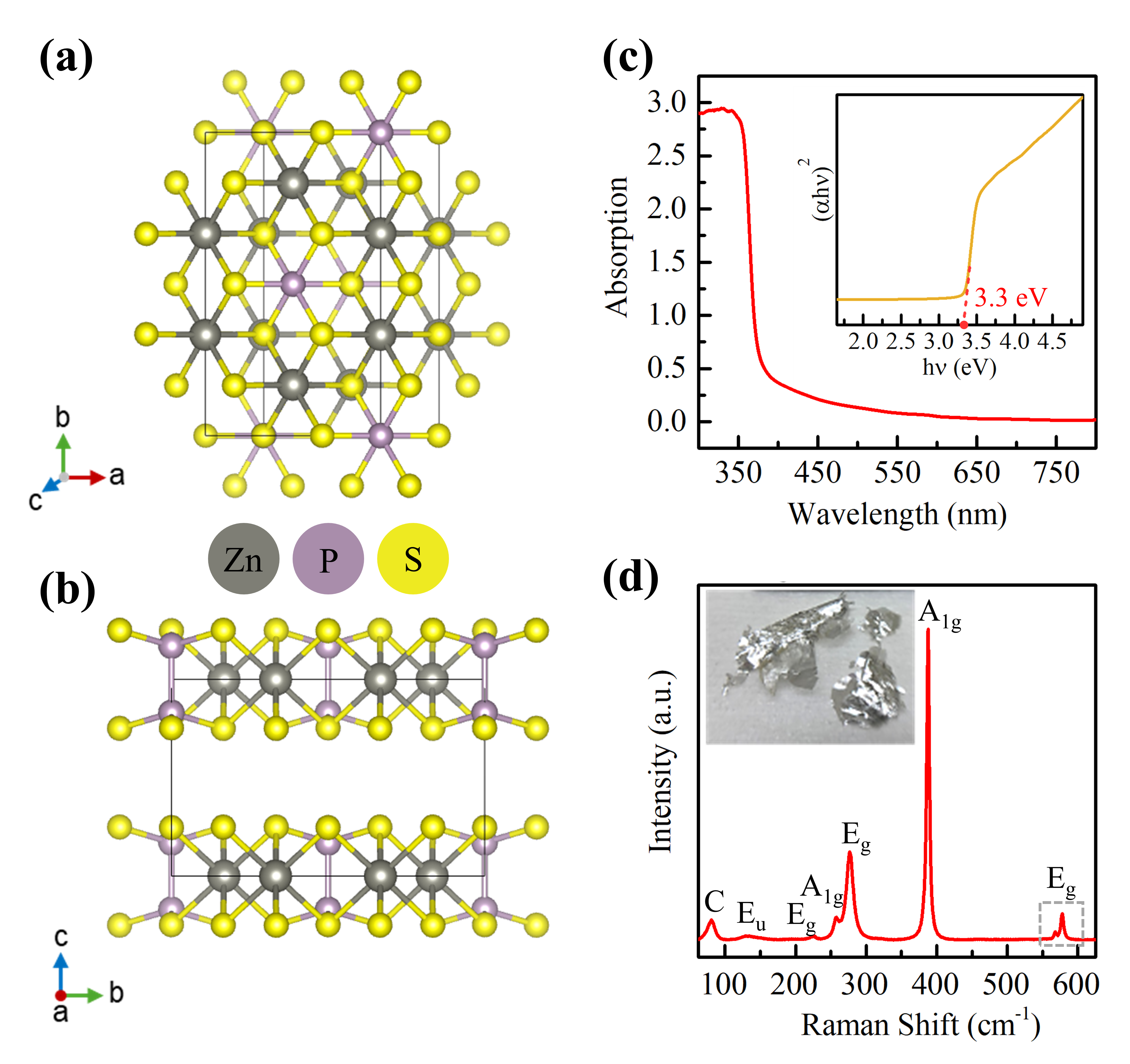}
    \caption{ZnPS$_3$ crystal lattice, visual appearance, and ambient characterization via absorption and Raman spectroscopy. Schematics of the (a) top and (b) side views of the crystals are shown. (c) Absorption spectra measured in the UV-visible spectral range. Inset shows the corresponding Tauc plot. (d) Raman spectra under ambient conditions. Inset shows a photograph of the bulk sample.  }
    \label{fig:schematic}
\end{figure}

Figure 1 introduces the ZnPS$_3$ material by demonstrating its structural configuration as well as its optical and vibrational properties. The characteristic feature of compounds in the MPT family is the presence of the [P$_2$S$_6$]$^{4-}$ anion sublattice within a single lamella of the layered crystal structure (Fig. \ref{fig:schematic}a,b). Though the overall structure within an individual lamella is essentially the same across this family, the symmetry and stacking arrangements differ depending on the composition. ZnPS$_3$ material is of type M$^{2+}$P$_2$S$_6$, where M is a transition metal, and belongs to the symmetry group $C2\textit/{m}$ \cite{susner2017metal}. The in-plane crystal structure comprises a regular pattern of octahedrally coordinated sites, filled by Zn$^{2+}$ ions and P--P dimers. These layers are held together by van der Waals forces. Fig. \ref{fig:schematic}c shows the measured optical absorption spectrum, revealing characteristic band-edge absorption in the UV-visible spectral range. 

ZnPS$_3$ exhibits the behavior of a direct band gap semiconductor, thus the absorption edge can be calculated using Tauc's analysis. The inset shows the corresponding plot where the absorption (Abs), as a function of wavelength ($\lambda$), can be directly correlated with the band gap energy ($E_g$) and the optical frequency ($\nu$) via the following relation \cite{ghobadi2013band}:

\begin{equation}
\text{Abs}(\nu)h\nu = B_1 \left(h\nu - E_g \right)^m + B_2,
\end{equation}

where \( B_1 = \left[ B \times \frac{d}{2.303} \right] \), with $d$ being the thickness of the material, $B$  -- a frequency-independent constant, and $h$ -- the Planck's constant. In turn, \( B_2 \) is a constant that accounts for the material reflectance. The exponent $m$ is set as $1/2$ for the allowed direct transitions. By extrapolating the linear part of the plot (red dashed line in the inset plot), the band gap energy $E_g$ was estimated to be 3.3 eV, in agreement with previous literature \cite{brec1979physical}. 

Next, Raman spectroscopy was used to characterize the vibrational modes of the sample under the ambient temperature and atmospheric pressure conditions, as shown in Fig. \ref{fig:schematic}d. Typical of layered MPTs, a spectral band appearing below 100 cm$^{-1}$ is attributed to cation vibrations, while a small peak near 150 cm$^{-1}$ is identified as E$_u$ mode, which is often present in MPT compounds. The Raman modes originating from P$_2$S$_6$ units correspond to either out-of-plane vibrations, identified as A$_{1g}$ (257.8 and 387.6 cm$^{-1}$) or to in-plane vibrations, identified as E$_g$ (226.9, 567.8 and 577.8 cm$^{-1}$). These modes are labeled in the spectrum shown in \ref{fig:schematic}d, and their corresponding energies are consistent with previous reports on Raman characterization of ZnPS$_3$ \cite{yan2021transforming}. \\

\begin{figure}[h]
\centering
\includegraphics[width=1\linewidth]{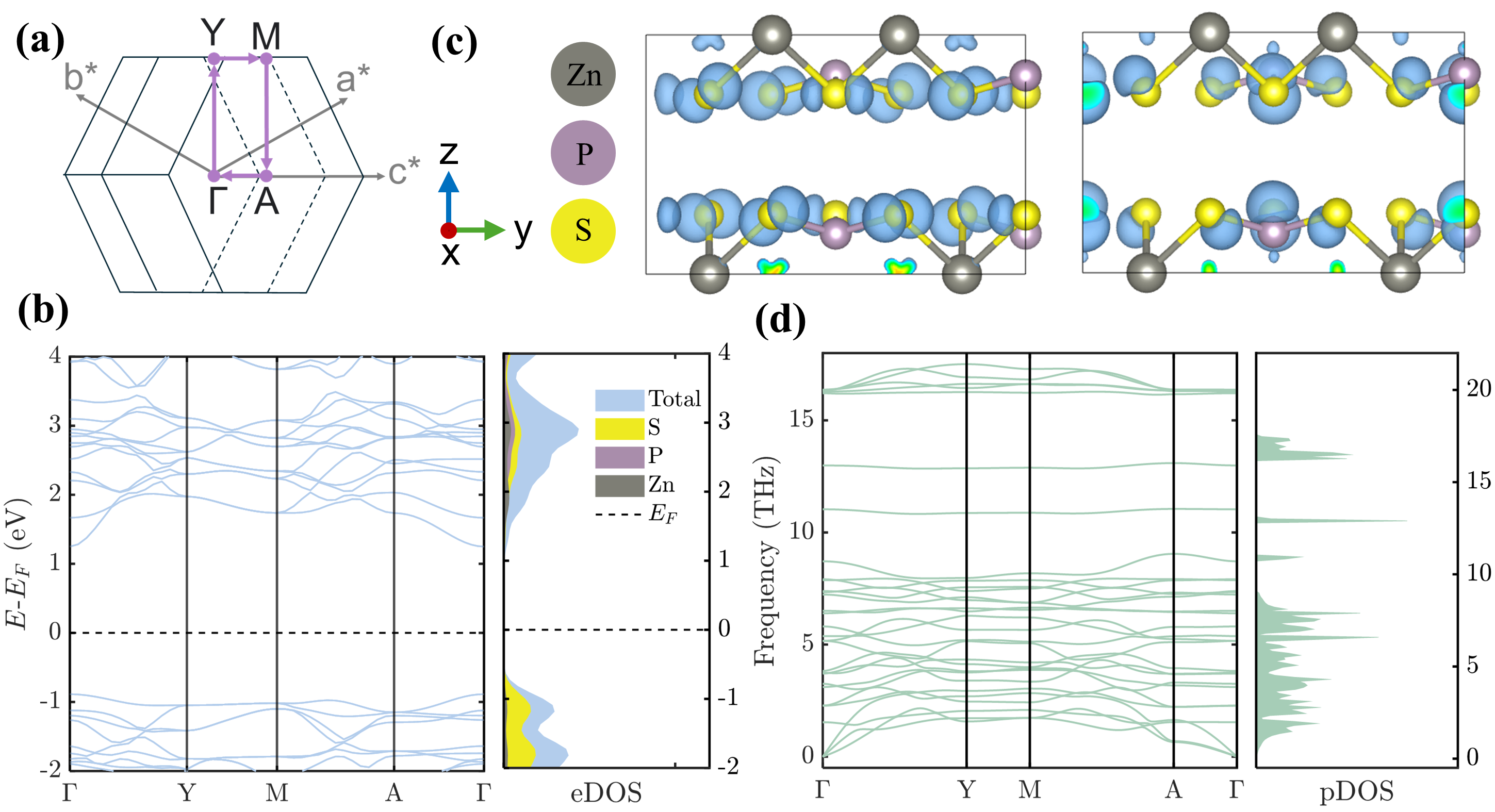}
\caption{Electronic and phononic ground state properties of ZnPS$_3$ predicted by DFT. (a) The Brillouin zone and high-symmetry points used for sampling. (b) Electronic band structure (left) and the corresponding density of states (eDOS, right). (c) Charge density isosurfaces of valence (left)  and conduction (right) bands at the gamma point. (d) Phonon dispersion (left) and corresponding density of states (pDOS, right).}
\label{fig:Band Structure}
\end{figure}

Figure \ref{fig:Band Structure} shows the electronic and phononic ground state properties of ZnPS$_3$ predicted by DFT, which agrees with previous reports \cite{ZnPS3_1,ZnPS3_2,ZnPS3_3,ZnPS3_4}. In Figure \ref{fig:Band Structure}a, the Brillouin Zone is shown along with the high-symmetry points used to build the wave vector path (\(\Gamma\) – Y – M – A – \(\Gamma\)) for sampling the electronic and phononic energy dispersion plots. Figure \ref{fig:Band Structure}b presents the electronic band structure on the left and the electronic density of states (eDOS) on the right, revealing an underestimated direct band gap of around 2.1 eV at the \(\Gamma\) point. The valence band lies slightly closer to the Fermi level ($E_F$) than the conduction band, as indicated by the eDOS plot, which shows an increased number of available states below the Fermi level. This proximity suggests that the material may exhibit p-type semiconductor behavior, where holes dominate the charge transport.  

Figure \ref{fig:Band Structure}c  illustrates the charge density contributions of the valence (left) and conduction (right) bands at the $\Gamma$ point. The isosurfaces of charge density in the valence band are primarily localized around S atoms, while in the conduction band, they extend to include some P atoms. Notably, there are no significant isosurfaces around Zn atoms in these bands, indicating that Zn atoms contribute minimally to the electronic states near the Fermi level, which suggests their lower reactivity in the ground state.
Figure \ref{fig:Band Structure}d displays the phonon energy dispersion and the corresponding phonon density of states (pDOS). Since no imaginary frequencies appear in the calculated spectrum, the material is dynamically stable. The phonon dispersion of ZnPS$_3$ exhibits distinct characteristics typical of layered materials. The dispersion curves are relatively flat, particularly in the optical phonon branches in the 10-15 THz range, indicating weak interlayer interactions, which arise because the atoms in adjacent layers are coupled only by van der Waals forces. These weak interactions result in minimal changes in vibrational energy as the wave vector varies, leading to nearly constant phonon frequencies across different points in the Brillouin zone.  This flatness suggests that the high-frequency modes, which are primarily associated with vibrations within the P$_2$S$_6$ octahedral units, experience minimal variation across different wave vectors due to the weak coupling between adjacent layers. The pDOS plot shows prominent peaks corresponding to these flat dispersion regions, emphasizing the dominance of intralayer P-S and P-P interactions. The transverse-acoustic (TA) branches exhibit a noticeable upward curvature as they move away from the \(\Gamma\) point, reflecting the anisotropic vibrational properties of the layered material. This curvature suggests that strong intralayer bonding dominates  weaker interlayer coupling, which is typical for layered semiconductors such as ZnPS$_3$.\\

\color{black}
\subsection{High-pressure characterization}
High-pressure characterization provides insights into structural and vibrational properties of ZnPS$_3$ as the material undergoes compression. By analyzing Raman spectra across a wide range of pressures, we can observe how the vibrational modes shift, indicating potential phase transitions and atomic rearrangements within the crystal lattice. This section delves into how these changes affect the structural stability and behavior of ZnPS$_3$, shedding light on its performance under extreme conditions.

\begin{figure}
    \centering
\includegraphics[width=1\linewidth]{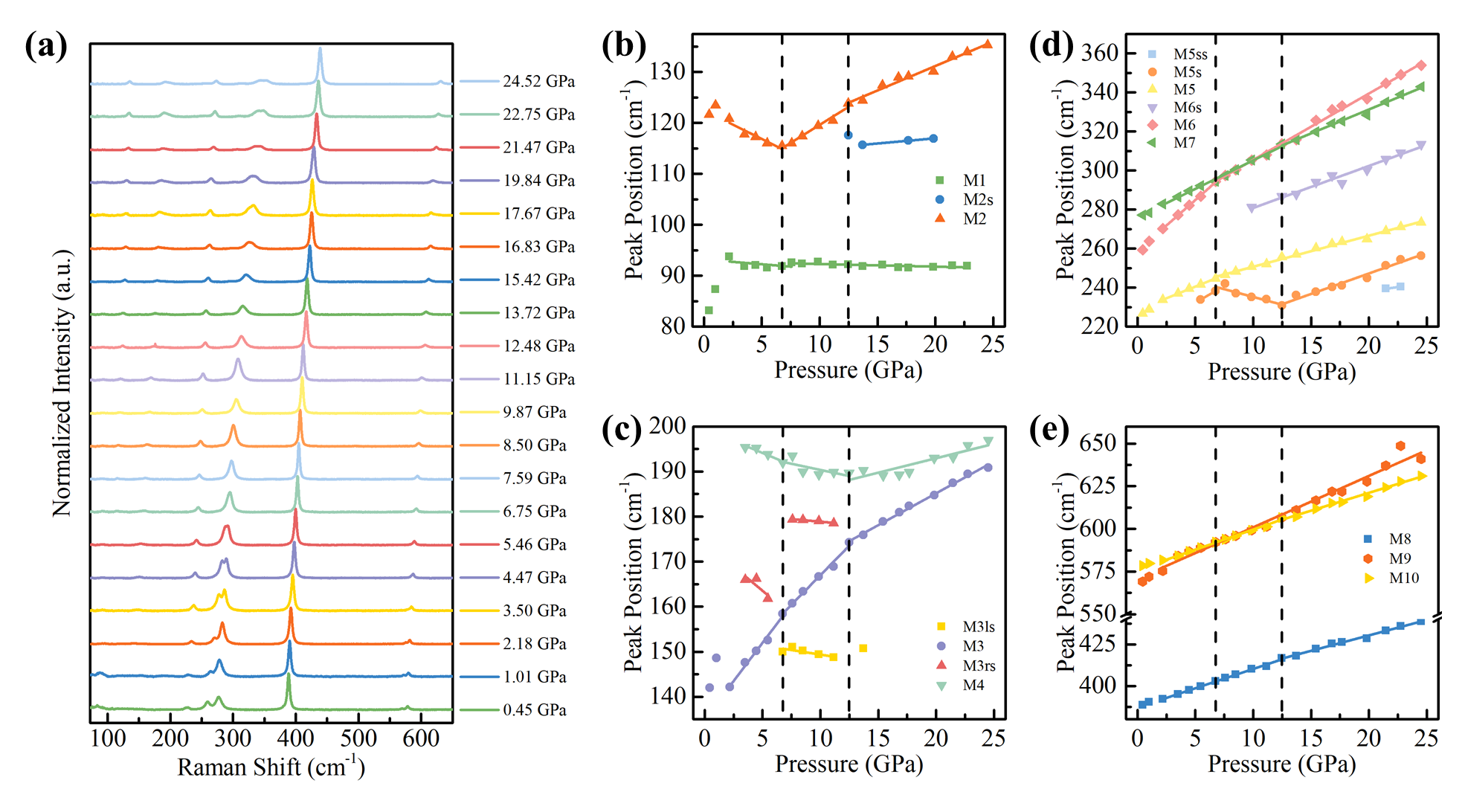}
    \caption{(a) Raman spectra of ZnPS$_3$ measured at high pressures up to 24.5 GPa. (b)-(e) Peak positions of the Raman modes as a function of applied pressure. Solid lines show linear fits to experimental data. Vertical dashed lines mark a change in the pressure coefficient (slope of the linear fit) as the modes evolve. In the legend, the subscript notation is avoided for readability, meaning that $M1$ is the same as $M_1$, $M2s$ is the same as $M_{2s}$, and so on.}
    \label{fig:HP-Raman}
\end{figure}

Figure 3 shows the Raman spectra of ZnPS$_3$ measured at pressures ranging from 0.45 to 24.75 GPa. At the lowest pressure (0.45 GPa), which is close to ambient, eight Raman modes were observed, located at 83 ($M_1$), 128 ($M_2$), 142 ($M_{3}$), 227 ($M_{5}$), 259 ($M_{6}$), 277 ($M_{7}$), 389 ($M_{8}$), 569 ($M_{9}$) and 578 ($M_{10}$) cm$^{-1}$. The peak positions were established by de-convoluting the spectra using Lorentzian curves, as shown in Figure S2 and Table S1, Supporting Information. A similar spectrum of modes has been observed at the start of a high-pressure compression cycle in MnPS$_3$ \cite{kozlenko2024high}, another metal thiophosphate having the same symmetry group as our material. Further, the observed vibrational modes are in agreement with the peaks reported in Fig. \ref{fig:schematic}d, except for the doubly-degenerate $E_u$ mode, which is now likely split into two distinct non-degenerate modes $M_2$ and $M_{3}$ as a result of the in-plane strain developed in the material (a single mode was found at 132 cm$^{-1}$ at ambient conditions, see description for Fig. \ref{fig:schematic}). In general, due to DAC instability at low pressures, the first two measurement points have been excluded in the calculation of linear pressure coefficients for the Raman modes (linear fits shown by solid lines in Fig.\ref{fig:HP-Raman}b-e exclude the first two points). The calculated coefficients, mode Grüneisen parameters, and $R^2$ values for the linear fits are reported in Table \ref{table:HP-Data}. With the increase in the lattice compression, the frequencies of all Raman modes shift, with anomalous behavior detected for lower-frequency modes $M_1$-$M_{6}$, indicated by a change in the pressure coefficients at ~6.75 GPa. Atomic rearrangements of Zn atoms and P$_2$S$_6$ group in the crystal lattice during compression result in a modified pressure dependence of each vibrational mode. Higher-frequency modes $M_{8}$-$M_{10}$  (associated with P$_2$S$_6$ molecular group vibrations) show no sign of anomalous behavior, indicating that the geometry of this group is not affected by the phase transition. Such a trend is consistent with past observations of Raman spectra of other metal thiophosphates with the same symmetry group \cite{kozlenko2024high}. \\

Several additional peaks were also identified as the phase transition evolved -- namely $M_{2s}$, $M_{3ls}$, $M_{3rs}$, $M_{4}$, $M_{5ss}$, $M_{5s}$ and $M_{6s}$. A majority of these peaks are shoulder peaks to existing vibrational modes which can be identified by the naming convention. The characteristics of each peak can be found in Table \ref{table:HP-Data}. Weak shoulder peaks $M_{3rs}$ and $M_{5s}$ emerge at 3.5 GPa, prior to the onset of the observed phase transition at 6.75 GPa, which may be due to orientation disorder in the lattice. An additional mode ($M_{4}$) also emerges at this pressure, the assignment of which remains ambiguous. However, estimating the linear pressure dependence gives an initial energy of this mode to be ~199 cm$^{-1}$, which indicates that this mode is either a symmetry-forbidden Raman $B_{3g}$ mode (observed at 185 cm$^{-1}$ in Rb$_2$P$_2$S$_6$ \cite{gjikaj2006rb2p2s6}) or a feature caused by rotation of the PS$_3$ group \cite{oliveira20232d}. At pressures equal to or higher than the predicted transition point, modes $M_{3ls}$, $M_{3rs}$, $M_{5ss}$, $M_{5s}$, and $M_{6s}$ emerge, with some of them disappearing/reappearing as the lattice evolves, indicating a change in the symmetry group. An additional number of Raman active modes confirm that the new symmetry group is one of lower order. Further, these peaks disappear before the end of the high-pressure cycle, except for $M_{6s}$. Some of the new modes disappear and then reappear before disappearing for the rest of the compression cycle, such as $M_{3rs}$, $M_{3ls}$, $M_{2s}$, and $M_{5ss}$.\\

Anomalous changes in the pressure coefficients of lower frequency Raman modes are also seen at high pressures above 12.5 GPa (see Fig. 3 and Table \ref{table:HP-Data}). Our current hypothesis is that this point either (1) marks the end of a phase stabilization range after which the transition with the onset at 6.75 GPa is fully complete, or (2) represents a secondary minor phase transition (e.g., a structural change within a particular symmetry group). There are observations in the Raman data that support either of these theories. For instance, a gradual and slow change in the linear pressure coefficients of modes $M_{4}$ (negative at low pressures, exhibiting lower values in the intermediate range between 6.75 GPa and 12.5 GPa, and becoming positive at higher pressures, see Fig. \ref{fig:HP-Raman}c and Table \ref{table:HP-Data}) support hypothesis (1). At the same time, sharp changes in the linear pressure coefficient of mode $M_{5s}$ at both 6.75 and 12.5 GPa, along with the emergence of mode $M_{2s}$ at 12.5 GPa support hypothesis (2). Within the scope of this work, we lean towards hypothesis (1) and support this claim using results from PL spectroscopy and DFT calculations in the later parts of this section. However, more work is necessary to confirm this claim as well as the nature of the phase transition(s) and mode assignments of the new Raman modes. Group theory calculations and high-pressure X-ray diffraction are potential techniques that can sufficiently support the deductions from this experiment but are outside of the scope of this study. An important aspect to be considered here is the stability of the high-pressure phase as well. We use Raman spectroscopy data from the decompression cycle (i.e pressure release and relaxation data collected in the same cycle as the compression series) to show that the modes at the endpoint match those of the starting point of the pressure cycle. This denotes the reversibility of the obtained high-pressure phase as all the higher-order induced modes vanish during decompression. Further, a lack of any significant hysteresis in the mode frequencies proves that the material phases are stable. The spectral evolution during pressure release and Raman mode comparison before and after the cycle are shown in Figure S2 and Table S2 in the Supporting Information. \\

\begin{table}[h!]
\centering
\small
\caption{Energies at ambient pressure ($E_0$) derived assuming the linear pressure dependence, pressure coefficients ($dE/dP$) in units of cm$^{-1}/$GPa, and $R^2$ parameters of different Raman modes obtained from the high-pressure Raman spectroscopy data. Under the column "Mode Assignment", the most probable assignments are listed in parenthesis by comparing the data with the Raman spectrum taken under ambient conditions, which was found consistent with past literature \cite{yan2021transforming}
*Gr\"{u}neisen parameters $\gamma_i$ are calculated using the DFT-predicted value of bulk modulus at ground state ($B_0 = 29.61$ GPa), shown in Fig. \ref{fig:DFT-QHA}c. \\}
\begin{tabular}{ccccccccccc}
\toprule
\multirow{2}{*}{\shortstack{Mode \\ Assignment}} & \multirow{2}{*}{$E_0$ (cm$^{-1}$)} & \multirow{2}{*}{Characteristics} & \multicolumn{3}{c}{$P < 6.75$ GPa} & \multicolumn{2}{c}{$6.75 < P < 12.48$ GPa} & \multicolumn{2}{c}{$P > 12.48$ GPa} \\
\cmidrule(lr){4-6} \cmidrule(lr){7-8} \cmidrule(lr){9-10}
& & & $dE/dP$ & $R^2$ & $\gamma_i$* & $\hspace{0.4cm}dE/dP$ & $R^2$ & $dE/dP$ & $R^2$ \\
\midrule
$M_1$ (C) & 92.56 & mw, br & -0.387 & 0.583 & -0.124 & \hspace{0.4cm}-0.053 & 0.515 & -0.053 & 0.515 \\
$M_{2s}$ & 112.93 & w, left shoulder & & & & & & 0.204 & 0.999 \\
$M_2$ (E$_u$) & 122.48 & mw, br & -1.122 & 0.897 & -0.271 & \hspace{0.4cm}0.961 & 0.979 & 1.413 & 0.972 \\
$M_{3}$ (E$_u$) & 135.00 & w, br & 3.398 & 0.988 & 0.745 &\hspace{0.4cm}2.620 & 0.988 & 1.414 & 0.995 \\
$M_{3ls}$ & 153.42 & w, left shoulder & & & & \hspace{0.34cm}-0.405 & 0.664 & & \\
$M_{3rs}$ & 174.31 & w, right shoulder & -2.156 & 0.723 & -0.366 & \hspace{0.34cm}-0.235 & 0.950 & & \\
$M_{4}$ & 199.49 & w, br & -1.072 & 0.932 & -0.159 & \hspace{0.34cm}-0.546 & 0.528 & 0.641 & 0.811 \\
$M_{5s}$ & 214.94 & w, left shoulder & 3.465 & 1.000 & 0.477 & \hspace{0.34cm}-1.592 & 0.815 & 2.099 & 0.977 \\
$M_{5ss}$ & 221.15 & w, left shoulder & & & & & & 0.852 & 1.000 \\
$M_{5}$ (E$_g$) & 228.58 & s, sh & 2.397 & 0.999 & 0.311 & \hspace{0.4cm}1.591 & 0.994 & 1.591 & 0.994 \\
$M_{6}$ (A$_{1g}$) & 258.75 & s, sh & 5.23 & 0.999 & 0.599 & \hspace{0.4cm}3.399 & 0.995 & 3.399 & 0.995 \\
$M_{6s}$ & 259.29 & w, left shoulder & & & & \hspace{0.4cm}2.154 & 0.970 & 2.154 & 0.970 \\
$M_{7}$ (E$_g$) & 276.03 & s, br & 2.917 & 0.994 & 0.313 & \hspace{0.4cm}2.454 & 0.990 & 2.454 & 0.990 \\
$M_{8}$ (A$_{1g}$) & 387.04 & vs, sh & 2.321 & 0.997 & 0.178 & \hspace{0.4cm}2.321 & 0.997 & 1.851 & 0.991 \\
$M_{9}$ (E$_g$) & 570.84 & mw, sh & 3.011 & 0.980 & 0.156 & \hspace{0.4cm}3.011 & 0.980 & 3.011 & 0.980 \\
$M_{10}$ (E$_g$) & 578.43 & mw, sh & 2.355 & 0.995 & 0.121 & \hspace{0.4cm}2.355 & 0.995 & 2.909 & 0.988 \\
\bottomrule
\end{tabular}
\label{table:HP-Data}
\end{table}

\begin{figure}
    \centering
    \includegraphics[width=1\linewidth]{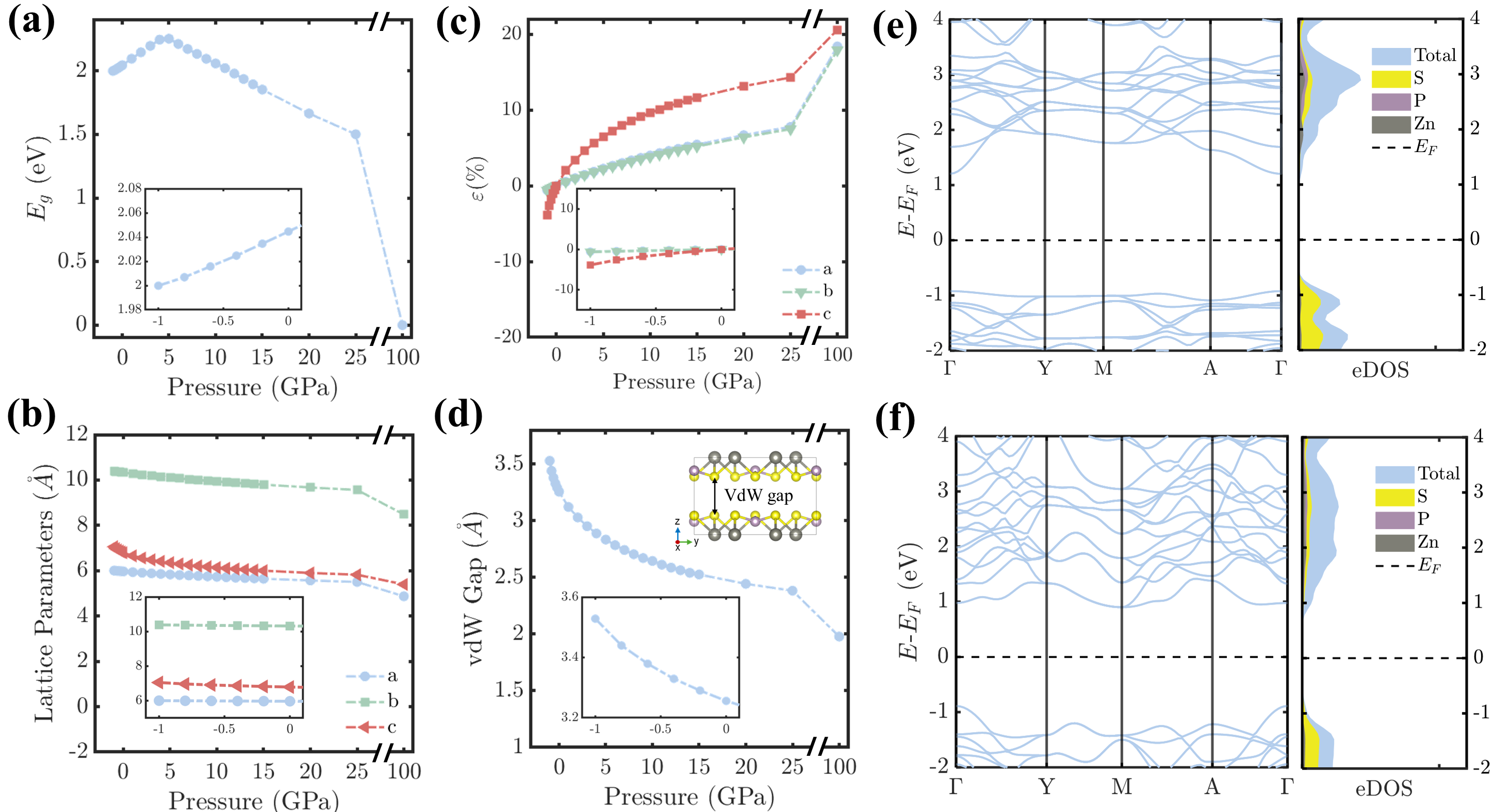}
    \caption{Electronic and structural properties of ZnPS$_3$ under high hydrostatic pressure conditions obtained by DFT. (a) Band gap evolution. (b) Changes of the lattice parameters (a,b,c) dimensions. (c) The stress-induced strain along the three principal directions. (d) The Van der Waals gap size evolution. (e) Band structure at -1 GPa (tension). (f) Band structure at 25 GPa (compression). The insets in (a-d) show zoomed-in portions of the corresponding plots within the -1-0 GPa pressure range.}
    \label{fig:DFT-Pressure}
\end{figure}

We conducted DFT calculations to gain deeper insights into the observed frequency shifts of the vibrational modes, as well as the emergence or disappearance of new peaks, which suggest changes in symmetry and potential phase transitions under increasing pressure. Our calculations revealed related changes in the electronic structure of the material, including its electronic band gap evolution and a transition from a semiconducting to a semi-metallic state. The DFT-predicted evolution of the electronic band gap under pressures ranging from $-$1 GPa to 25 GPa is shown in Figure \ref{fig:DFT-Pressure}a.  When the system is subjected to negative pressure, i.e., tensile stress, the band gap consistently decreases as the magnitude of the pressure increases. Under compressive stress, the system exhibits a distinct behavior: the band gap initially widens, reaching a maximum at pressures about 5 GPa, and then decreases as the pressure increases further. Notably, the band gap values calculated at pressure values of 5 GPa and 10 GPa are greater than that at the ground state, a phenomenon previously reported for this material in a monolayer form \cite{ZnPS3_monolayer_Eg}. The initial increase in the band gap width under compression is attributed to modifications in the bond lengths and angles within the crystal lattice, which elevate the energy levels of the conduction band more significantly than those of the valence band. However, as the compressive strain continues to increase, the band gap width decreases. This reduction occurs as the material reaches a point where the increased orbital overlap favors band convergence, potentially leading to a complete phase transition caused by a significant symmetry breaking. These data align with the experimental observations of a phase transition between 6.75-12.5 GPa and provide insight into the underlying physical mechanisms driving this transition.

Figure \ref{fig:DFT-Pressure}b shows the evolution of the lattice parameters under different stress conditions. As expected, the unit cell dimensions increase under tensile stress and decrease under compressive stress. The strain corresponding to these changes is depicted in Figure \ref{fig:DFT-Pressure}c. Under tensile stress, the strain reaches 0.6\% for the in-plane \(a\) and \(b\) parameters and 14.3\% for the \(c\) parameter. This points out that the \(c\) parameter, associated with the out-of-plane direction, is more susceptible to pressure-induced changes, which is consistent with the layered nature of ZnPS$_3$. Figure \ref{fig:DFT-Pressure}d illustrates the van der Waals distance, which increases under tensile stress and decreases under compressive stress, following a consistent trend. These pressure-induced changes in the lattice parameters and van der Waals distances play a crucial role in determining the electronic properties of ZnPS$_3$. As the material is subjected to higher pressures, the resulting lattice compression affects its electronic band structure, leading to significant modifications in the conductive behavior of the material. The band structures calculated at select pressure values of -1 GPa and 25 GPa are presented in Figures \ref{fig:DFT-Pressure}e and \ref{fig:DFT-Pressure}f, respectively. The electronic band structures and eDOS at various pressures considered in this study are also summarized in Figure S10, available in the Supporting Information.

\begin{figure}
    \centering
    \includegraphics[width=1\linewidth]{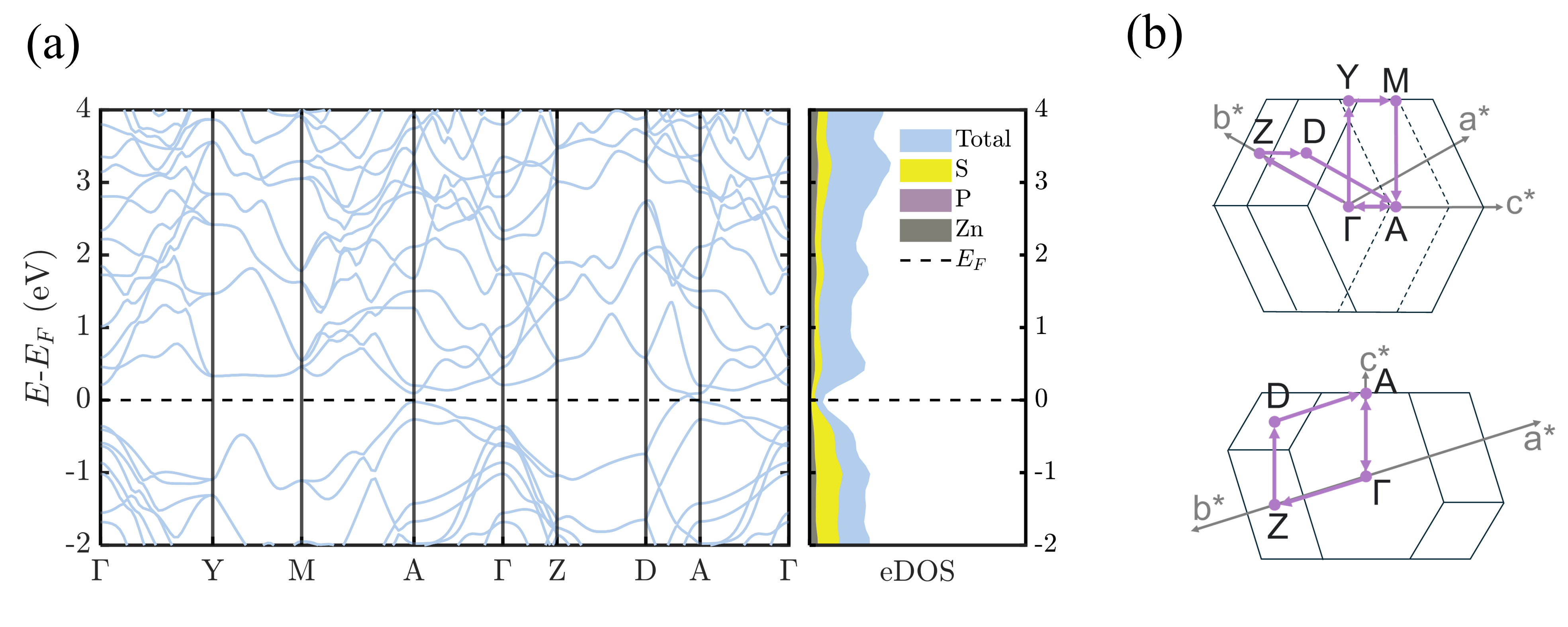}
    \caption{\vs{Electronic properties of ZnPS$_3$ at 100 GPa: (a) band structure (left) and eDOS (right). The band structure reveals the absence of a band gap, with a notable band overlap along the D-A path, indicating a semi-metallic state. The eDOS plot highlights the contributions from sulfur (S), phosphorus (P), and zinc (Zn) atoms near the Fermi level (dashed line). (b) Two views of the Brillouin zone illustrating the selected k-path used to sample the band structure.}}. 
    \label{fig:100GPa}
\end{figure}

The band structure at -1 GPa closely resembles that of the ground state, with the eDOS plot showing more states available below the Fermi energy. In contrast, at 25 GPa, the band structure reveals a narrowing band gap as the conduction band shifts towards the Fermi level. The eDOS plot at this pressure shows a higher concentration of states above the Fermi level, indicating a transition towards a more metallic state. As pressure continues to increase, the impact on the electronic structure becomes even more pronounced. The electronic band structure of ZnPS$_3$ at 100 GPa, as depicted in Fig. \ref{fig:100GPa}, shows an overlap between the valence and conduction bands \vs{along the D-A path}, indicating the absence of a band gap and suggesting that ZnPS$_3$ has entered a semi-metallic state. The eDOS plot corroborates this observation by displaying non-zero eDOS values at the Fermi level ($E_F$). This semi-metallic behavior arises from the pressure-induced band overlap, where the high pressure compresses the crystal structure and shifts the energy levels of both the valence and conduction bands.

\color{black}

\begin{figure}[h]
    \centering
    \includegraphics[width=1\linewidth]{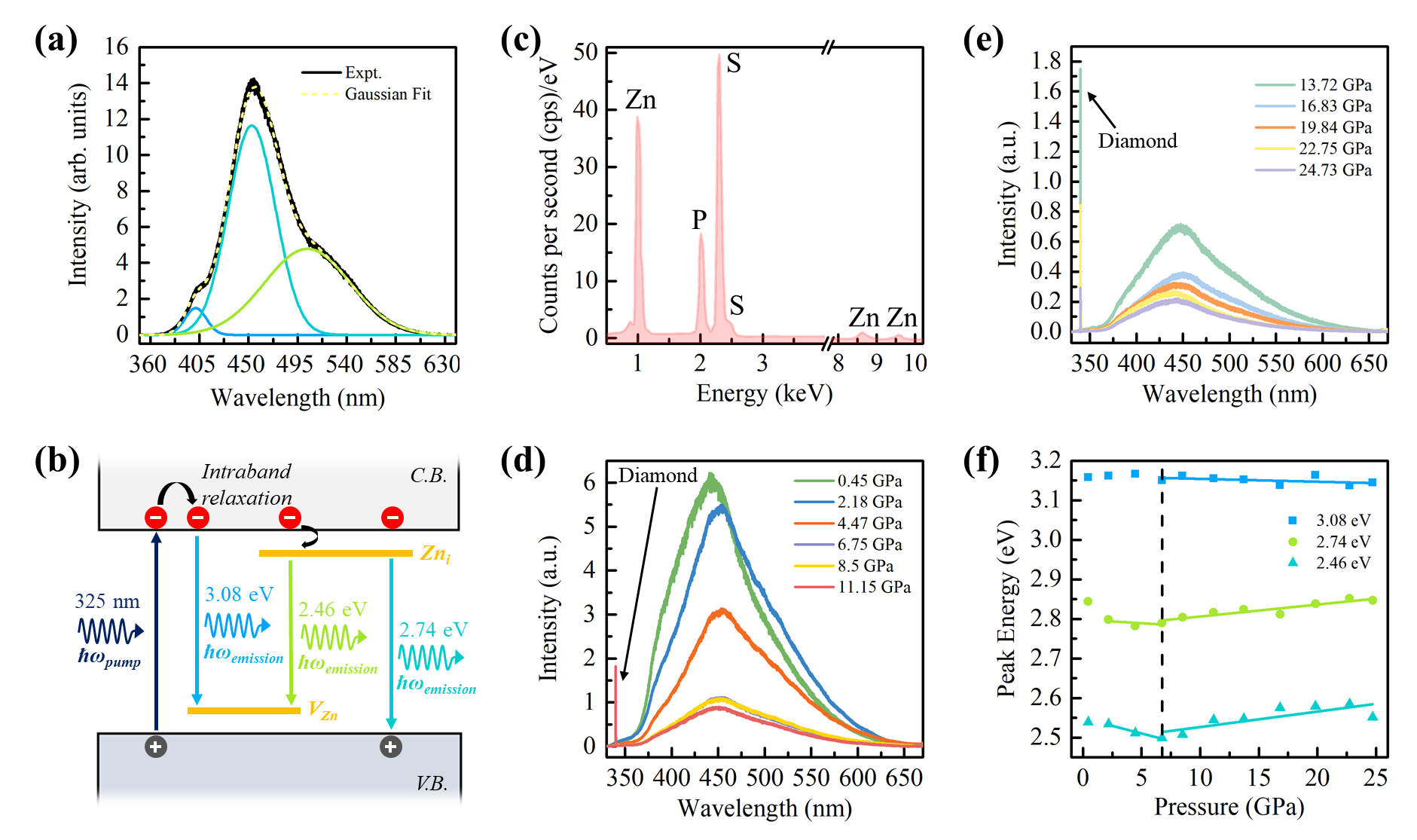}
    \caption{Pressure-dependent photoluminescence spectra of ZnPS$_3$: (a) Ambient PL spectra under a near-UV (325 nm) laser excitation. Gaussian de-convolution reveals three luminescent bands as shown in solid blue, turquoise, and green lines. The overall sum of the fitted Gaussian spectra of the three bands is shown as a yellow dashed line and matches well with experimental data (black line). (b) Band schematic showing expected upward and downward radiative transitions to/from Zn point defects (V$_{Zn}$) and Zn interstitials (Zn$_i$) leading to the observed PL emission (non-radiative processes not shown for simplicity). (c) EDS spectra at ambient conditions, demonstrating the presence of Zn vacancies in the bulk sample (refer to Table S1, supporting information). (d), (e) PL spectra as a function of DAC-induced pressure on ZnPS$_3$. The Raman line of diamonds under 325 nm excitation is marked separately for each spectrum. (f) Peak position energies of each of the three deconvoluted Gaussian curves shown in (a) as a function of pressure. }
    \label{fig:HP-PL}
\end{figure}

Experimental measurements of the photoluminescence spectra have been performed to supplement the DFT predictions of the material electronic properties evolution under pressure.  Figure \ref{fig:HP-PL} introduces the origin of the PL emission in ZnPS$_3$ and describes the compressive pressure effect on the PL intensity and frequency spectrum of the material. Figure \ref{fig:HP-PL}a shows (to the best of our knowledge) the first report of the room temperature photoluminescence measurement for this compound. The PL spectra were obtained under a near-UV excitation of 325 nm and can be de-convoluted into three Gaussian bands for establishing the positions of the luminescence peaks, as shown by solid blue, turquoise, and green colored plots. The fit, represented by the bright yellow dashed line, closely matches the experimental spectrum shown by the solid black line, demonstrating strong agreement between the model and the experimental data. These luminescent bands peak at 403, 489 nm, and 504 nm respectively, indicating violet/blue-green emissions in the sample, commonly reported in other zinc-based compounds such as ZnS \cite{zhang2013synthesis,wageh2013band,zhang2005low,wu2006generation,kar2005controlled} and ZnO \cite{zhao2016photoluminescence,galdamezphotoluminescence,jangir2017investigation}. 

We propose that the observed broadband emission spectrum is a manifestation of defect-assisted photoluminescence. The multi-band PL mechanism is explained via an electron band schematic shown in Fig. \ref{fig:HP-PL}b. The violet band at 403 nm (3.08 eV) is likely due to radiative transitions from the conduction band to the levels associated with zinc vacancies \cite{wageh2013band}, whose presence is confirmed using the energy-dispersive X-ray spectroscopy (EDS, Fig. \ref{fig:HP-PL}e and Table S2 in supporting information). The blue-green emission features peaking at 489 nm and 504 nm can be attributed to the presence of zinc interstitials and zinc vacancies, but cannot be confirmed without additional measurements such as scanning transmission microscopy to confirm the exact nature of the defect (for instance, interstitial vs point defects under laser excitation to explicitly show PL emission originating from either or both). Here, we propose that the 489 nm peak (2.74 eV) arises as a result of radiative recombination between the levels corresponding to zinc intersititals and the valence band, while the 504 nm (2.46 eV) peak can be attributed to recombination between zinc interstitials and zinc vacancy defect states (Fig. \ref{fig:HP-PL}b). The investigation of the origin of these defects and the mechanisms of controlling them to influence the PL emission is a broad topic that remains outside the scope of the present work.

\begin{figure}[h]
    \centering
    \includegraphics[width=1\linewidth]{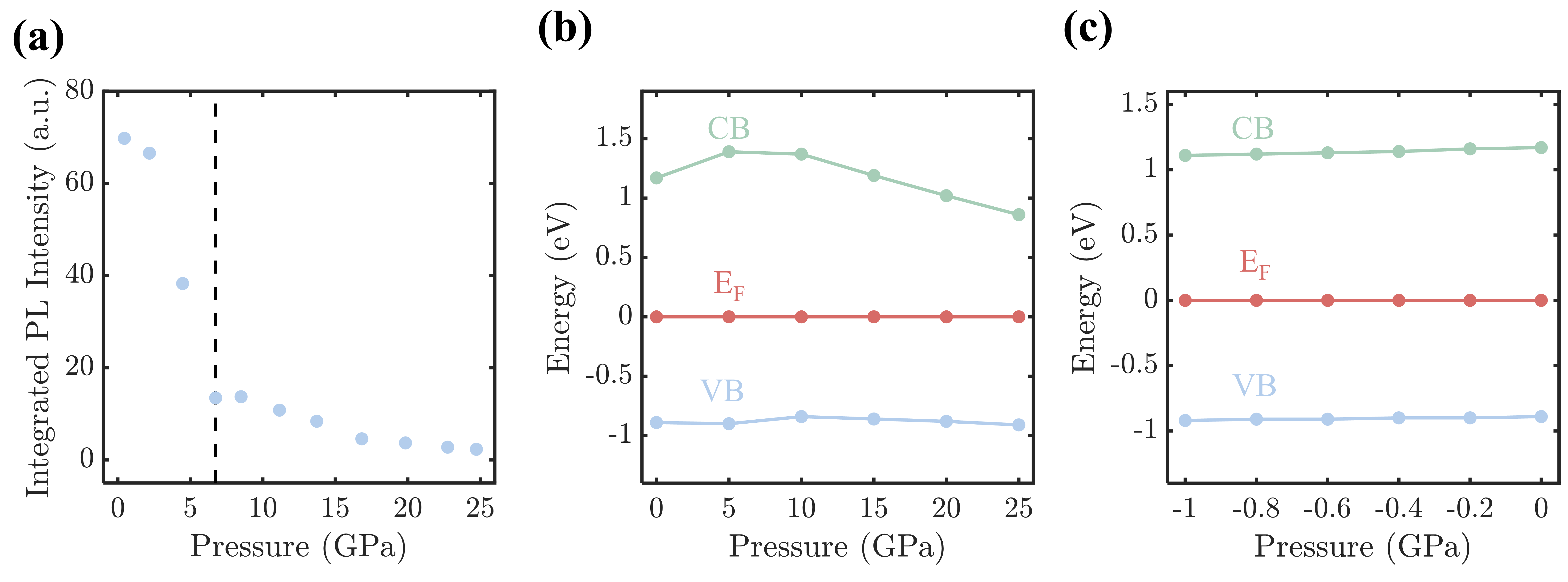}
    \caption{\vs{Pressure-dependent properties of ZnPS$_3$: (a) Integrated photoluminescence (PL) intensity as a function of pressure, showing significant quenching above 6.75 GPa. (b) Band edge positions relative to the Fermi level (E$_F$) for pressures ranging from 0 to 25 GPa, with noticeable shifts of the conduction band (CB) and slight changes of the valence band (VB). (c) Band edge positions in the negative pressure range (-1 to 0 GPa), where minimal changes are observed.}}   
    \label{fig:IPL}
\end{figure}

We focus our experimental study on probing the tunability of PL emission under high compressive stress, and the results are summarized in Figs. \ref{fig:HP-PL}d-f. As shown by Fig. \ref{fig:HP-PL}d,e, PL emission quenches with the increase in pressure.\vs{This is further supported by the integrated photoluminescence intensities (IPL) in Fig. \ref{fig:IPL}a, which track the total emitted light as a function of pressure, representing the overall radiative recombination efficiency. The IPL shows a significant decrease above 6.75 GPa, corresponding to the onset of the phase transition predicted by DFT calculations and Raman spectroscopy.} \am{The observed PL quenching can likely be attributed to defect dynamics, including: (i) the annihilation of vacancy states, and (ii) the enhancement of non-radiative recombination pathways at defect centers. The latter mechanism is supported by DFT-calculated pDOS under high pressure. As shown in Fig. S12 (Supporting Information), the shift of phonon modes to higher frequencies with increasing pressure suggests lattice stiffening, which amplifies phonon interactions and promotes non-radiative relaxation processes. This leads to excited electrons losing energy to the lattice via phonons, rather than through photon emission. The broadening of the pDOS peaks further implies increased phonon-phonon interactions, offering additional pathways for non-radiative decay, thereby contributing to the PL quenching. Furthermore, the shift to higher phonon frequencies reduces the energy gap between electronic states and phonons, which enhances the efficiency of non-radiative transitions, causing the thermalization of excited states and, consequently, reduced PL intensity.} 

\vs{The pressure dependence of the peak positions for the three PL spectral lines is shown in Fig. 6f. Linear curve fitting, represented by solid lines, was used to calculate the peak energy shift coefficients. The first emission measurement at 1.01 GPa was excluded from the fit due to the initial instability of the DAC. The 3.08 eV emission peak shows behavior that closely follows the trend of the conduction band in Fig. \ref{fig:IPL}b. While the valence band edge remains highly stable under pressure, the conduction band increases in energy, peaking at 5 GPa, and then begins to decrease. This explains the initial blueshift in the peak position, which corresponds to an increase in the band gap. This behavior suggests that the zinc vacancy states are as energetically stable as the valence band edge. In Fig. \ref{fig:HP-PL}f, no linear fitting was performed below 5 GPa for 3.08 eV emission peak due to the limited number of data points (only two measurements). However, above 5 GPa, the fitting produced a coefficient of -0.716 meV/GPa, indicating a redshift in the emission peak. This redshift is consistent with the narrowing of the band gap, which is primarily attributed to the conduction band shifting closer to the Fermi level, as seen in Fig. \ref{fig:IPL}b. }

\vs{The evolution of the 2.74 eV and 2.46 eV emission peaks shows a redshift below 6.75 GPa (indicated by the dashed line), with shift coefficients of -1.850 meV/GPa and -7.690 meV/GPa, respectively (Fig. \ref{fig:HP-PL}f). At lower pressures, the lattice is less compressed, allowing zinc interstitials to be more mobile, particularly within the van der Waals gaps where they are less localized. This increased mobility enables zinc interstitials to interact more freely with the surrounding lattice without forming strong bonds, resulting in relatively higher energy for the interstitial states. Zinc defects have been shown to have a propensity to form complexes, according to prior literature \cite{villafuerte2020zinc,fioretti2018exciton}. The enhanced mobility of zinc interstitials allows them to move more easily through the lattice and potentially encounter zinc vacancies. When zinc interstitials come into close proximity with zinc vacancies, the defects may recombine, leading to the annihilation of the vacancies. This recombination reduces the overall defect concentration, stabilizing the remaining interstitial states and lowering their energy. The ordering of the lattice, with fewer disruptions caused by free interstitials, contributes to the observed redshift as the energy gap between the Zn interstitial states and the valence band decreases. Above 6.75 GPa, the 2.74 eV and 2.46 eV emission peaks show a blueshift, with shift coefficients of 3.020 meV/GPa and 3.920 meV/GPa, respectively. At higher pressures, increased lattice compression forces zinc interstitials closer to zinc vacancies or other atoms, leading to stronger interactions and potentially forming defect complexes. These stronger interactions raise the energy of the zinc interstitials, widening the energy gap between the interstitial states and either the valence band or zinc vacancy states, causing the observed blueshift. Additionally, the strain from lattice compression may enhance the splitting of electronic states, further contributing to the increase in energy.}

PL measurements were also made when the pressure in the DAC was released in the same cycle from 22.75 GPa to 0.15 GPa. These results are shown in Fig. S3, demonstrating partial recovery of PL emission as the sample was allowed to relax.

\color{black}

\subsection{Cryogenic Measurements}

\begin{figure}
    \centering
    \includegraphics[width=0.75\linewidth]{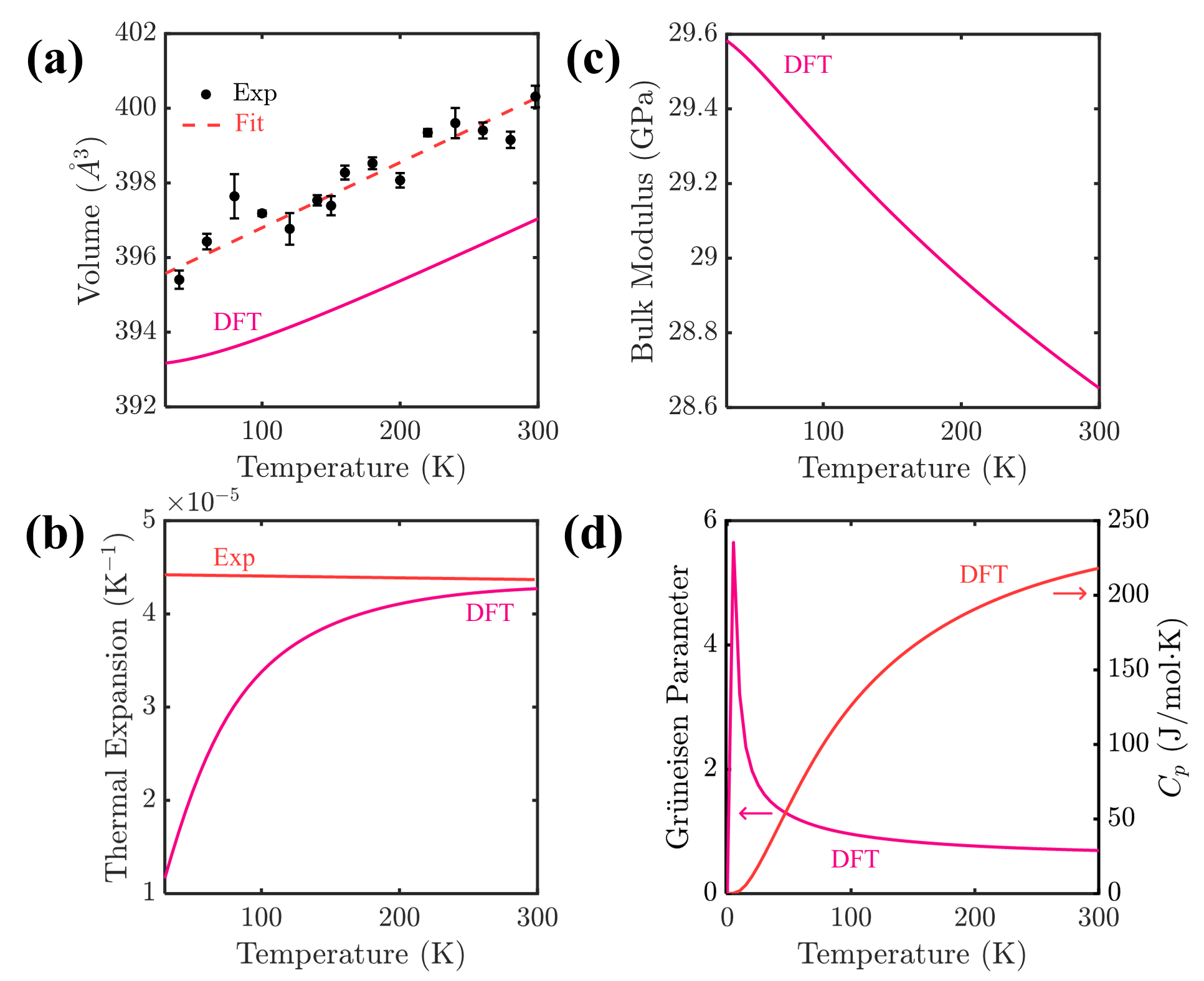}
    \caption{Temperature dependent properties of ZnPS$_3$ derived from QHA. (a) Volume, (b) thermal expansion, (c) bulk modulus,  (d) average Gr\"{u}neisen parameter and heat capacity. The legend `Exp' denotes the experimentally calculated corresponding quantity in (a) and (b).}
    \label{fig:DFT-QHA}
\end{figure}

Cryogenic measurements and simulations are crucial for understanding material behavior at low temperatures, providing key insights into its stability and properties in extreme conditions. Figure \ref{fig:DFT-QHA} presents the results of DFT calculations under the quasi-harmonic approximation (QHA) for the temperature-dependent properties of ZnPS$_3$, with a focus on the low-temperature range from 0 to 300 K. These calculations were performed by determining the phonon frequencies of the material under various hydrostatic pressures. These pressure-dependent phonon frequencies were then used to compute the Helmholtz free energy as a function of the volume and temperature. By minimizing this free energy at each temperature, equilibrium volumes were obtained, allowing the conversion of pressure-dependent properties into temperature-dependent ones.

Figures \ref{fig:DFT-QHA}a and \ref{fig:DFT-QHA}b compare these QHA-derived properties with the experimental data, showing a gradual increase in volume with temperature, which closely matches experimental trends, although slightly underestimates the thermal expansion coefficient. Figures \ref{fig:DFT-QHA}c and \ref{fig:DFT-QHA}d provide further insights into the mechanical properties of ZnPS$_3$ at low temperatures. The bulk modulus, shown in panel (c), decreases as the temperature increases, indicating that the material becomes softer as thermal energy disrupts its lattice structure. This higher-temperature material softening is typical, as increased thermal motion weakens the interatomic forces, leading to a reduction in the material's resistance to compression. 

\vs{The Grüneisen parameter quantifies how the vibrational frequencies, and hence the phonon modes, shift in response to changes in volume or pressure. As shown in Fig. \ref{fig:DFT-QHA}d, the average Grüneisen parameter increases as temperature decreases, indicating that phonon frequencies become more sensitive to volume changes at lower temperatures. As the temperature approaches 0 K, the phonon modes shift to higher energies, reflecting the stiffening of the lattice as atomic bonds become stronger and less flexible. At the same time, the heat capacity (C$_p$) also evolves with decreasing temperature, as shown in Fig. \ref{fig:DFT-QHA}d. At higher temperatures, more phonon modes are populated, as thermal energy excites a greater number of vibrational states. This increase in the phonon population leads to a higher C$_p$, as more energy is required to raise the temperature of the system. However, as temperature decreases, the population of phonons diminishes because fewer vibrational modes are thermally excited. The decrease in C$_p$ reflects the diminishing capacity of the material to store thermal energy as the phonon population decreases with temperature. These observations are consistent with the  behavior of the thermal expansion and bulk modulus (Fig.\ref{fig:DFT-QHA}b,c) and reinforces the overall understanding that the material has high structural stability at lower temperatures.}

\begin{figure}[h]
    \centering
    \includegraphics[width=0.75\linewidth]{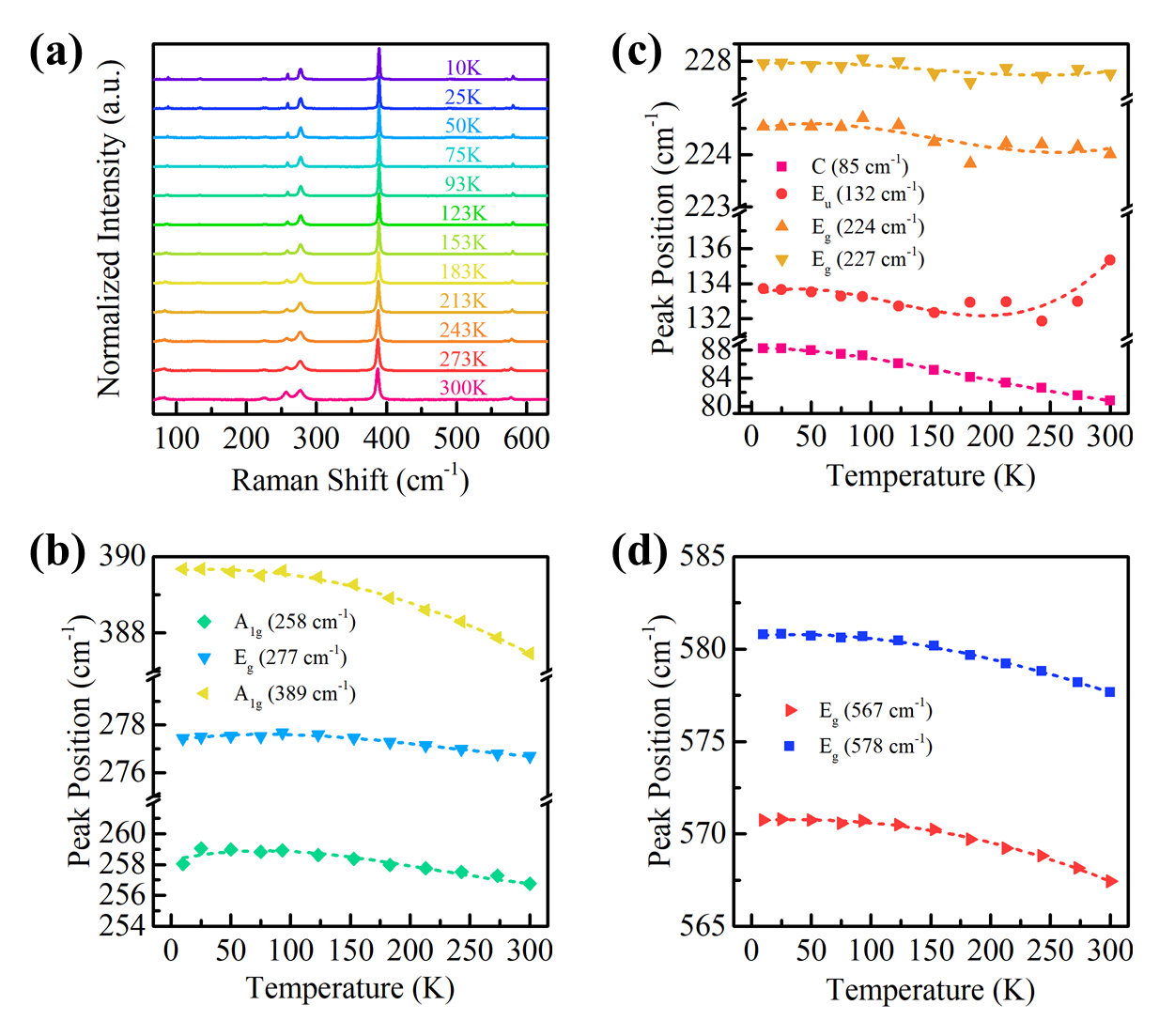}
    \caption{(a) Temperature-dependent Raman spectra measured from 300K to 10K. (b)-(d) Peak energies of all Raman modes as a function of temperature. Notations used in legends are explained in Fig. \ref{fig:schematic}d. Dashed lines show the theoretical fit to the data according to equation \ref{eq:energy_change}}
    \label{fig:Temp-Raman}
\end{figure}

Figure \ref{fig:Temp-Raman}a shows Raman cryogenic measurements and phonon modes of this material under ambient pressure. No phase transitions were reported as the temperature was dropped down from room temperature to 10K. We plot the peak positions for each vibrational mode as a function of temperature in Fig. \ref{fig:Temp-Raman}b-d. The phonon modes shift to higher energies as temperature reduces. Tighter chemical bonds in the material can be expected on account of thermal contraction, thus leading to an increase in the energy of vibrations. The nonlinear dependence of Raman peak positions on temperature can be empirically described by \cite{su2014dependence}: 

\begin{equation}
    E(\Delta T) = E_0 + \chi_1 \Delta T + \chi_2 (\Delta T)^2 + \chi_3 (\Delta T)^3
    \label{eq:energy_change}
\end{equation}

where \( E_0 \) is the frequency at room temperature, \( \Delta T \) is the temperature change relative to room temperature, and \( \chi_1 \) is the first-order temperature coefficient. The coefficients used for fitting the experimental data can be found in the Supporting Information (see Table S3, Supporting Information). The second (\( \chi_2 \)), third (\( \chi_3 \)), or higher-order temperature effects indicate the degree of non-linearity and are usually small. Our analysis of the Raman spectra of ZnPS$_3$ reveals that these coefficients (as well as the linear coefficient of fit) have very high values, with \( \chi_1 \) on the order of 10$^{-3}$, \( \chi_2 \) on the order of 10$^{-5}$, and \( \chi_3 \) on the order of $\sim$ 10$^{-7}$--10$^{-8}$ cm$^{-1}/$K$^{-1}$.  These values even exceed by an order of magnitude the highest values reported previously for monolayer and bulk MoS$_2$  \cite{su2014dependence}, where coefficients (\( \chi_2 \)) and third (\( \chi_3 \)) were found to be on the order of $10^{-5}$ and $10^{-8}$--10$^{-9}$ cm$^{-1}/$K$^{-1}$ for E$_g$ and A$_{1g}$ modes, respectively. While in MoS$_2$ the strong nonlinearity was attributed to a strain-induced inhomogeneity in mechanically exfoliated films, we believe that the observed nonlinearity in ZnPS$_3$ is an aspect of the strong thermal expansion found in this material. 

To verify this, we experimentally determined the crystal volume changes as a function of temperature in the 15-300K range using low-temperature X-ray diffraction (XRD). The experimental data are plotted in Fig. \ref{fig:DFT-QHA}a, alongside the DFT results, and show good agreement. The crystal volume was calculated using a Pawley fitting regime on the XRD data in HighScore Plus \cite{degen2014highscore} and was found to follow a linear relation with temperature (fit shown by solid aqua line). The respective XRD spectra at each temperature are shown in Fig. S6 (Supporting Information). The mean coefficient of thermal expansion was computed to be ($4.369 \pm 0.393) \times 10^{-5}$. Supplementary Fig. S4 shows the full width at half maximum (FWHM) for all the Raman peaks observed in Fig. \ref{fig:Temp-Raman}. FWHM is directly related to the lifetime of phonon modes, which in turn is influenced by two main phenomena: (i) phonon-phonon scattering -- which is the characteristic anharmonic decay of a phonon into two or more phonons, and (ii) phonon-carrier scattering -- caused by the perturbation in the crystal's translational symmetry due to the presence of defects. It is difficult to separate the relative contributions of these mechanisms to the overall lifetime and further studies would be required to investigate their individual contributions. In general, a monotonic decrease in the FWHM for all peaks is observed. The decrease occurs due to the increased thermal occupancy and interaction of phonons at higher temperatures \cite{beechem2008temperature}, which in turn increases the rate of phonon-phonon and carrier-phonon scattering.

\begin{figure}
    \centering
    \includegraphics[width=0.8\linewidth]{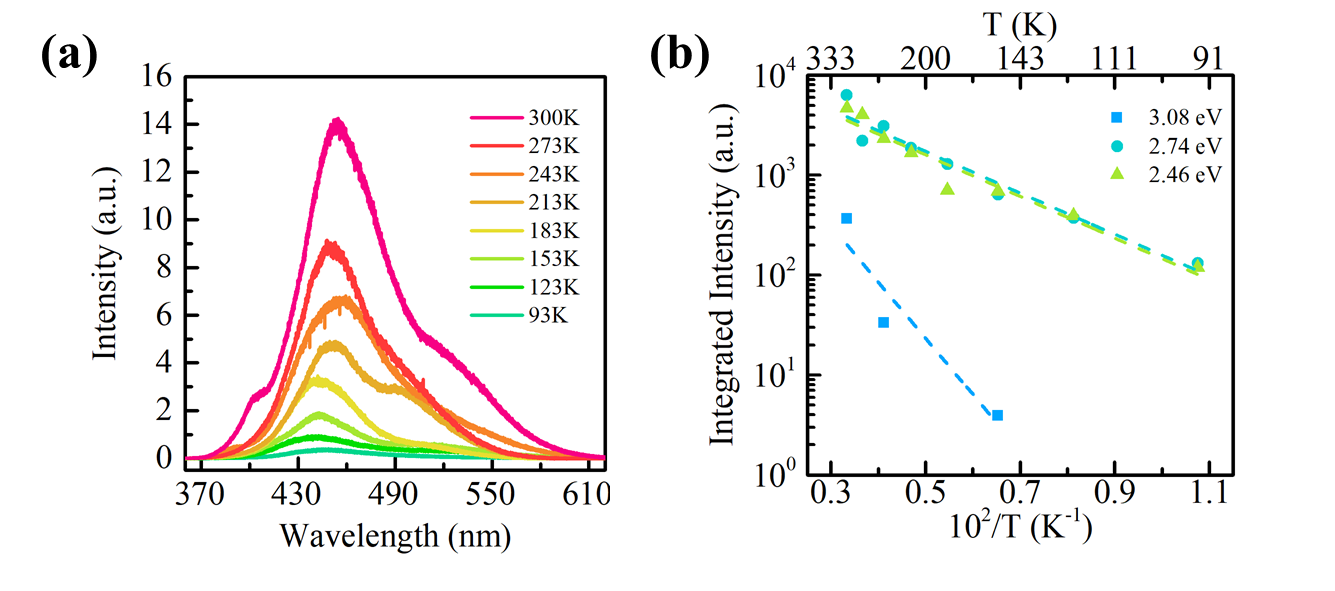}
    \caption{(a) PL spectra of ZnPS$_3$ at temperatures ranging from 300K to 93K. (b) Integrated PL intensities corresponding to deconvoluted luminescent bands (shown in Fig. \ref{fig:HP-PL}a) as a function of temperature. Dashed lines in (b) are for visual aid and do not represent any data fit. }
    \label{fig:Temp-PL}
\end{figure}

\am{In Figure \ref{fig:Temp-PL}, we investigate the influence of temperature on the PL emission in  ZnPS$_3$. As in the case of the high-pressure study, we observe PL quenching as the temperature is reduced from 300K to 93K. At any temperature, PL intensities are affected by competing radiative vs non-radiative processes. At room temperature (300K), carriers (electrons and holes) have sufficient thermal energy, which can activate non-radiative recombination processes. These thermally activated processes are often mediated by defects or material impurities, as seen in several semiconductors and lead to energy loss without the emission of photons. As the temperature decreases, the thermally activated non-radiative recombination pathways are suppressed because the carriers do not have enough energy to overcome the activation barrier for these processes. This would generally increase PL as non-radiative recombination decreases. However, the trend in our experimental data shows a decrease in PL, hence there must be other competing effects at play. }

\am{The unusually high Grüneisen parameter, as shown in Fig. \ref{fig:DFT-QHA}d, provides further evidence of the negative thermal quenching of photoluminescence. This shows that though phonon populations decrease with decreasing temperature, the remaining phonons may be more sensitive and can facilitate more efficient nonradiative recombination process even at low temperatures. Secondly, the Raman FWHM trends confirm the presence of enhanced phonon-phonon and phonon-carrier interactions. Thus, the eventual recombination of trapped carriers can be assisted by single-phonon- or multi-phonon-assisted processes, all of which would ultimately result in energy dissipation in the form of heat. Further, at lower temperatures, free carrier populations are sparse, and the remaining ones may not have sufficient mobility to recombine with the other carriers occupying the defect states, reducing the likelihood of any radiative processes.} 

\am{The individual integrated PL intensities (obtained by integrating each individual deconvoluted Gaussian curve across the spectrum shown in Fig. \ref{fig:HP-PL}a for the three luminescent bands are shown in Fig. \ref{fig:Temp-PL}b. As can be seen, PL bands caused by recombination of zinc interstitials quench at a different rate and slow down, with respect to decreasing temperature, in comparison to the 3.08 eV luminescent band caused by recombination of excited electrons with holes trapped by zinc vacancies. The peaks at 2.74 eV and 2.48 eV quench at a similar rate as the temperature is dropped. This can be explained by the valence band stability under temperature changes, similarly to the case of high pressure. Zn vacancies are shallow traps lying close to the valence band, which remain stable at all temperatures, not affecting the nature of quenching. On the other hand, zinc interstitials recombine with the valence band on zinc vacancies, but become less mobile with decreasing temperature, hence causing PL quenching at similar rates for these two luminescent bands. Interestingly, the luminescent band at 3.08 eV is not seen at all temperatures. Rather, it disappears and re-appears during the cycle. One way to explain this is the fact that Zn interstitials lie close to the conduction band and act as shallow trap states for the electrons excited after pumping with the near-UV laser. We posit that these states lie so close to the conduction band that even in a low thermal energy environment, the excited carriers can relax into interstitial states with minimal energy barriers, and thereafter they recombine either radiatively or nonradiatively. Hence, most excited carriers are efficiently trapped by interstitials, reducing their probability to recombine with Zn vacancies. This is indicated by the higher quenching rate of the 3.08 eV PL band, however, being a probabilistic process like any, it is difficult to determine the exact nature of this process and address the reason behind its reappearance. The capture rates of these defects acting as traps is an important aspect to be considered. Further, PL lifetime measurements can shed more light on the exact trapping/de-trapping mechanisms involved. While not common, the negative thermal quenching observed in this material is known to be exhibited by other semiconductors, including Zn-containing compounds such as ZnO \cite{he2008negative} and GaN:Si,Zn \cite{Reshchikov_etal}. A more detailed study of temperature-dependent ZnPS$_3$ PL in a broader range of temperatures, including those above ambient, might be interesting but is beyond the scope of our work.}

\section{Conclusion and Future Outlook}
\begin{figure}[h]
    \centering
    \includegraphics[width=0.75\linewidth]{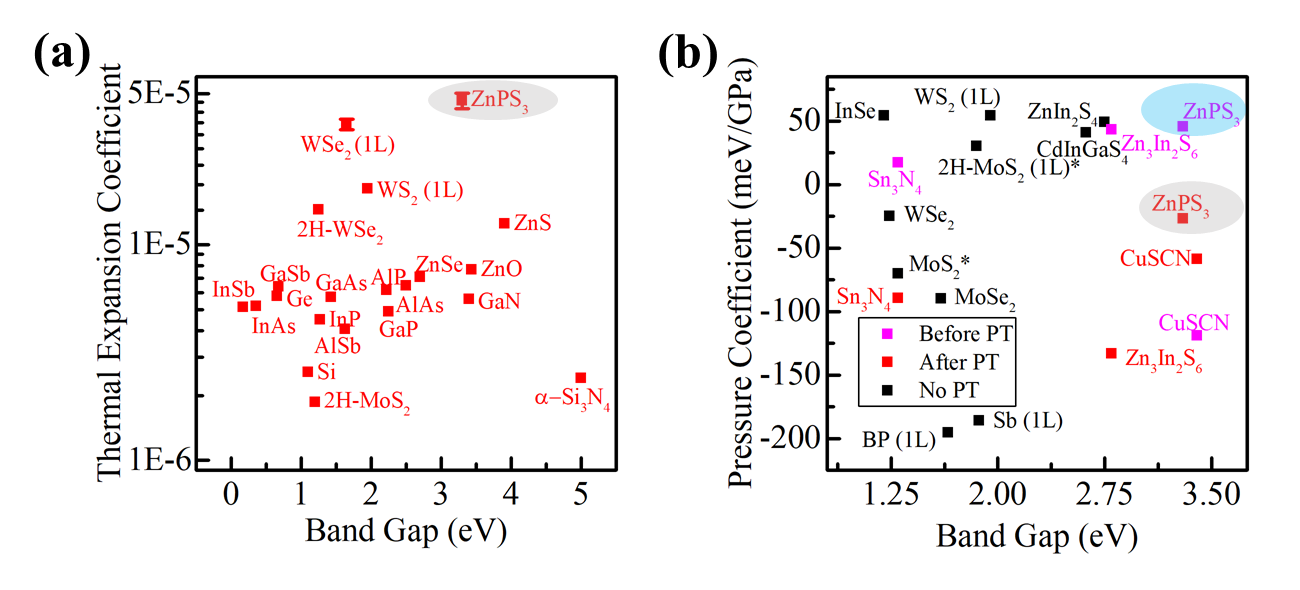}
    \caption{(a) Thermal expansion coefficients at room temperature for different bulk and 2D semiconductors and (b) Linear pressure coefficients of band gap energy (describing band gap change with pressure) for different semiconductors, plotted against their respective band gaps. Corresponding data and references can be found in Tables \ref{table:1} and \ref{table:2}.}
    \label{fig:Comparison}
\end{figure}

We have characterized the vibrational, optical, and electronic properties of ZnPS$_3$ in extreme high-pressure and cryogenic environments. A key takeaway from our measurements, which can prove to be pivotal for practical applications, is depicted in Fig.\ref{fig:Comparison}. We compare two properties that quantify the material's tunability with changes in its physical environment -- (i) the coefficient of thermal expansion (TEC), which is a manifestation of the structural changes in the lattice with respect to temperature, and (ii) the electronic band gap tunability, described by a linear coefficient in units of meV/GPa, which is indicative of changes in the electronic properties under compressive strain. The results obtained in this work are compared against the literature data for other semiconductors, and plotted against the electronic band gap as a common metric, in Figs. \ref{fig:Comparison}a and  \ref{fig:Comparison}b, corresponding to TEC and linear pressure coefficient, respectively. The data for other materials have been taken from references enlisted in Tables \ref{table:1} and \ref{table:2}, which also contain relevant data for calculations of TEC and pressure coefficients. The exact ranges of pressure and temperature within which these values were obtained (either via experiment or theory), and previous literature references for each material are shown in Tables \ref{table:1} and \ref{table:2}.

ZnPS$_3$ exhibits a remarkable combination of low-pressure tunability in its band gap along with a high TEC -- one of the highest among semiconductors and, to the best of our knowledge, the highest among 2D materials at 300K or lower temperatures. In Fig. \ref{fig:Comparison}b, we observe a sign change in the linear band gap coefficient after the phase transition at 6.75 GPa (reported in Fig. \ref{fig:HP-Raman}), but its absolute value remains relatively low, compared against materials such MoSe$_2$, Zn$_3$In$_2$S$_6$, and CuSCN, which show higher tunability in their band gaps either before, or after the phase transition or in both scenarios. We note that monolayer WSe$_2$ is another candidate that can potentially exhibit this characteristic, given its high TEC. Currently, pressure tunability data is available for only bulk WSe$_2$ to the best of our knowledge, hence it's not possible to provide an exact comparison at this point. The TEC for ZnPS$_3$ is marginally higher than that for a monolayer WSe$_2$ (Table \ref{table:1}, Fig. \ref{fig:Comparison}a) and its pressure coefficient is comparable to bulk WSe$_2$ (Table \ref{table:2}, \ref{fig:Comparison}b). Similarly, MoSe$_2$ has a lower pressure band gap tunability post its phase transition relative to ZnPS$_3$, but data on the thermal expansion of this material is not available in the literature given our current understanding. \\

A high TEC opens doors to a lot of potential applications. Having a high lattice deformation in an extremely wide temperature range can make this material a useful temperature sensor. The PL emission in the sample can be used as an active probe for temperature detection. Using 2D materials for temperature sensors in wearables is a hotspot for researchers today \cite{lim2023flexible}. Materials with a high TEC can also enhance the efficiency and responsiveness of thermal switches \cite{wehmeyer2017thermal}. Subject to further investigation of material stability and hysteresis under thermal cycling and mechanical loading, another potential application can be in the realm of MEMS-based microactuators. The cyclic reliability of TEC enhances the performance and sensitivity of a microactuator with precise thermal control. The most promising application, however, is in battery systems as this has been recently demonstrated in \cite{lv2024stable}. Many battery materials are being investigated for their performance at high pressures in the search for properties such as higher conductivity, or transitions into new phases that can better support battery design and performance in applications under extreme conditions \cite{yang2017oxygen,wang2021crystalline}. ZnPS$_3$ shows excellent stability under high-pressure conditions as shown in this research, with no phase transitions in the 15-100 GPa range. The relatively high band gap (3.3 eV) as measured in this study also provides the advantage of low charge leakage, which can be maintained at higher pressures as evidenced by the low band gap tunability -- an aspect advantageous from both an energy storage and battery systems perspective.
Similarly, probing the PL at various pressures is an additional functionality that can be used simultaneously for sensing, thus showing potential for applications where pressure monitoring is important such as aerospace and underwater systems. This work thereby offers fundamental insights into the photonic and electronic changes in ZnPS$_3$ under high pressure and low temperatures, providing a sophiscated roadmap for its applications under extreme conditions. It also charts further directions of research on this material, materials from the MPS$_3$ family, and other Zn-containing compounds. \\

There are further avenues that are yet to be explored. We prove the presence of defects via EDS in this material to explain the origin of PL, however, defect control is an important property missing from our work. This can be established with various physical and chemical techniques and is a long-standing research domain in 2D materials \cite{ahmad2013effect,donarelli2013tunable,liu2017temperature,fuchtbauer2013morphology,ueng1990defect}. In general, defect engineering reports on the metal thiophosphate family are lacking, with the only known reports being Zhong et al's work \cite{zhong2023intercalation} on enhancing the electrochemical performance of MnPS$_3$. The exact nature of defects (point/interstitial) in ZnPS$_3$, along with the potential presence of defect clusters/complexes, needs to be investigated via advanced techniques such as EPR (Electron Paramagnetic Resonance) and STM, which will provide deeper insights into the PL properties. On the computational side, simulations comprising point defects in the lattice structure can provide more insights into the exact changes in band structure, that can be then correlated with PL emission data. The confirmation of a single vs. double phase transition remains pending, which can be accomplished with additional high pressure characterization techniques. For future applications, changes in both the electronic and ionic conductivity are critical to study in various temperature, pressure, and other physical parameters. The versatility and intriguing properties demonstrated here further highlight the plethora of opportunities that lie within the MTP family of materials that can be exploited to enrich the rapidly growing community of 2D electronics.

\begin{table}[h!]
%\begin{table}[hbt!]
\centering
\small
\caption{Thermal expansion coefficients, their corresponding measurement techniques and temperature range, along with room temperature band gap values for various bulk and 2D semiconductors. \\}
\resizebox{\textwidth}{!}{
\begin{tabular}{cccccc}
\toprule
\multirow{2}{*}{\textbf{Material}} & \multirow{2}{*}{\textbf{Band Gap (eV)}} & \multirow{2}{*}{\shortstack{\textbf{Thermal Expansion} \\ \textbf{Coefficient}}} & \multirow{2}{*}{\shortstack{\textbf{Temperature} \\ \textbf{Range (K)}}} & \multirow{2}{*}{\shortstack{\textbf{Measurement} \\ \textbf{Technique}}} \\
 & & & & & \\
\midrule
GaN & 3.40 \cite{sharma2023gallium} & $5.59 \times 10^{-6} \cite{ravindra2017radiative}$ & 300-900K & X-ray diffraction \\
GaP & 2.25 \cite{kittel2018introduction} & $0.12-4.89 \times 10^{-6}$ \cite{haruna1986negative} & 60-293K & The Bond method \cite{bond1960precision} \\
GaAs & 1.43 \cite{kittel2018introduction} & $0.2-5.71 \times 10^{-6}$ \cite{novikova1966thermal} & 60-300K & Low temperature dilatometry\\
GaSb & 0.68 \cite{kittel2018introduction} & $0.5-6.37 \times 10^{-6}$ \cite{novikova1966thermal} & 60-300K & Low temperature dilatometry\\
Ge & 0.66 \cite{kittel2018introduction} & $0.07-5.75 \times 10^{-6}$ \cite{gibbons1958thermal} & 4.2-300K & Interferometric Method by Nix \& Macnair \cite{nix1941interferometric} \\
AlP & 2.50 \cite{ehsanfar2017first} & $0.28-6.44 \times 10^{-6}$ \cite{kagaya1987specific} & 50-300K & First Principles Calculation \\
AlAs & 2.22 (indirect) \cite{mao2014alattice} & $0.65-6.16 \times 10^{-6}$ \cite{kagaya1987specific} & 50-300K & First Principles Calculation \\
AlSb & 1.63 (indirect) \cite{yang2017structural} & $0.49-4.05 \times 10^{-6}$ \cite{novikova1966thermal} & 100-300K & Low temperature dilatometry \\
InAs & 0.36 \cite{kittel2018introduction} & $0.3-5.17 \times 10^{-6}$ \cite{sirota1985temperature} & 80-310K & X-ray spectroscopy \\
InP & 1.27 \cite{kittel2018introduction} & $0.03-4.38 \times 10^{-6}$ \cite{sirota1985temperature} & 90-310K & X-ray spectroscopy \\
InSb & 0.17 \cite{kittel2018introduction} & $1.5-5.13 \times 10^{-6}$ \cite{novikova1966thermal} & 80-300K & Low temperature dilatometry \\
CdTe & 1.44 \cite{kittel2018introduction} & $0.35-4.94 \times 10^{-6}$ \cite{novikova1966thermal} & 80-300K & Low temperature dilatometry \\
Si & 1.1 \cite{kittel2018introduction} & $1.02-2.56 \times 10^{-6}$ \cite{middelmann2015thermal} & 178-293K & Imaging Twyman-Green interferometry \\
$\alpha$-Si$_3$N$_4$ & 5.0 \cite{zhang2005optical} & $2.41 \times 10^{-6}$ \cite{niihara1986thermal} & 293-1273K & X-ray diffraction \\
ZnO & 3.44 \cite{kittel2018introduction} & $7.65 \times 10^{-6}$ \cite{reeber1970lattice} & 100-300K & Powder X-ray diffraction \\
ZnS & 3.91 \cite{kittel2018introduction} & $1.25 \times 10^{-5}$ \cite{reeber1967thermal} & 60-300K & Back‐reflection Debye‐Scherrer X‐ray camera \\
ZnSe & 2.7 \cite{hile2022zinc} & $1.29-7.05 \times 10^{-6}$ \cite{novikova1966thermal} & 80-300K & Low temperature dilatometry \\
2H-MoS$_2$ & 1.2 (indirect) \cite{radisavljevic2011single} & $0.49-1.86 \times 10^{-6}$ \cite{murray1979thermal} & 10-300K & X-ray diffraction \\
2H-WSe$_2$ & 1.25 (indirect) \cite{sharma1999optical} & $1.37-1.45 \times 10^{-5}$ \cite{murray1979thermal} & 10-300K & X-ray diffraction \\
WSe$_2$ (1L) & 1.65 \cite{liu2015chemical} & $(3.6 \pm 0.2) \times 10^{-5} \cite{kucinski2024direct}$ & 291-837K & Four-dimensional scanning transmission electron microscopy \\
WS$_2$ (1L) & 1.95 \cite{deng2018stability} & $0.26-1.81 \times 10^{-5} \cite{huang2016quantitative}$ & 100-300K & Raman spectroscopy \\
ZnPS$_3$ [This Work] & 3.3 & $(4.369 \pm 0.39) \times 10^{-5}$ & 15-300K & Powder X-ray diffraction \\
\bottomrule
\end{tabular}
}
\label{table:1}
\end{table}

%%%%%%%%%%%%%%%%%%%%%

%%%%%%%%%%%%

\begin{table}[h!]
%\begin{table}[hbt!]
\centering
\small
\caption{Ambient band gaps, linear pressure coefficients, and corresponding experimental/computational techniques for various semiconductors. PT = Phase Transition. \\ 
$^*$Band gap has a quadratic dependence on pressure. Hence, linear pressure coefficient was calculated from a quadratic fit to the data. \\}
\resizebox{\textwidth}{!}{
\begin{tabular}{c c c c c c c}
\toprule
\multirow{2}{*}{\textbf{Material}} & 
\multirow{2}{*}{\shortstack{\textbf{Ambient} \\ \textbf{Band Gap (eV)}}} & 
\multirow{2}{*}{\shortstack{\textbf{Linear Pressure} \\ \textbf{Coefficient (meV/GPa)}}} & 
\multirow{2}{*}{\shortstack{\textbf{Pressure} \\ \textbf{Range}}} & 
\multirow{2}{*}{\shortstack{\textbf{Experiment/} \\ \textbf{Theory?}}} & 
\multirow{2}{*}{\textbf{Method}} & 
\multirow{2}{*}{\shortstack{\textbf{Additional Comments} \\ \textbf{(if any)}}} \\
& & & & & & \\
\midrule
\multirow{2}{*}{InSe} & 
\multirow{2}{*}{1.20 \cite{olguin2003effect}} & 
\multirow{2}{*}{54.00} & 
\multirow{2}{*}{3-8 GPa} & 
\multirow{2}{*}{Theory \cite{olguin2003effect}} & 
\multirow{2}{*}{LAPW} & 
\multirow{2}{*}{\shortstack{No PT reported, but direct to indirect \\ band gap PT reported in [] at 5 GPa}} \\
 & & & & & & \\
ZnIn$_2$S$_4$ & 2.75 \cite{toyoda1993hydrostatic} & $49.00$ & 0-3.5 GPa & Expt \cite{toyoda1993hydrostatic} & Absorption spectroscopy & -- \\
\multirow{2}{*}{Zn$_3$In$_2$S$_6$} & 
\multirow{2}{*}{2.8 \cite{susilo2022tunable}} & 
43.00 (Before PT) & 0-12.88 GPa
 & 
\multirow{2}{*}{Expt \cite{susilo2022tunable}} & 
\multirow{2}{*}{Absorption spectroscopy} & \multirow{2}{*}{\shortstack{Semiconductor to metallic PT \\ at $\sim$ 13 GPa.}}
\\
 &  
 & 
$-133.00$ (After PT) & 
13.95-37 GPa & 
 &  
 & 
 \\
CdIn$_2$S$_4$ & 2.62 (indirect) \cite{toyoda1993hydrostatic} & 41.00 & 0-3.5 GPa & Expt \cite{toyoda1993hydrostatic} & Absorption spectroscopy & --\\ 
\multirow{2}{*}{WSe$_2$} & 
\multirow{2}{*}{1.24 \cite{shen2017linear}} & 
\multirow{2}{*}{-25.00} & 
\multirow{2}{*}{0-50.9 GPa} & 
\multirow{2}{*}{Expt \cite{shen2017linear}} & 
\multirow{2}{*}{Absorption spectroscopy} & 
\multirow{2}{*}{\shortstack{Band gap closes at 50.9 GPa, followed \\ by an isostructural transition at 51.7 GPa.}} \\
 & & & & & & \\
MoSe$_2$ & 
1.60 \cite{zhao2015pressure} & 
-90.00$^*$ & 
0-40.7 GPa & 
Expt \cite{zhao2015pressure} & 
IR spectroscopy & Metallizes between 28-40 GPa.
\\
BP (1L) & 1.65 \cite{dai2020benchmark} & -195.43 & 0-8 GPa & Theory \cite{dai2020benchmark} & VASP-HSE06 (V-H) & -- \\ 
Sb (1L) & 1.87 (indirect) \cite{dai2020benchmark} & $-185.99$ & 0-4 GPa & Theory \cite{dai2020benchmark} & VASP-HSE06 (V-H) & PT predicted between 3-5 GPa. \\ 
MoS$_2$ & 1.30 \cite{nayak2014pressure} & -70.00$^*$ & 0-21 GPa & Theory \cite{nayak2014pressure} & VASP-PBE-GGA & Experiments prove PT to metal at $\sim$ 21 GPa. \\ 
\multirow{2}{*}{Sn$_3$N$_4$} & 
\multirow{2}{*}{1.30 (indirect) \cite{kearney2018pressure}} & 
17.00 (Before PT) & 
50-102 GPa & 
\multirow{2}{*}{Expt \cite{kearney2018pressure}} & 
\multirow{2}{*}{Absorption spectroscopy} & \multirow{2}{*}{Direct band gap opens at 50 GPa.}
\\
 &  
 & 
-89.50 (After PT) & 
128.5-231 GPa &  
 &  
 & 
\\
\multirow{2}{*}{CuSCN} & 
\multirow{2}{*}{3.40 \cite{yang2022pressure}} & 
-118.98 (Before PT) & 
0-3.9 GPa & 
\multirow{2}{*}{Expt \cite{yang2022pressure}} & 
\multirow{2}{*}{IR spectroscopy} & \multirow{2}{*}{PT from $\alpha$-$\beta$ phase.}
\\
 &  
 & 
$-58.68$ (After PT) & 
3.9-12.7 GPa &  
 &  
 & 
 \\
2H-MoS$_2$ (1L) & 1.85 \cite{nayak2015pressure} & 30.00$^*$ & 0-16 GPa & Expt \cite{nayak2015pressure} & PL spectroscopy & Metallization predicted at $\sim$ 68 GPa. \\ 
WS$_2$ (1L) & 1.95 \cite{kim2017towards} & 54.00 & 0-4 GPa & Expt \cite{kim2017towards} & PL spectroscopy & Direct to indirect PT calculated at $\sim$ 21 GPa. \\ 
\multirow{2}{*}{ZnPS$_3$ [This Work]} & 
\multirow{2}{*}{3.3} & 
45.52 (Before PT) & 
0-5 GPa & 
\multirow{2}{*}{Theory} & 
\multirow{2}{*}{VASP-PBE-GGA} & \multirow{2}{*}{\shortstack{PT at 6.75 GPa. Metallization \\ predicted at 100 GPa.}}
\\
 &  
 & 
$-27.08$ (After PT) & 
5-25 GPa &  
 &  
 & 
 \\
\bottomrule \\
\end{tabular}
}
\label{table:2}
\end{table}

\clearpage

\section*{Acknowledgements}
This work has been supported by the MIT-Poland Lockheed Martin Seed Fund, the ARO MURI Grant No. W911NF-19-1-0279, and MIT-Tec de Monterrey program. Abhishek Mukherjee appreciates the support provided by the Siebel Scholarship and the MIT Mathworks Engineering Fellowship. Michael A. Susner acknowledges the support of the Air Force Office of Scientific Research (AFOSR) Grant No. LRIR 23RXCOR003 and AOARD MOST Grant No. F4GGA21207H002 as well as general support from the Air Force Materials and Manufacturing (RX) and Aerospace Systems (RQ) Directorates. This work was also partially supported by the Polish National Science Center SHENG-2 Grant No. 2021/40/Q/ST5/00336. The authors acknowledge the MIT SuperCloud and Lincoln Laboratory Supercomputing Center for providing HPC resources that have contributed to the research results reported within this article. Additionally, the authors appreciate the facilities provided by the MIT Materials Research Laboratory, Institute of Soldier Nanotechnologies, and MIT.nano for conducting this research.

\section*{Conflict of Interest}
The authors declare no conflict of interest.

\clearpage

\captionsetup[figure]{labelfont=bf,labelformat=simple,labelsep=period,name={Fig. S}}

\captionsetup[table]{labelfont=bf,labelformat=simple,labelsep=period,name={Table S}}

%Header
\pagestyle{fancy}
\thispagestyle{empty}
\rhead{ \textit{ }} 

% Update your Headers here
\fancyhead[LO]{}
% \fancyhead[RE]{Firstauthor and Secondauthor} % Firstauthor et al. if more than 2 - must use \documentclass[twoside]{article}
  
%% Title
% \title{Supporting Information \\ \vspace{0.5cm} Thermal and dimensional stability of photocatalytic material ZnPS$_3$ under extreme environmental conditions
% }
% % Adjust author font to be bold
% \renewcommand\Authfont{\bfseries\normalsize}  % Make author names bold
% % Adjust affiliation font to be normal (non-bold) and left-aligned
% \renewcommand\Affilfont{\normalfont\raggedright}

% \author[1,$\S$]{Abhishek Mukherjee}
% \author[1,2,$\S$]{Vivian J. Santamar\'{i}a-Garc\'{i}a}
% \author[3]{Damian Wlodarczyk}
% \author[3]{Ajeesh K. Somakumar}
% \author[3]{Piotr Sybilski}
% \author[4]{Ryan Siebenaller}
% \author[5]{Emmanuel Rowe}
% \author[3]{Saranya Narayanan}
% \author[4]{Michael A. Susner}
% \author[2]{L. Marcelo Lozano-Sanchez}
% \author[3]{Andrzej Suchocki}
% \author[6]{Julio L. Palma}
% \author[1,*]{Svetlana V. Boriskina}

% \affil[1]{Massachusetts Institute of Technology, Cambridge, MA 02139, USA.}
% \affil[2]{Tecnologico de Monterrey, Escuela de Ingeniería y Ciencias, Monterrey 64849, Mexico.}
% \affil[3]{Institute of Physics, Polish Academy of Sciences, Warsaw 02-668, Poland.}
% \affil[4]{Air Force Research Laboratory, Wright-Patterson AFB, OH 45433, USA.}
% \affil[5]{National Research Council, Washington, D. C. 20001, USA.}
% \affil[6]{Department of Chemistry, Penn State University, Lemont Furnace, PA 15456, USA.}
% \affil[$\S$]{Co-first authors}
% \affil[*]{sborisk@mit.edu}

% \maketitle

% \appendix
\section*{{\Large{Supporting Information: Thermal and dimensional stability of photocatalytic material ZnPS$_3$ under extreme environmental conditions}}}

\renewcommand\thesection{S\arabic{section}}
\setcounter{section}{0}
\setcounter{figure}{0}
\setcounter{table}{0}

\section{Material Characterization}
\begin{figure}[h]
    \centering
    \includegraphics[width=0.7\linewidth]{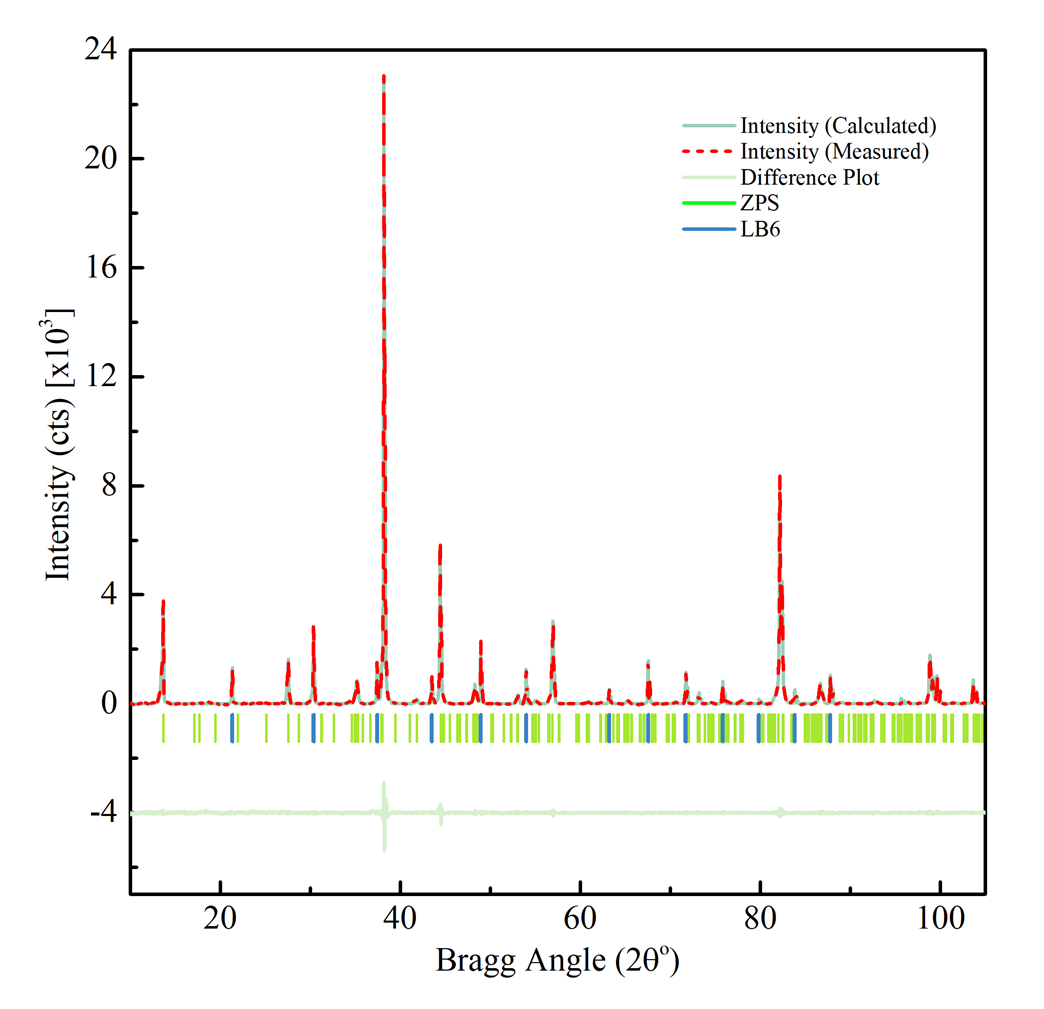}
    \caption{XRD data and corresponding Pawley fit of ZnPS$_3$ at room temperature. The aqua line represents the calculated fit, the red dashed line is the collected data, and the light orange line represents the difference between the two. The green and pink dashes mark the expected Bragg positions for ZnPS$_3$ (ZPS) and LaB$_6$ (LB6), respectively. Here, we used Cu K$\alpha = 1.5406$ \AA.}
    \label{fig:XRD}
\end{figure}

X-ray diffraction was done with a powdered sample to verify the structure with previous reports. Additionally, a powdered sample of the thermal standard LaB$_6$ (Lanthanum Hexaboride) was added with ZnPS$_3$ to have a reference material with known thermal expansion to account for specimen displacement error during Pawley fitting of XRD spectra. The thermal expansion for LaB$_6$ was determined to be 6.4 $\pm$ 0.8 $\times$ 10$^{-6}$, in close agreement with the known value of 7 $\times$ 10$^{-6}$ \cite{chen2004structural}.  The experiments were performed with a Cu source at all temperatures. The ambient-conditions spectrum and its corresponding fit can be seen in Fig. \ref{fig:XRD}, and cryogenic measurements are shown in Fig. \ref{fig:temp-XRD}. The Pawley fit agrees well with the expected monoclinic structure ${C2/m}$ as can been. The lattice constants, at ambient conditions, were found to be as follows:  $a =$ 5.9799 \AA, $b = $10.3473 \AA, $c = $6.806 \AA, and $\beta = $ 106.9760$^\circ$. These are in good agreement with previous literature \cite{martinolich2019solid}.

Additionally, bulk crystals were characterized using EDS to verify the presence of Zinc defects. The corresponding data is shown in Table \ref{table:EDS}.

% Adjust the space between columns
\setlength{\tabcolsep}{4pt} % Default is 6pt, adjust as needed
\vspace*{0.75cm}
\begin{table}[h]
\centering
\caption{EDS on ZnPS$_3$ sample. Data was taken on and mapped over a sample containing 2-3 crystallites.}
\label{table:EDS}
\resizebox{\textwidth}{!}{%
\begin{tabular}{>{\centering\arraybackslash}p{2.5cm} >{\centering\arraybackslash}p{2.5cm} >
{\centering\arraybackslash}p{2.5cm} >{\centering\arraybackslash}p{2.5cm} >{\centering\arraybackslash}p{2.5cm} >{\centering\arraybackslash}p{2.5cm}}
\toprule
\textbf{Element} & \textbf{Apparent Concentration} & \textbf{k Ratio} & \textbf{Weight\%} & \textbf{Weight\% $\sigma$} & \textbf{Atomic \%} \\
\midrule
P & 19.63 & 0.10982 & 11.64 & 0.24 & 7.16 \\
S & 37.58 & 0.32371 & 36.67 & 0.68 & 21.79 \\
Zn & 10.14 & 0.06925 & 9.80 & 0.23 & 5.31 \\
\bottomrule \\
\end{tabular}%
}
%\addvspace{20pt}
\end{table}

\begin{figure}
    \centering
    \includegraphics[width=1\linewidth]{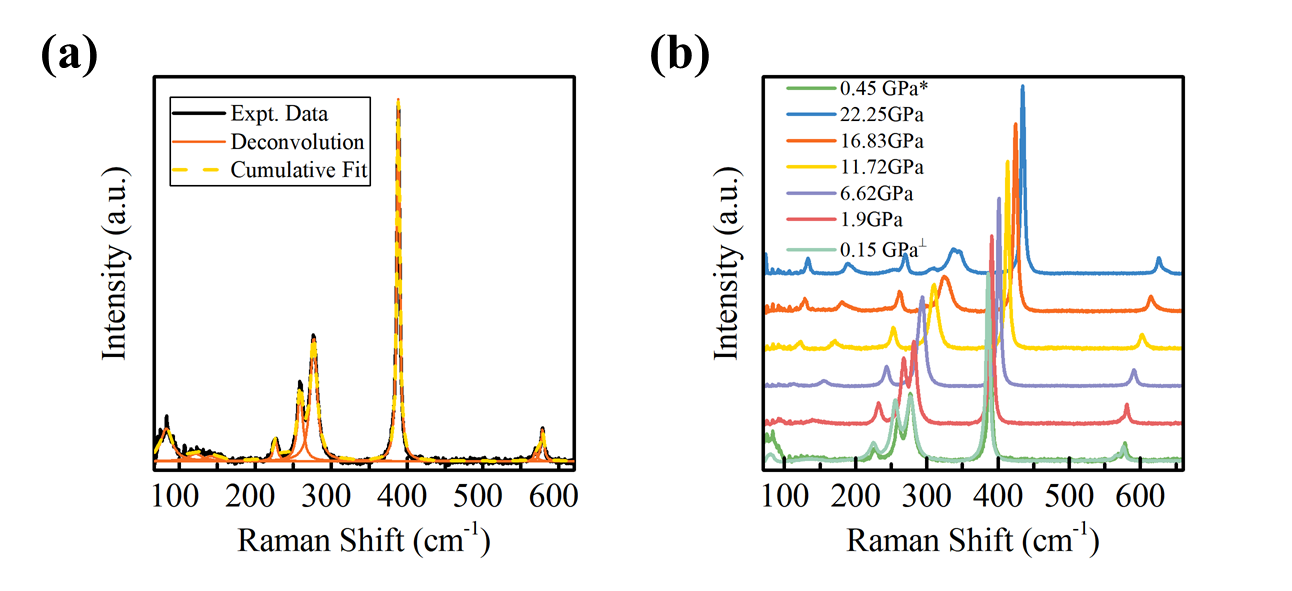}
    \caption{(a) Deconvolution of the Raman spectra at 0.45 GPa using Lorentzian curves. (b) Raman spectra evolution during pressure release from 22.25 to 0.15 GPa, compared with the starting point of compression at 0.45 GPa. * and $\perp$ denote the beginning and end of the pressure cycle, respectively. The spectra at 0.45 GPa and 0.15 GPa agree reasonably well, as shown by comparing the Raman modes obtained from line fitting in Table S\ref{table:raman_modes}.}
    \label{fig:enter-label}
\end{figure}

\begin{table}[h!]
\centering
\caption{Raman mode frequencies at ambient temperature reported for the start and end points of the pressure cycle. Modes \(M_4\) and other shoulder peaks originated during mid-cycle and were not observed here.}
\label{table:raman_modes}
\begin{tabular}{cccccccccc}
\toprule
\textbf{Assigned Mode (cm\(^{-1}\))} & \(M_1\) & \(M_2\) & \(M_3\) & \(M_5\) & \(M_6\) & \(M_7\) & \(M_8\) & \(M_9\) & \(M_{10}\) \\
\midrule
\textbf{Cycle Start (0.45 GPa)} & 83.14 & 121.66 & 141.99 & 226.75 & 259.33 & 277.13 & 388.56 & 569.16 & 578.43 \\
\textbf{Cycle End (0.15 GPa)} & 80.18 & 129.72 & 142.24 & 224.48 & 255.64 & 277.09 & 385.98 & 567.26 & 576.18 \\
\bottomrule \\
\end{tabular}
\end{table}

%\begin{figure}
%    \centering
%    \includegraphics[width=0.4\linewidth]{Fig2.png}
%    \caption{Integrated PL intensity calculated from PL spectral measurements in the range 350-700 nm during compression cycle from 0.45 GPa to 24.5 GPa.}
%    \label{fig:enter-label}
%\end{figure}

\clearpage
\newpage
\begin{figure}
    \centering
    \includegraphics[width=0.8\linewidth]{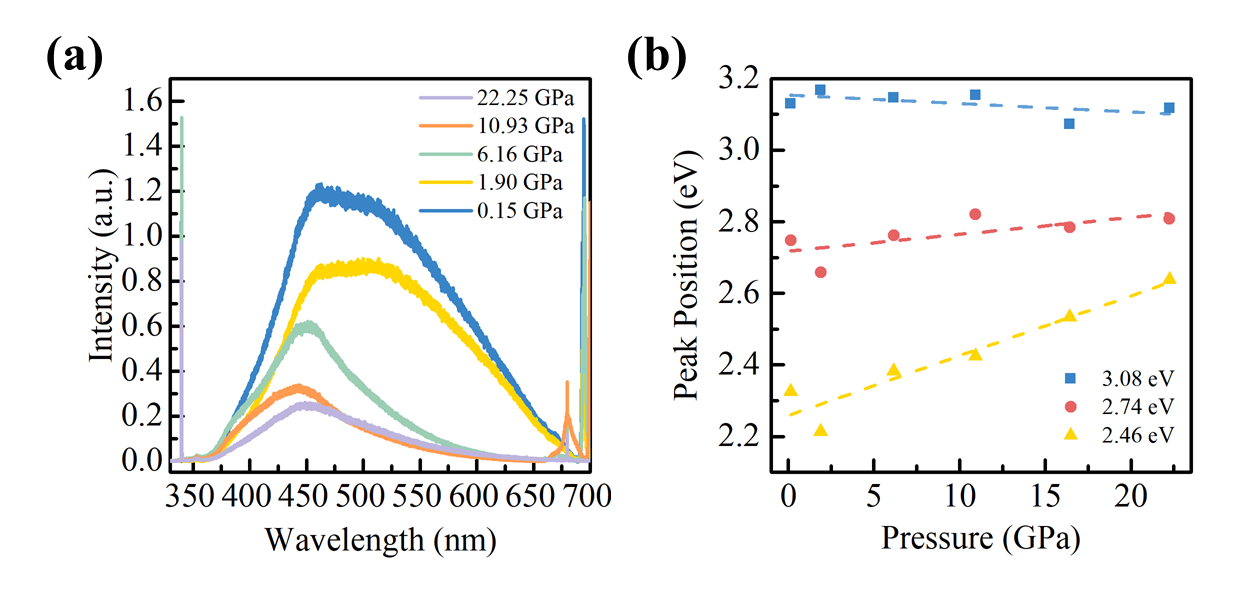}
    \caption{(a) PL spectra as a function of decreasing pressure (relaxation cycle) from 22.25 GPa to 0.15 GPa. (b) Corresponding peak positions of the three luminescent bands described in the main text. }
    \label{fig:enter-label}
\end{figure}

\begin{figure}
    \centering
    \includegraphics[width=0.8\linewidth]{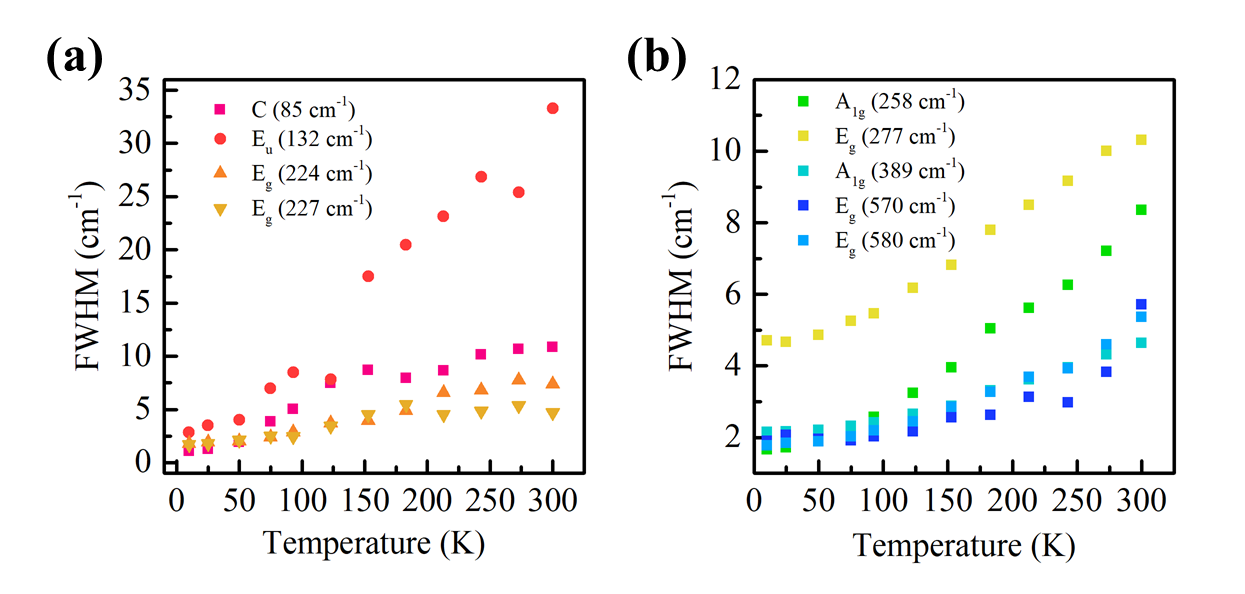}
    \caption{Full Width at Half Maximum (FWHM) of the Raman modes (shown in Figs. 1,8 in main text) as a function of temperature.}
    \label{fig:enter-label}
\end{figure}

\begin{figure}
    \centering
    \includegraphics[width=0.8\linewidth]{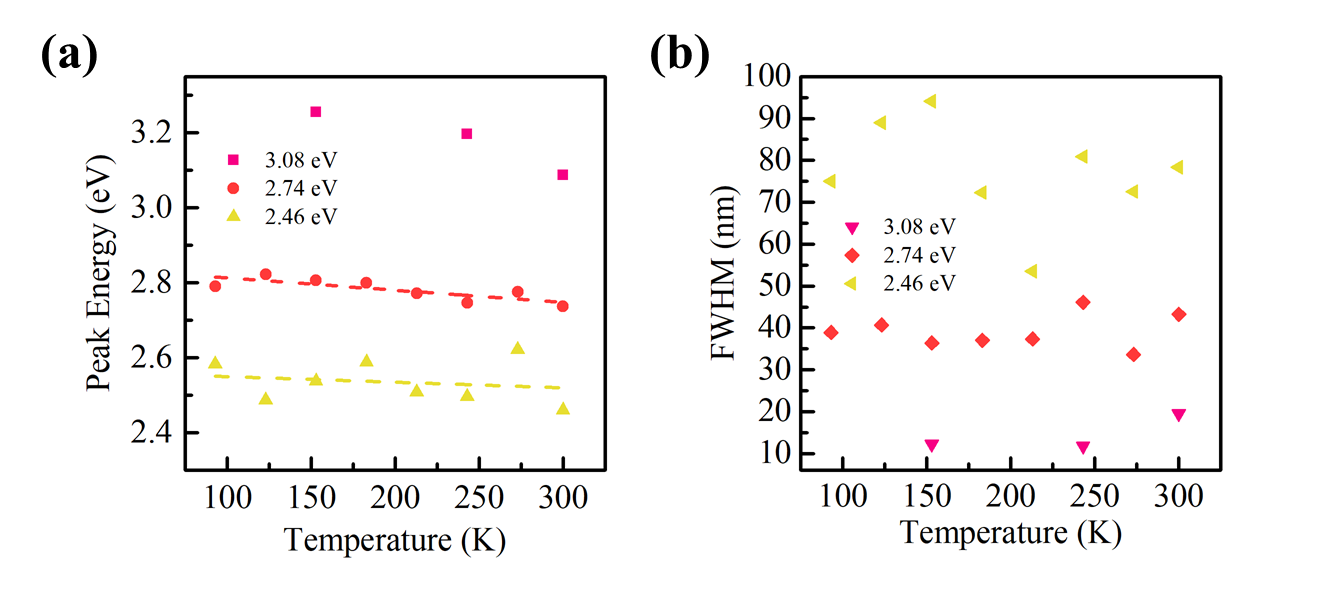}
    \caption{(a) Peak energies and (b) Peak FWHM values corresponding to the deconvoluted luminescent bands (shown in Fig. 9a of main text) as a function of temperature. Dashed lines in (a) show the corresponding linear fits to the data.}
    \label{fig:enter-label}
\end{figure}

\begin{table}[h!]
\centering
\caption{Fitted Coefficients and their respective errors derived from temperature-dependent Raman spectroscopy (Fig. 6 in main text).}
\label{tab:coefficients}
\begin{tabular}{ccccc}
\toprule
Mode & $E_0$ & $\chi_1$ & $\chi_2$ & $\chi_3$ \\
\midrule
C   & 80.826 & -0.023 $\pm$ 0.003  & (9.293 $\pm$ 2.587) $\times 10^{-5}$  & (2.846 $\pm$ 0.625) $\times 10^{-8}$ \\
E$_u$    & 135.342 & 0.069 $\pm$ 0.01  & (4.580 $\pm$ 0.958) $\times 10^{-4}$  & (8.340 $\pm$ 2.317) $\times 10^{-7}$ \\
E$_g$   & 224.126 & 0.004 $\pm$ 0.003  & (5.373 $\pm$ 3.007 $\times 10^{-5}$  & (1.214 $\pm$ 0.727) $\times 10^{-7}$ \\
E$_g$    & 227.718 & 0.004 $\pm$ 0.003  & (4.354 $\pm$ 2.763) $\times 10^{-5}$  & (9.146 $\pm$ 6.681) $\times 10^{-8}$ \\
A$_{1g}$    & 256.764 & -0.009 $\pm$ 0.004  & (5.162 $\pm$ 4.136) $\times 10^{-5}$ & (2.106 $\pm$ 1.000) $\times 10^{-7}$ \\
E$_g$   & 276.679 & (-4.520 $\pm$ 0.862) $\times 10^{-3}$  & (1.643 $\pm$ 0.869) $\times 10^{-5}$  &(8.065 $\pm$ 0.210) $\times 10^{-7}$ \\
A$_{1g}$   & 387.459 & -0.016 $\pm$ 0.001  & (-2.753 $\pm$ 1.035) $\times 10^{-5}$  & (0.643 $\pm$ 2.503) $\times 10^{-8}$ \\
E$_g$   & 567.442 & -0.026 $\pm$ 0.002  & (-5.626 $\pm$ 1.468) $\times 10^{-5}$  & (-1.529 $\pm$ 3.547) $\times 10^{-8}$ \\
E$_g$   & 577.660 & -0.021 $\pm$ 0.001  & (-2.238 $\pm$ 1.145) $\times 10^{-5}$  & (4.193 $\pm$ 2.770) $\times 10^{-8}$ \\
\bottomrule \\
\end{tabular}
\end{table}

\begin{figure}
    \centering
    \includegraphics[width=0.7\linewidth]{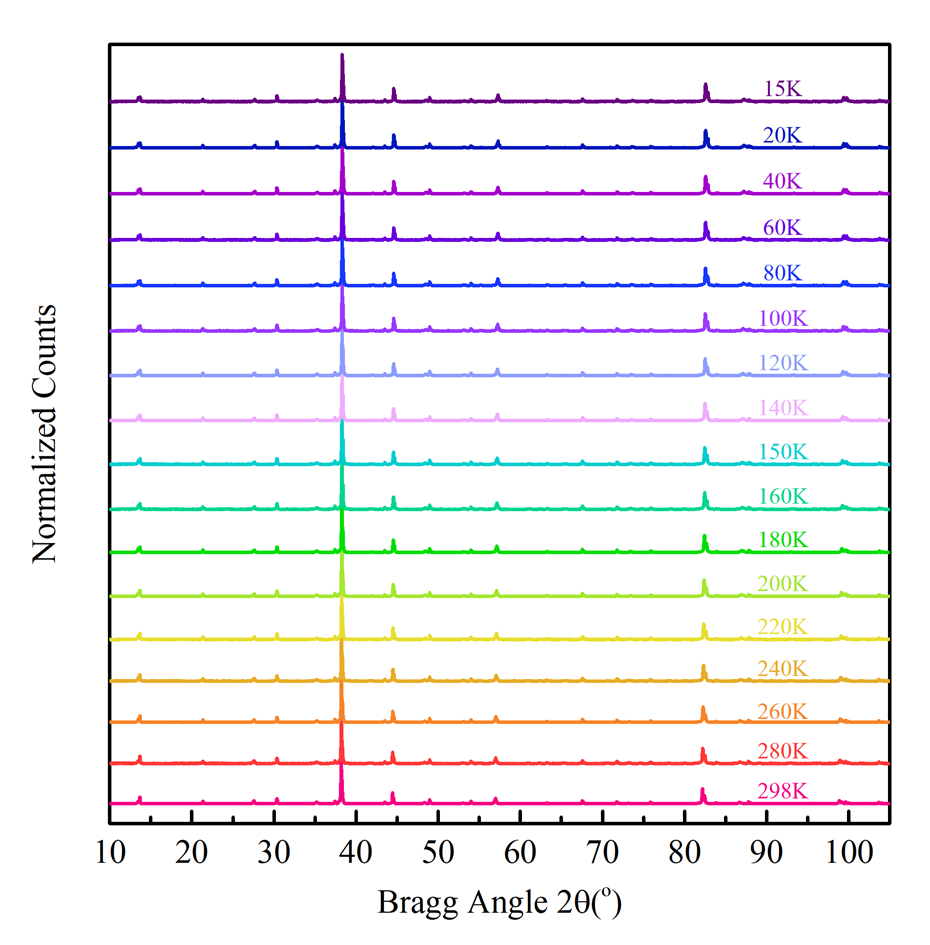}
    \caption{XRD Spectra for LaB$_6$ + ZnPS$_3$ powdered samples from 300K to 15K.}
    \label{fig:temp-XRD}
\end{figure}

\begin{figure}
    \centering
    \hspace*{-0.06\textwidth} % Adjust the value as needed to shift the figure left
    \includegraphics[width=1.1\linewidth]{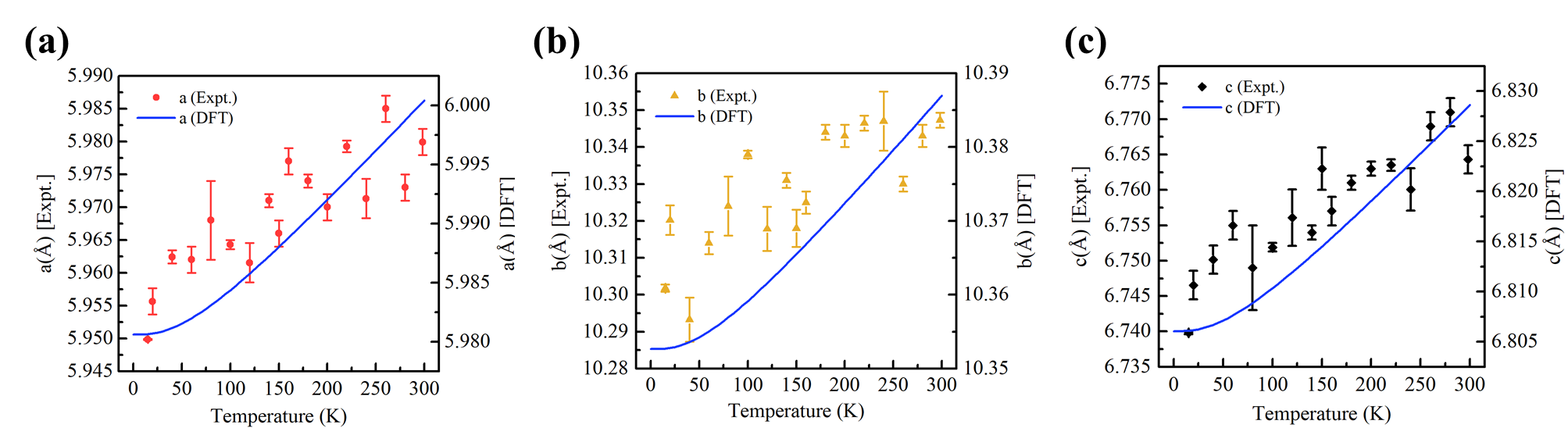}
    \caption{Lattice constants derived from Pawley fitting on XRD data (left y-axis) and DFT (right y-axis) as a function of temperature.}
    \label{fig:enter-label}
\end{figure}

\begin{figure}
    \centering
    \includegraphics[width=0.4\linewidth]{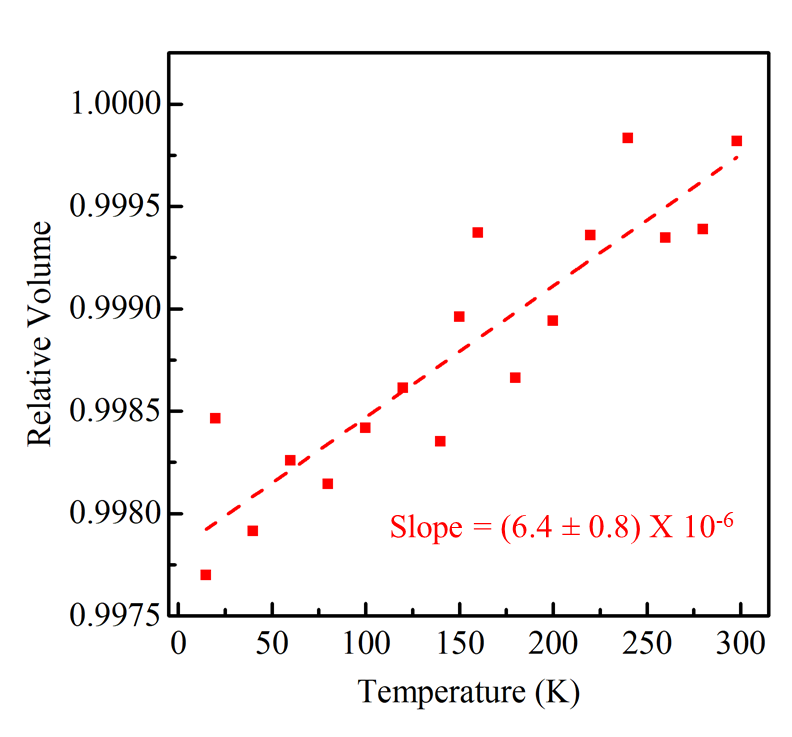}
    \caption{Thermal expansion determination of LaB$_6$.}
    \label{fig:LaB6}
\end{figure}

\clearpage
\newpage
\section{First principles calculations}
\subsection{Electronic Properties}
\begin{figure} [h]
    \centering
    \includegraphics[width=0.75\linewidth]{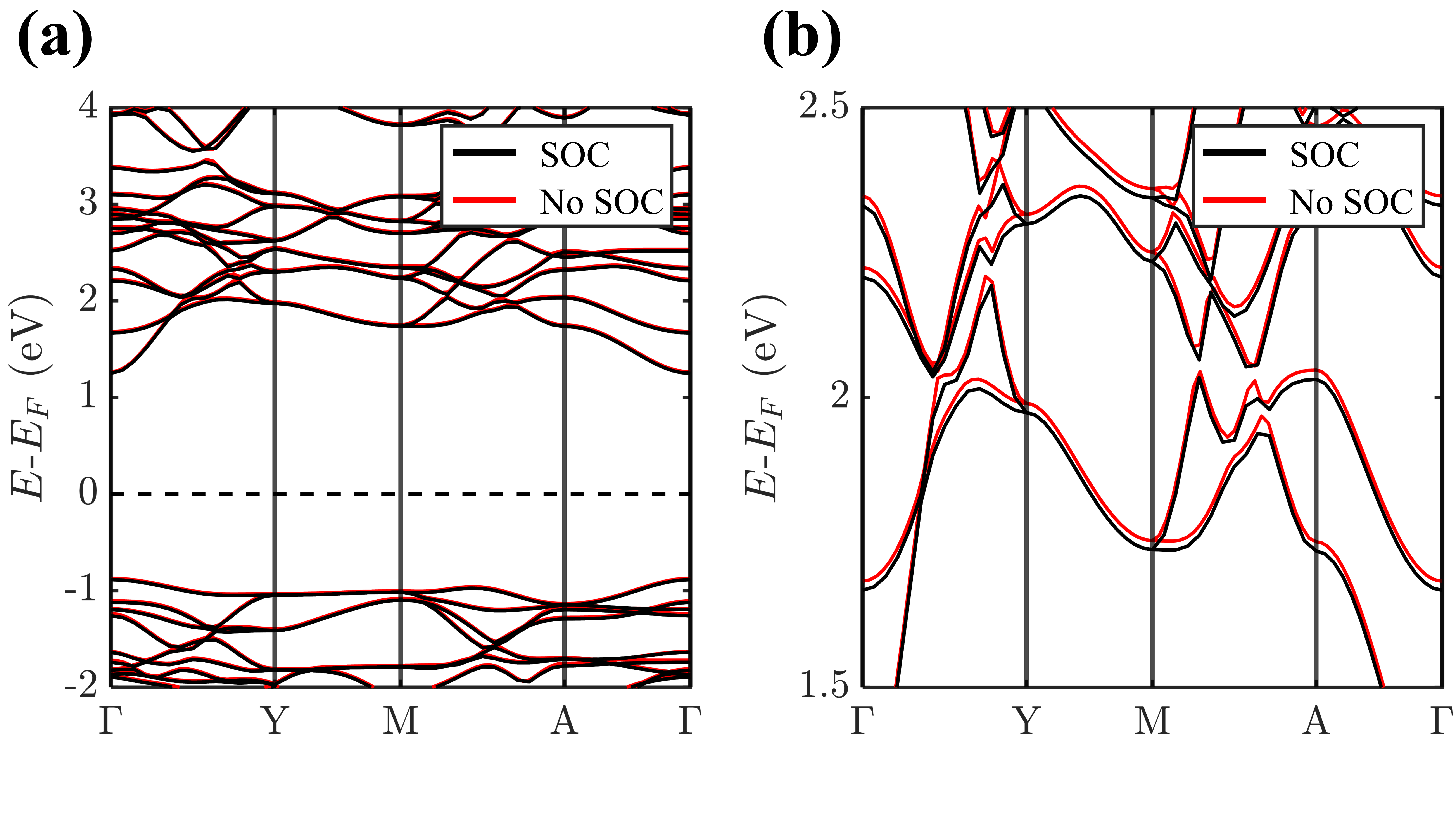}
    \caption{Comparison of the electronic band structure of ZnPS$_3$ with and without spin-orbit coupling (SOC). (a) The electronic band structure shows no significant changes between calculations with SOC (black lines) and without SOC (red lines). (b) A zoomed-in view of the band structure highlighting the region where both calculations align closely, further demonstrating the minimal impact of SOC.}
    \label{fig:SOC}
\end{figure}

\clearpage
\newpage
\begin{figure}
    \centering
    \includegraphics[width=1\linewidth]{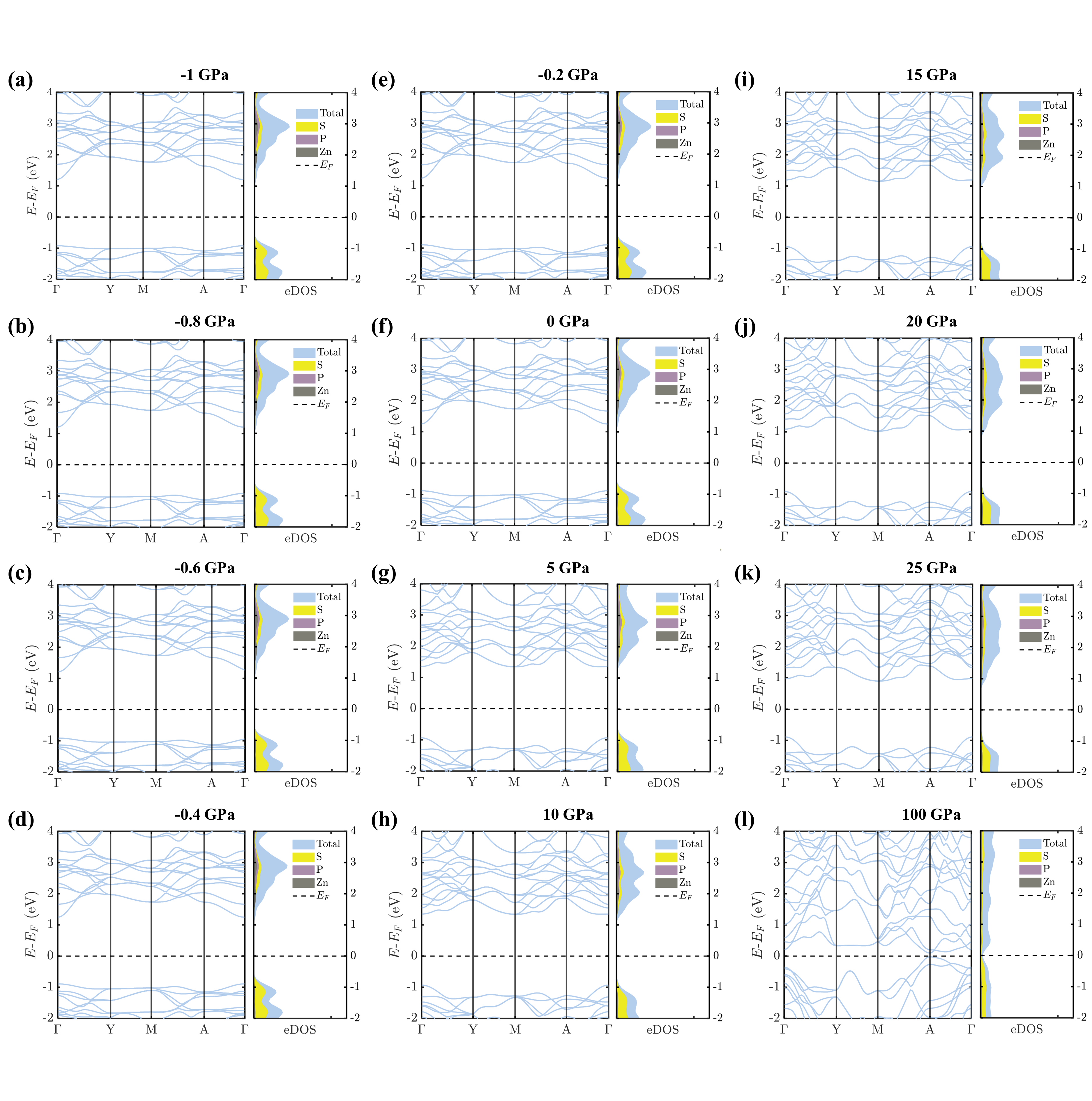}
    \caption{Electronic band structure and density of states at (a) -1 GPa, (b) -0.8 GPa, (c) -0.6 GPa, (d) -0.4 GPa, (e) -0.2 GPa, (f) 0 GPa, (g) 5 GPa, (h) 10 GPa, (i) 15 GPa, (j) 20 GPa, (k) 25 GPa, and (l) 100 GPa.}
    \label{eBS_all}
\end{figure}

\clearpage
\newpage
\subsection{Vibrational Frequencies}
\begin{figure}[h]
    \centering
    \includegraphics[width=1\linewidth]{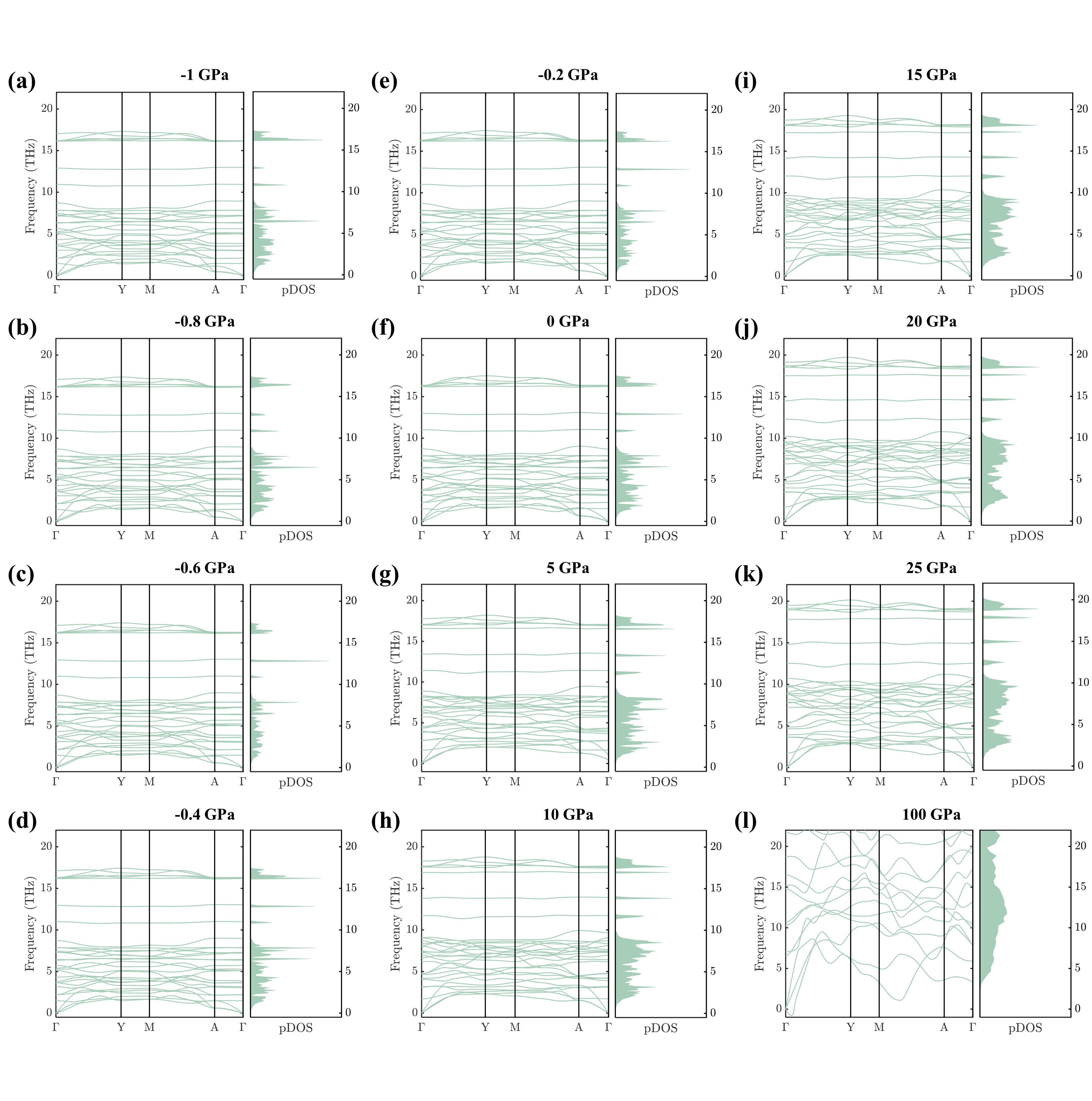}
    \caption{ Phonon dispersion and density of states at (a) -1 GPa, (b) -0.8 GPa, (c) -0.6 GPa, (d) -0.4 GPa, (e) -0.2 GPa, (f) 0 GPa, (g) 5 GPa, (h) 10 GPa, (i) 15 GPa, (j) 20 GPa, and (k) 25 GPa.}
    \label{pBS_all}
\end{figure}

\begin{figure}
    \centering
    \includegraphics[width=.85\linewidth]{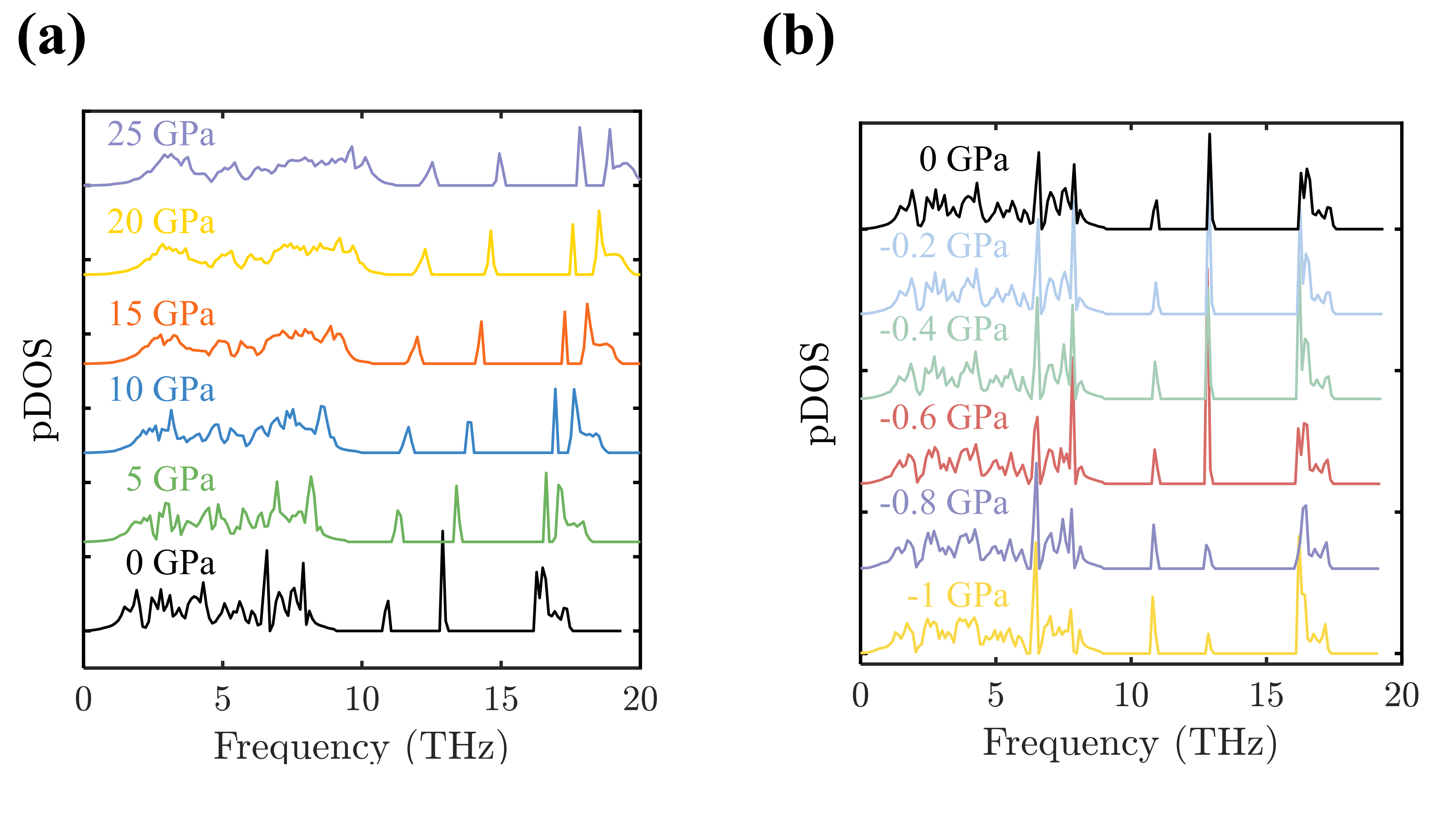}
    \caption{ Phonon density of states evolution under (a) compressive and (b) tensile hydrostatic pressure}
    \label{pBS_all}
\end{figure}

\begin{table}[h!]
\centering
\caption{DFT Character table at multiple pressures}
\label{tab:Character Table Multiple Pressures}
\begin{tabular}{cccccccccccc}
\toprule
\multicolumn{3}{c}{-1.0 GPa} & \multicolumn{3}{c}{-0.8 GPa} & \multicolumn{3}{c}{-0.6 GPa} & \multicolumn{3}{c}{-0.4 GPa}\\
\cmidrule(lr){1-3} \cmidrule(lr){4-6} \cmidrule(lr){7-9}\cmidrule(lr){10-12}
Mode & THz & $cm^{-1}$ & Mode & THz & $cm^{-1}$ & Mode & THz & $cm^{-1}$ & Mode & THz & $cm^{-1}$\\
\midrule
Bg	&	1.44	&	48.03	&	Bg	&	1.45	&	48.23	&	Au	&	0.00	&	0.00	&	Bu	&	0.00	&	0.00	\\
Ag	&	2.08	&	69.38	&	Bg	&	2.13	&	70.92	&	Bg	&	1.48	&	49.47	&	Bu	&	0.00	&	0.00	\\
Bg	&	2.08	&	69.41	&	Ag	&	2.13	&	71.05	&	Bg	&	2.17	&	72.32	&	Au	&	0.00	&	0.00	\\
Bu	&	2.97	&	99.17	&	Bu	&	3.00	&	100.07	&	Ag	&	2.17	&	72.45	&	Bg	&	1.49	&	49.80	\\
Au	&	3.03	&	101.00	&	Au	&	3.09	&	102.94	&	Bu	&	3.03	&	100.97	&	Bg	&	2.20	&	73.52	\\
Bg	&	3.54	&	118.01	&	Bg	&	3.57	&	119.22	&	Au	&	3.13	&	104.47	&	Ag	&	2.21	&	73.75	\\
Ag	&	3.55	&	118.48	&	Ag	&	3.60	&	120.12	&	Bg	&	3.61	&	120.38	&	Bu	&	3.05	&	101.80	\\
Bu	&	3.74	&	124.72	&	Bu	&	3.75	&	125.15	&	Ag	&	3.64	&	121.48	&	Au	&	3.18	&	105.94	\\
Bu	&	4.99	&	166.55	&	Bu	&	5.03	&	167.88	&	Bu	&	3.77	&	125.89	&	Bg	&	3.64	&	121.48	\\
Au	&	5.01	&	167.22	&	Au	&	5.05	&	168.45	&	Bu	&	5.07	&	169.08	&	Ag	&	3.68	&	122.78	\\
Bg	&	5.13	&	171.12	&	Bg	&	5.18	&	172.85	&	Au	&	5.08	&	169.52	&	Bu	&	3.79	&	126.35	\\
Au	&	5.63	&	187.73	&	Au	&	5.67	&	189.00	&	Bg	&	5.23	&	174.59	&	Bu	&	5.10	&	170.22	\\
Bg	&	6.30	&	210.21	&	Bg	&	6.32	&	210.85	&	Au	&	5.70	&	190.23	&	Au	&	5.11	&	170.48	\\
Ag	&	6.36	&	212.25	&	Ag	&	6.40	&	213.38	&	Bg	&	6.34	&	211.48	&	Bg	&	5.28	&	176.19	\\
Ag	&	7.15	&	238.50	&	Bu	&	7.20	&	240.00	&	Ag	&	6.43	&	214.38	&	Au	&	5.74	&	191.47	\\
Bu	&	7.19	&	239.77	&	Ag	&	7.20	&	240.27	&	Bu	&	7.20	&	240.27	&	Bg	&	6.36	&	212.08	\\
Au	&	7.23	&	241.27	&	Au	&	7.26	&	242.07	&	Ag	&	7.25	&	241.90	&	Ag	&	6.46	&	215.32	\\
Bg	&	7.82	&	260.81	&	Bg	&	7.83	&	261.18	&	Au	&	7.28	&	242.80	&	Bu	&	7.21	&	240.57	\\
Ag	&	7.84	&	261.51	&	Ag	&	7.86	&	262.01	&	Bg	&	7.84	&	261.51	&	Ag	&	7.30	&	243.44	\\
Bu	&	8.70	&	290.13	&	Bu	&	8.70	&	290.13	&	Ag	&	7.87	&	262.48	&	Au	&	7.30	&	243.50	\\
Ag	&	10.94	&	364.95	&	Ag	&	10.96	&	365.59	&	Bu	&	8.70	&	290.23	&	Bg	&	7.85	&	261.85	\\
Bu	&	12.89	&	430.06	&	Bu	&	12.91	&	430.70	&	Ag	&	10.98	&	366.25	&	Ag	&	7.88	&	262.98	\\
Bu	&	16.13	&	538.07	&	Ag	&	16.15	&	538.67	&	Bu	&	12.93	&	431.40	&	Bu	&	8.70	&	290.30	\\
Ag	&	16.13	&	538.14	&	Bu	&	16.16	&	539.04	&	Ag	&	16.16	&	539.17	&	Ag	&	11.00	&	366.85	\\
Ag	&	16.16	&	538.94	&	Ag	&	16.19	&	539.94	&	Bu	&	16.19	&	540.04	&	Bu	&	12.95	&	432.00	\\
Au	&	16.18	&	539.67	&	Au	&	16.22	&	541.11	&	Ag	&	16.22	&	541.01	&	Ag	&	16.18	&	539.67	\\
Bg	&	16.19	&	539.87	&	Bg	&	16.23	&	541.21	&	Au	&	16.26	&	542.44	&	Bu	&	16.22	&	541.01	\\
	&		&		&		&		&		&	Bg	&	16.26	&	542.48	&	Ag	&	16.25	&	542.08	\\
	&		&		&		&		&		&		&		&		&	Bg	&	16.30	&	543.61	\\
	&		&		&		&		&		&		&		&		&	Au	&	16.30	&	543.64	\\
\bottomrule
\end{tabular}
\end{table}

\begin{table}[h!]
\centering
\caption{DFT Character table at multiple pressures}
\label{tab:Character Table Multiple Pressures_1}
\begin{tabular}{cccccccccccc}
\toprule
\multicolumn{3}{c}{-0.2 GPa} & \multicolumn{3}{c}{0.0 GPa} & \multicolumn{3}{c}{5.0 GPa} & \multicolumn{3}{c}{10.0 GPa}\\
\cmidrule(lr){1-3} \cmidrule(lr){4-6} \cmidrule(lr){7-9}\cmidrule(lr){10-12}
Mode & THz & $cm^{-1}$ & Mode & THz & $cm^{-1}$ & Mode & THz & $cm^{-1}$ & Mode & THz & $cm^{-1}$\\
\midrule
Bu	&	0.00	&	0.00	&	Au	&	0.00	&	0.00	&	Bu	&	0.00	&	0.00	&	Bu	&	0.00	&	0.00	\\
Bu	&	0.00	&	0.00	&	Bg	&	1.54	&	51.20	&	Bu	&	0.00	&	0.00	&	Bu	&	0.00	&	0.00	\\
Au	&	0.00	&	0.00	&	Bg	&	2.27	&	75.72	&	Au	&	0.00	&	0.00	&	Au	&	0.00	&	0.00	\\
Bg	&	1.52	&	50.80	&	Ag	&	2.28	&	75.99	&	Bg	&	1.64	&	54.64	&	Bg	&	1.71	&	57.07	\\
Bg	&	2.24	&	74.58	&	Bu	&	3.11	&	103.57	&	Bg	&	2.81	&	93.86	&	Bg	&	3.13	&	104.27	\\
Ag	&	2.25	&	74.95	&	Au	&	3.25	&	108.37	&	Ag	&	2.85	&	95.07	&	Ag	&	3.19	&	106.24	\\
Bu	&	3.08	&	102.60	&	Bg	&	3.71	&	123.59	&	Bu	&	3.60	&	120.15	&	Bu	&	3.91	&	130.39	\\
Au	&	3.22	&	107.24	&	Ag	&	3.75	&	125.05	&	Au	&	3.86	&	128.76	&	Au	&	4.19	&	139.66	\\
Bg	&	3.67	&	122.52	&	Bu	&	3.82	&	127.52	&	Bu	&	4.07	&	135.79	&	Bu	&	4.22	&	140.66	\\
Ag	&	3.72	&	123.92	&	Au	&	5.16	&	172.25	&	Bg	&	4.34	&	144.90	&	Ag	&	4.73	&	157.78	\\
Bu	&	3.81	&	127.15	&	Bu	&	5.17	&	172.35	&	Ag	&	4.35	&	145.00	&	Bg	&	4.78	&	159.41	\\
Bu	&	5.14	&	171.29	&	Bg	&	5.38	&	179.32	&	Au	&	5.73	&	191.03	&	Au	&	6.18	&	206.21	\\
Au	&	5.14	&	171.39	&	Au	&	5.81	&	193.73	&	Bu	&	5.85	&	195.27	&	Bu	&	6.42	&	214.08	\\
Bg	&	5.33	&	177.79	&	Bg	&	6.40	&	213.31	&	Bg	&	6.33	&	211.18	&	Bg	&	6.96	&	232.13	\\
Au	&	5.77	&	192.60	&	Ag	&	6.51	&	217.05	&	Au	&	6.59	&	219.79	&	Au	&	7.27	&	242.53	\\
Bg	&	6.38	&	212.68	&	Bu	&	7.23	&	241.20	&	Bg	&	6.89	&	229.66	&	Ag	&	7.37	&	245.87	\\
Ag	&	6.48	&	216.22	&	Au	&	7.34	&	244.80	&	Ag	&	7.00	&	233.59	&	Bg	&	7.48	&	249.34	\\
Bu	&	7.22	&	240.87	&	Ag	&	7.38	&	246.30	&	Bu	&	7.45	&	248.47	&	Bu	&	7.65	&	255.08	\\
Au	&	7.32	&	244.17	&	Bg	&	7.87	&	262.51	&	Au	&	7.75	&	258.61	&	Au	&	8.11	&	270.39	\\
Ag	&	7.34	&	244.87	&	Ag	&	7.91	&	263.95	&	Bg	&	8.09	&	269.99	&	Bg	&	8.35	&	278.43	\\
Bg	&	7.86	&	262.18	&	Bu	&	8.71	&	290.67	&	Ag	&	8.29	&	276.66	&	Ag	&	8.70	&	290.20	\\
Ag	&	7.90	&	263.45	&	Ag	&	11.03	&	368.02	&	Ag	&	8.30	&	276.92	&	Bu	&	9.04	&	301.64	\\
Bu	&	8.71	&	290.50	&	Bu	&	12.99	&	433.27	&	Bu	&	8.87	&	295.74	&	Ag	&	9.05	&	301.98	\\
Ag	&	11.02	&	367.45	&	Ag	&	16.21	&	540.71	&	Ag	&	11.42	&	380.80	&	Ag	&	11.73	&	391.37	\\
Bu	&	12.97	&	432.63	&	Bu	&	16.28	&	542.94	&	Bu	&	13.42	&	447.51	&	Bu	&	13.80	&	460.19	\\
Ag	&	16.20	&	540.21	&	Ag	&	16.31	&	544.14	&	Ag	&	16.57	&	552.85	&	Ag	&	16.90	&	563.82	\\
Bu	&	16.25	&	541.94	&	Bg	&	16.36	&	545.84	&	Bu	&	16.92	&	564.26	&	Bu	&	17.45	&	581.94	\\
Ag	&	16.28	&	543.08	&	Au	&	16.37	&	546.04	&	Ag	&	17.00	&	567.03	&	Ag	&	17.58	&	586.31	\\
Bg	&	16.33	&	544.74	&		&		&		&	Bg	&	17.04	&	568.33	&	Bg	&	17.59	&	586.64	\\
Au	&	16.34	&	544.88	&		&		&		&	Au	&	17.09	&	569.96	&	Au	&	17.68	&	589.61	\\
\bottomrule
\end{tabular}
\end{table}

\begin{table}[h!]
\centering
\caption{DFT Character Table at multiple pressures}
\label{tab:Character Table Multiple Pressures_2}
\begin{tabular}{cccccccccccc}
\toprule
\multicolumn{3}{c}{15.0 GPa} & \multicolumn{3}{c}{20.0 GPa} & \multicolumn{3}{c}{25.0 GPa} & \multicolumn{3}{c}{100.0 GPa}\\
\cmidrule(lr){1-3} \cmidrule(lr){4-6} \cmidrule(lr){7-9}\cmidrule(lr){10-12}
Mode & THz & $cm^{-1}$ & Mode & THz & $cm^{-1}$ & Mode & THz & $cm^{-1}$ & Mode & THz & $cm^{-1}$\\
\midrule
Bg	&	3.35	&	111.58	&	Bg	&	1.71	&	56.87	&	Bu	&	0.00	&	0.00	&	Bg	&	2.08	&	69.31	\\
Ag	&	3.41	&	113.61	&	Bg	&	3.50	&	116.81	&	Bg	&	1.71	&	57.01	&	Bu	&	2.58	&	85.89	\\
Bu	&	4.03	&	134.56	&	Ag	&	3.56	&	118.75	&	Bg	&	3.64	&	121.52	&	Ag	&	4.18	&	139.56	\\
Bu	&	4.34	&	144.73	&	Bu	&	4.01	&	133.86	&	Ag	&	3.68	&	122.72	&	Bg	&	4.26	&	142.20	\\
Au	&	4.40	&	146.70	&	Bu	&	4.48	&	149.30	&	Bu	&	3.93	&	131.06	&	Ag	&	4.58	&	152.91	\\
Ag	&	5.01	&	167.02	&	Au	&	4.55	&	151.64	&	Bu	&	4.61	&	153.61	&	Bu	&	5.10	&	170.18	\\
Bg	&	5.08	&	169.32	&	Ag	&	5.21	&	173.79	&	Au	&	4.66	&	155.51	&	Bg	&	5.29	&	176.52	\\
Au	&	6.57	&	219.15	&	Bg	&	5.28	&	176.12	&	Ag	&	5.36	&	178.82	&	Au	&	5.39	&	179.62	\\
Bu	&	6.88	&	229.56	&	Au	&	6.91	&	230.56	&	Bg	&	5.42	&	180.69	&	Au	&	9.31	&	310.45	\\
Bg	&	7.36	&	245.54	&	Bu	&	7.28	&	242.80	&	Au	&	7.22	&	240.87	&	Bg	&	9.84	&	328.09	\\
Ag	&	7.67	&	255.91	&	Bg	&	7.66	&	255.41	&	Bu	&	7.63	&	254.44	&	Bu	&	10.35	&	345.07	\\
Bu	&	7.84	&	261.61	&	Ag	&	7.94	&	264.88	&	Bg	&	7.90	&	263.42	&	Bu	&	10.61	&	353.78	\\
Au	&	7.87	&	262.48	&	Bu	&	8.04	&	268.12	&	Ag	&	8.20	&	273.49	&	Ag	&	10.94	&	365.05	\\
Bg	&	8.07	&	269.05	&	Au	&	8.39	&	279.99	&	Bu	&	8.23	&	274.66	&	Au	&	11.06	&	369.02	\\
Au	&	8.42	&	280.93	&	Bg	&	8.55	&	285.20	&	Au	&	8.84	&	294.80	&	Ag	&	12.65	&	422.06	\\
Bg	&	8.69	&	289.83	&	Au	&	8.72	&	290.70	&	Bg	&	8.95	&	298.51	&	Bg	&	12.70	&	423.66	\\
Ag	&	9.09	&	303.34	&	Bg	&	9.15	&	305.04	&	Au	&	9.01	&	300.61	&	Au	&	13.28	&	442.87	\\
Bu	&	9.26	&	308.75	&	Ag	&	9.46	&	315.65	&	Bg	&	9.67	&	322.46	&	Bu	&	13.57	&	452.61	\\
Ag	&	9.69	&	323.06	&	Bu	&	9.50	&	316.79	&	Bu	&	9.76	&	325.53	&	Ag	&	13.87	&	462.62	\\
Ag	&	12.02	&	400.88	&	Ag	&	10.23	&	341.34	&	Ag	&	9.81	&	327.06	&	Bg	&	15.08	&	502.88	\\
Bu	&	14.15	&	472.03	&	Ag	&	12.28	&	409.58	&	Ag	&	10.72	&	357.48	&	Ag	&	15.24	&	508.45	\\
Ag	&	17.22	&	574.33	&	Bu	&	14.49	&	483.20	&	Ag	&	12.52	&	417.76	&	Bu	&	17.66	&	589.01	\\
Bu	&	17.91	&	597.51	&	Ag	&	17.52	&	584.40	&	Bu	&	14.80	&	493.71	&	Ag	&	20.91	&	697.55	\\
Bg	&	18.08	&	602.95	&	Bu	&	18.34	&	611.69	&	Ag	&	17.81	&	594.04	&	Bu	&	22.01	&	734.21	\\
Ag	&	18.09	&	603.35	&	Bg	&	18.52	&	617.83	&	Bu	&	18.73	&	624.67	&	Bg	&	22.43	&	748.12	\\
Au	&	18.21	&	607.35	&	Ag	&	18.55	&	618.86	&	Bg	&	18.93	&	631.47	&	Ag	&	22.81	&	760.93	\\
&		&		&	Au	&	18.70	&	623.66	&	Ag	&	18.98	&	633.10	&	Au	&	23.30	&	777.30	\\
&		&		&		&		&		&	Au	&	19.15	&	638.84	&		&		&		\\
\bottomrule
\end{tabular}
\end{table}

\clearpage
\newpage
\subsection{Ab-initio thermodynamics}
In this study, the quasi-harmonic approximation (QHA) is used to evaluate the volume-dependent thermodynamic properties of materials under varying temperature conditions. The methodology involves performing phonon calculations at multiple volumes around the equilibrium volume of the material within the harmonic approximation. 

The second-order interatomic force constants and the dynamical matrix are determined for each volume. By diagonalizing the dynamical matrix, the phonon frequencies and corresponding displacement vectors at specific wave vectors are obtained. These frequencies are subsequently utilized to calculate the phonon density of states (pDOS), $g\left(\nu\right)$, as well as the phonon dispersion curves.

When phonon frequencies over the Brillouin zone are known, the vibrational contribution to the free energy, $F_{vib}$, of the system is given by:
 \begin{equation}
F_{vib}\left(V,T\right)=\int_{0}^{\infty} g\left(\nu\right)\left[\frac{h\nu}{2}+k_{B}T\ln{\left(1-\exp{\left(-\frac{h\nu}{k_{B}T}\right)}\right)}\right]d\nu,
\end{equation}

where $k_{B}$, $h$ and $\nu$ are the Boltzmann constant, the Planck constant, and the vibrational frequency, respectively. 

The electronic contribution to the free energy as a function of volume and temperature, $F_{\text{elec}}$, is calculated as
\begin{equation}
F_{elec}\left(V,T\right)=\Delta E_{elec}\left(V,T\right)-TS_{elec}\left(V,T\right),
\end{equation}

where $\Delta E_{elec}\left(V,T\right)$ represents the contribution to the electronic energy due to temperature variations and $S_{elec}\left(V,T\right)$ denotes the electronic entropic contribution to the free energy. At low temperatures, $\Delta E_{elec}\left(V, T\right)$ is minimal and can be disregarded. Both quantities can be calculated using the electronic density of states (eDOS).

Then, the Helmholtz free energy, $F\left(V,T\right)$ is calculated at different volumes and temperatures:

\begin{equation}
F\left(V,T\right)=E_{0K}+F_{vib}\left(V,T\right)+F_{elec}\left(V,T\right),
\end{equation}

where $E_{0K}$ is the total energy of the system at 0 K and given volume, V.   

After calculating $F\left(V,T\right)$ at various volumes and temperatures, extracting the thermodynamic data becomes straightforward using the equations of state. In this study, we employed the Vinet equation \cite{VinetRose}, which is given by:

\begin{equation}
F\left(V\right)=F_{eq}+\frac{BV_{eq}}{C^{2}}\left[1-\left(1+C\left(\left(\frac{V}{V_{eq}}\right)^{1/3}-1\right)\right)\exp{\left(C\left(1-\left(\frac{V}{V_{eq}}\right) ^{\frac{1}{3}}\right)\right)}\right],
\end{equation}

where $B$ and $C$ are fitting parameters and $V_{eq}$ is the equilibrium volume. The bulk modulus, $K$, and its pressure derivative $K'$, are respectively given by
\begin{equation}
K=\frac{B}{9} ,
\end{equation}

\begin{equation}
K'=\frac{2C}{3} +1 .
\end{equation}

 The Gibbs free energy, $G\left(T,p\right)$, is found by minimizing $F\left(V,T\right)$ with respect to the volume $V$ at a target temperature $T$:

\begin{equation}
G\left(T,p\right)=\genfrac{}{}{0pt}{}{min}{V}\left[F\left(V,T\right)+pV\right]
\end{equation}

Once the volume as a function of temperature has been calculated, the volumetric thermal expansion coefficient, $\alpha_{V}$, can be obtained. This coefficient describes the rate of volume increase with temperature and is defined as:
\begin{equation}
\alpha_{V}\left(T\right)=\frac{1}{V\left(T\right)}\frac{\partial V \left(T\right)}{\partial T}.
\end{equation}

Finally, the quasi-harmonic approximation (QHA) facilitates the derivation of the thermodynamic Grüneisen parameter $\gamma$. This parameter quantifies the variation of phonon frequencies with volume, thereby indirectly indicating the effect of temperature on the lattice dynamics.

\begin{equation}
\gamma \left(T\right) = \frac{V\left(T\right)\alpha_{V}\left( T\right)B\left(T\right)}{C_{V}\left(T\right)}
\end{equation}

\subsection{Vinet equation fitting}
\begin{figure} [h]
    \centering
    \includegraphics[width=1\linewidth]{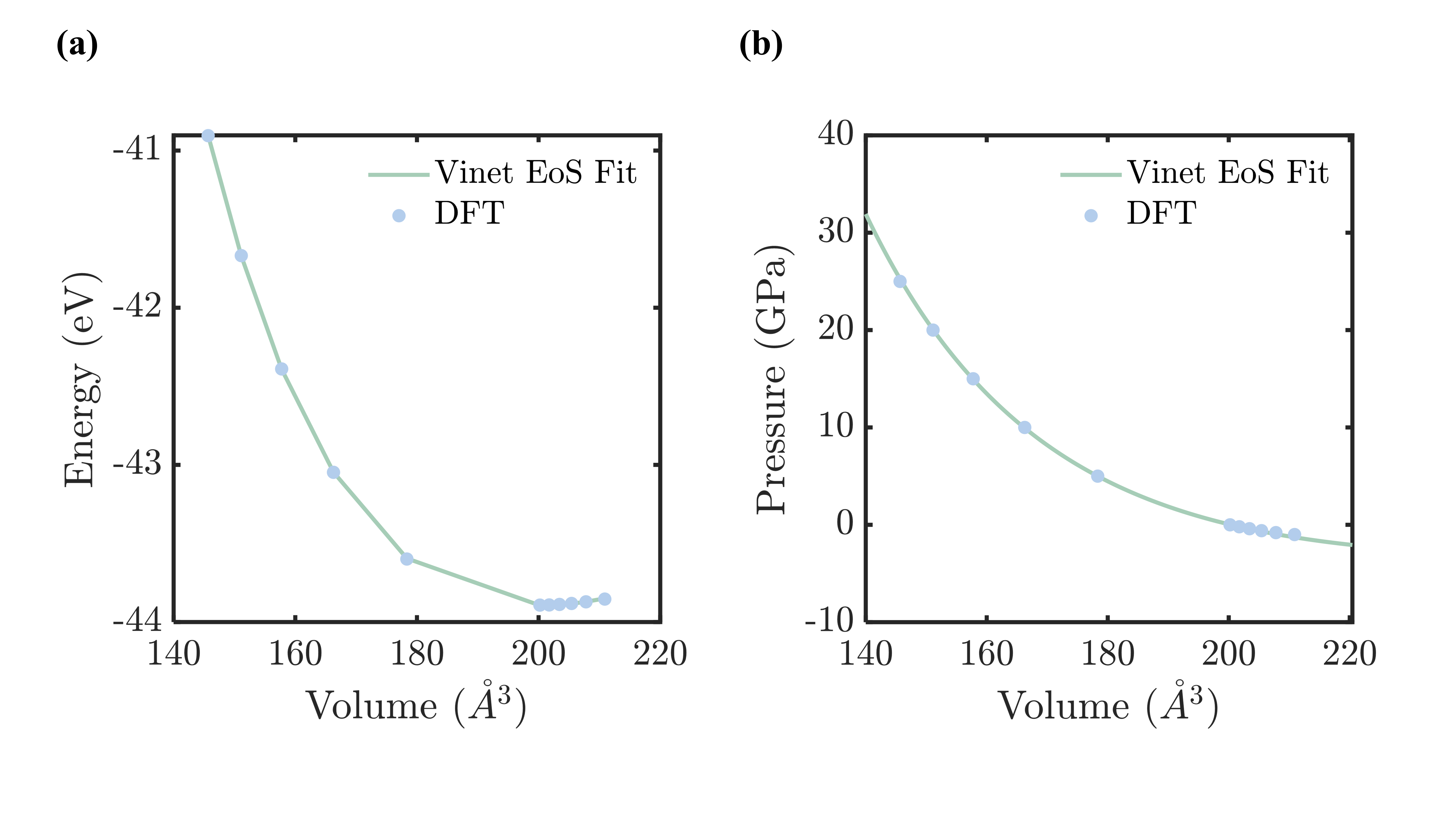}
    \caption{ Vinet equation of state fitting: a) Total energy versus volume,  and b) pressure versus volume.}
    \label{fig:EoS}
\end{figure}

The fitted parameters resulting from the Vinet equation are shown in Table \ref{tab:Vinet}.
The resulted fitting is shown in Figure \ref{fig:EoS}.

\begin{table}[h!]
\centering
\caption{Vinet equation fitted parameters.}
\label{tab:Vinet}
\begin{tabular}{cccccc}
\toprule
$F_{eq}$ & $V_{eq}$ & B & C & K & $K'$\\
\midrule 
$eV$ & $A^3$ & $eV/A^3$ & & GPa & GPa\\
\midrule
-43.894 & 200.396 & 1.665 & 8.163 & 29.632 & 6.442\\
\bottomrule \\
\end{tabular}
\end{table}

\clearpage
\newpage
\section{POSCAR files} 
\begin{table}[h]
\centering
\caption{-1 GPa POSCAR file}
\label{tab:-1 GPa POSCAR file }
\begin{tabular}{lll}
%\begin{tabularx}{\textwidth}{XXX}
-10kBar/-1GPa 	&		&		\\
1	&		&		\\
6.00316520320050	&	-0.00000000278551	&	-0.00072837189484	\\
0.00000001188232	&	10.39532523842780	&	0.00000000457888	\\
-1.99128675342093	&	0.00000002251754	&	6.76836160088136	\\
Zn	&	P	&	S	\\
4	&	4	&	12	\\
Direct	&		&		\\
0.50000000169319	&	0.16718084756900	&	0.00000000353767	\\
0.49999999830681	&	0.83281915243100	&	-0.00000000353767	\\
0.00000000169319	&	0.66718084756900	&	0.00000000353767	\\
-0.00000000169319	&	0.33281915243100	&	-0.00000000353767	\\
0.05447591103021	&	-0.00000000032136	&	0.16373062165981	\\
0.94552408896979	&	0.00000000032136	&	0.83626937834019	\\
0.55447591103021	&	0.49999999967864	&	0.16373062165981	\\
0.44552408896979	&	0.50000000032136	&	0.83626937834019	\\
0.24386582170820	&	0.83532867262604	&	0.23933828173351	\\
0.75613418003742	&	0.83532867291284	&	0.76066171847349	\\
0.25042496476031	&	0.00000000099224	&	0.76169401810598	\\
0.74957503523969	&	-0.00000000099224	&	0.23830598189402	\\
0.75613417829180	&	0.16467132737396	&	0.76066171826649	\\
0.24386581996258	&	0.16467132708716	&	0.23933828152651	\\
0.74386582170820	&	0.33532867262604	&	0.23933828173351	\\
0.25613418003742	&	0.33532867291284	&	0.76066171847349	\\
0.75042496476031	&	0.50000000099224	&	0.76169401810598	\\
0.24957503523969	&	0.49999999900776	&	0.23830598189402	\\
0.25613417829180	&	0.66467132737396	&	0.76066171826649	\\
0.74386581996258	&	0.66467132708716	&	0.23933828152651	\\
\end{tabular}
\end{table}

\begin{table}[h]
\centering
\caption{-0.8 GPa POSCAR file}
\label{tab:-0.8 GPa POSCAR file }
\begin{tabular}{lll}
8kBar/0.8GPa	&		&		\\
1	&		&		\\
5.99550216060941	&	-0.00000000235891	&	-0.00028245869606	\\
0.00000000939226	&	10.38075877408890	&	0.00000000493977	\\
-1.99199599040231	&	0.00000000077824	&	6.67982326585020	\\
Zn	&	P	&	S	\\
4	&	4	&	12	\\
Direct	&		&		\\
0.50000000088753	&	0.16728406032717	&	0.00000000240544	\\
0.49999999911247	&	0.83271593967283	&	-0.00000000240544	\\
0.00000000088753	&	0.66728406032717	&	0.00000000240544	\\
-0.00000000088753	&	0.33271593967283	&	-0.00000000240544	\\
0.05519791578508	&	-0.00000000000425	&	0.16583908206296	\\
0.94480208421492	&	0.00000000000425	&	0.83416091793704	\\
0.55519791578509	&	0.49999999999575	&	0.16583908206296	\\
0.44480208421492	&	0.50000000000425	&	0.83416091793704	\\
0.24501613566484	&	0.83516205216810	&	0.24249148836936	\\
0.75498386575924	&	0.83516205188560	&	0.75750851192291	\\
0.24986409257503	&	0.00000000035417	&	0.75893511620540	\\
0.75013590742497	&	-0.00000000035417	&	0.24106488379460	\\
0.75498386433516	&	0.16483794783190	&	0.75750851163064	\\
0.24501613424076	&	0.16483794811440	&	0.24249148807709	\\
0.74501613566484	&	0.33516205216810	&	0.24249148836936	\\
0.25498386575924	&	0.33516205188560	&	0.75750851192291	\\
0.74986409257503	&	0.50000000035417	&	0.75893511620540	\\
0.25013590742497	&	0.49999999964583	&	0.24106488379460	\\
0.25498386433516	&	0.66483794783190	&	0.75750851163064	\\
0.74501613424076	&	0.66483794811440	&	0.24249148807709	\\
\end{tabular}
\end{table}

\begin{table}[h!]
\centering
\caption{-0.6 GPa POSCAR file}
\label{tab:-0.6 GPa POSCAR file }
\begin{tabular}{lll}
-6kBar/-0.6GPa  	&		&		\\
1	&		&		\\
5.98822174380890	&	0.00000000943131	&	-0.00172515812326	\\
0.00000000254838	&	10.36729552080250	&	0.00000002604626	\\
-1.99117461000483	&	0.00000000522798	&	6.62001396312547	\\
Zn	&	P	&	S	\\
4	&	4	&	12	\\
Direct	&		&		\\
0.50000000174859	&	0.16736937654044	&	0.00000000403451	\\
0.49999999825141	&	0.83263062345956	&	0.99999999596549	\\
0.00000000174859	&	0.66736937654044	&	0.00000000403451	\\
0.99999999825141	&	0.33263062345956	&	0.99999999596549	\\
0.05572908710092	&	0.00000000008545	&	0.16728962478448	\\
0.94427091289908	&	-0.00000000008545	&	0.83271037521552	\\
0.55572908710092	&	0.50000000008545	&	0.16728962478448	\\
0.44427091289908	&	0.49999999991455	&	0.83271037521552	\\
0.24589805327482	&	0.83500464469962	&	0.24468145372312	\\
0.75410194957121	&	0.83500464420090	&	0.75531854679540	\\
0.24954054676151	&	0.00000000078745	&	0.75707675981995	\\
0.75045945323849	&	0.99999999921255	&	0.24292324018005	\\
0.75410194672518	&	0.16499535530038	&	0.75531854627688	\\
0.24589805042879	&	0.16499535579910	&	0.24468145320460	\\
0.74589805327482	&	0.33500464469962	&	0.24468145372312	\\
0.25410194957121	&	0.33500464420090	&	0.75531854679540	\\
0.74954054676151	&	0.50000000078745	&	0.75707675981995	\\
0.25045945323849	&	0.49999999921255	&	0.24292324018005	\\
0.25410194672518	&	0.66499535530038	&	0.75531854627688	\\
0.74589805042879	&	0.66499535579910	&	0.24468145320460	\\
\end{tabular}
\end{table}

\begin{table}[h!]
\centering
\caption{-0.4 GPa POSCAR file}
\label{tab:5 GPa POSCAR file }
\begin{tabular}{lll}
-4kBar/-0.4GPa	&		&		\\
1	&		&		\\
5.98150234621447	&	-0.00000001419704	&	-0.00014492285534	\\
-0.00000002587161	&	10.35469130472150	&	-0.00000001354678	\\
-1.99166269718929	&	-0.00000000262423	&	6.56976535382774	\\
Zn	&	P	&	S	\\
4	&	4	&	12	\\
Direct	&		&		\\
0.50000000131688	&	0.16745533582942	&	0.00000000121150	\\
0.49999999868312	&	0.83254466417058	&	-0.00000000121150	\\
0.00000000131688	&	0.66745533582942	&	0.00000000121150	\\
-0.00000000131688	&	0.33254466417058	&	-0.00000000121150	\\
0.05610350817210	&	-0.00000000061244	&	0.16849436886768	\\
0.94389649182790	&	0.00000000061244	&	0.83150563113232	\\
0.55610350817210	&	0.49999999938756	&	0.16849436886768	\\
0.44389649182790	&	0.50000000061244	&	0.83150563113232	\\
0.24654382050263	&	0.83486187774263	&	0.24650265399551	\\
0.75345617949454	&	0.83486187738916	&	0.75349734616105	\\
0.24940137092997	&	0.00000000010810	&	0.75558144111324	\\
0.75059862907003	&	-0.00000000010810	&	0.24441855888676	\\
0.75345617949737	&	0.16513812225737	&	0.75349734600449	\\
0.24654382050546	&	0.16513812261084	&	0.24650265383895	\\
0.74654382050263	&	0.33486187774263	&	0.24650265399551	\\
0.25345617949454	&	0.33486187738916	&	0.75349734616105	\\
0.74940137092997	&	0.50000000010810	&	0.75558144111324	\\
0.25059862907003	&	0.49999999989190	&	0.24441855888676	\\
0.25345617949737	&	0.66513812225737	&	0.75349734600449	\\
0.74654382050546	&	0.66513812261084	&	0.24650265383895	\\
\end{tabular}
\end{table}

\begin{table}[h!]
\centering
\caption{-0.2 GPa POSCAR file}
\label{tab:-0.2 GPa POSCAR file}
\begin{tabular}{lll}
-2kBar/-0.2GPa  	&		&		\\
1	&		&		\\
5.97478632338851	&	0.00000000240354	&	0.00005903030624	\\
0.00000000510831	&	10.34264710101870	&	0.00000000726623	\\
-1.99104005199052	&	0.00000000018956	&	6.52911715518037	\\
Zn	&	P	&	S	\\
4	&	4	&	12	\\
Direct	&		&		\\
0.50000000021253	&	0.16752811139730	&	0.00000000037130	\\
0.49999999978747	&	0.83247188860270	&	-0.00000000037130	\\
0.00000000021253	&	0.66752811139730	&	0.00000000037130	\\
-0.00000000021253	&	0.33247188860270	&	-0.00000000037130	\\
0.05644568066446	&	0.00000000015117	&	0.16947687125064	\\
0.94355431933554	&	-0.00000000015117	&	0.83052312874936	\\
0.55644568066446	&	0.50000000015117	&	0.16947687125064	\\
0.44355431933554	&	0.49999999984883	&	0.83052312874936	\\
0.24714878090040	&	0.83472516274224	&	0.24799395734242	\\
0.75285121918821	&	0.83472516288663	&	0.75200604277066	\\
0.24928396035212	&	-0.00000000000416	&	0.75434632871722	\\
0.75071603964788	&	0.00000000000416	&	0.24565367128278	\\
0.75285121909960	&	0.16527483725776	&	0.75200604265758	\\
0.24714878081179	&	0.16527483711337	&	0.24799395722934	\\
0.74714878090040	&	0.33472516274224	&	0.24799395734242	\\
0.25285121918821	&	0.33472516288663	&	0.75200604277066	\\
0.74928396035212	&	0.49999999999584	&	0.75434632871722	\\
0.25071603964788	&	0.50000000000416	&	0.24565367128278	\\
0.25285121909960	&	0.66527483725776	&	0.75200604265758	\\
0.74714878081179	&	0.66527483711337	&	0.24799395722934	\\
\end{tabular}
\end{table}

\begin{table}[h!]
\centering
\caption{0 GPa POSCAR file}
\label{tab:0 GPa POSCAR file }
\begin{tabular}{lll}
0kBar/0GPa 	&		&		\\
1	&		&		\\
5.96817652349330	&	0.00000000281205	&	-0.00003657551813	\\
0.00000000569872	&	10.33110320908160	&	0.00000000485761	\\
-1.98977992817616	&	0.00000000185044	&	6.49389121551074	\\
Zn	&	P	&	S	\\
4	&	4	&	12	\\
Direct	&		&		\\
0.50000000027641	&	0.16756960205447	&	0.00000000037491	\\
0.49999999972359	&	0.83243039794553	&	-0.00000000037491	\\
0.00000000027641	&	0.66756960205447	&	0.00000000037491	\\
-0.00000000027641	&	0.33243039794553	&	-0.00000000037491	\\
0.05675466064062	&	0.00000000016596	&	0.17033330699731	\\
0.94324533935938	&	-0.00000000016596	&	0.82966669300269	\\
0.55675466064062	&	0.50000000016596	&	0.17033330699731	\\
0.44324533935938	&	0.49999999983404	&	0.82966669300269	\\
0.24769902271447	&	0.83459340482013	&	0.24927414634162	\\
0.75230097733767	&	0.83459340492078	&	0.75072585378777	\\
0.24919425262324	&	-0.00000000000987	&	0.75325646345325	\\
0.75080574737676	&	0.00000000000987	&	0.24674353654675	\\
0.75230097728553	&	0.16540659517987	&	0.75072585365838	\\
0.24769902266233	&	0.16540659507922	&	0.24927414621223	\\
0.74769902271447	&	0.33459340482013	&	0.24927414634162	\\
0.25230097733767	&	0.33459340492078	&	0.75072585378777	\\
0.74919425262324	&	0.49999999999013	&	0.75325646345325	\\
0.25080574737676	&	0.50000000000987	&	0.24674353654675	\\
0.25230097728553	&	0.66540659517987	&	0.75072585365838	\\
0.74769902266233	&	0.66540659507922	&	0.24927414621223	\\
\end{tabular}
\end{table}

\begin{table}[h!]
\centering
\caption{5 GPa POSCAR file}
\label{tab:5 GPa POSCAR file }
\begin{tabular}{lll}
50kBar/5GPa	&		&		\\
1	&		&		\\
5.83315299357284	&	-0.00000000234313	&	0.00001103655674	\\
-0.00000000251527	&	10.10566225217940	&	-0.00000000415427	\\
-1.94240190822316	&	-0.00000000008452	&	6.04952015931070	\\
Zn	&	P	&	S	\\
4	&	4	&	12	\\
Direct	&		&		\\
0.49999999999196	&	0.16847594644284	&	0.00000000007541	\\
0.00000000000804	&	0.33152405355716	&	-0.00000000007541	\\
-0.00000000000804	&	0.66847594644284	&	0.00000000007541	\\
0.50000000000804	&	0.83152405355716	&	-0.00000000007541	\\
0.06052691232376	&	-0.00000000000331	&	0.18133846014091	\\
0.93947308767624	&	0.00000000000331	&	0.81866153985909	\\
0.56052691232376	&	0.49999999999669	&	0.18133846014091	\\
0.43947308767624	&	0.50000000000331	&	0.81866153985909	\\
0.75537892176677	&	0.33205925498362	&	0.26579680011805	\\
0.24462107807023	&	0.33205925496907	&	0.73420319973340	\\
0.25039489705917	&	0.00000000005741	&	0.73943344911032	\\
0.24960510294083	&	0.49999999994259	&	0.26056655088968	\\
0.74462107823323	&	0.16794074501638	&	0.73420319988195	\\
0.25537892192977	&	0.16794074503093	&	0.26579680026660	\\
0.25537892176677	&	0.83205925498362	&	0.26579680011805	\\
0.74462107807023	&	0.83205925496907	&	0.73420319973340	\\
0.75039489705917	&	0.50000000005741	&	0.73943344911032	\\
0.74960510294083	&	-0.00000000005741	&	0.26056655088968	\\
0.24462107823323	&	0.66794074501638	&	0.73420319988195	\\
0.75537892192978	&	0.66794074503093	&	0.26579680026660	\\
\end{tabular}
\end{table}

\begin{table}[h!]
\centering
\caption{10 GPa POSCAR file}
\label{tab:10 GPa POSCAR file }
\begin{tabular}{lll}
100kBar/10GPa	&		&		\\
1	&		&		\\
5.72976950858816	&	-0.00000000076130	&	0.00443766055989	\\
-0.00000000219485	&	9.93758683335474	&	0.00000000454866	\\
-1.90336961117475	&	-0.00000000506585	&	5.83793475984477	\\
Zn	&	P	&	S	\\
4	&	4	&	12	\\
Direct	&		&		\\
0.49999999976163	&	0.16896136629576	&	-0.00000000029965	\\
0.00000000023837	&	0.33103863370424	&	0.00000000029965	\\
-0.00000000023837	&	0.66896136629576	&	-0.00000000029965	\\
0.50000000023837	&	0.83103863370424	&	0.00000000029965	\\
0.06244231266172	&	-0.00000000000491	&	0.18652393693238	\\
0.93755768733828	&	0.00000000000491	&	0.81347606306762	\\
0.56244231266172	&	0.49999999999509	&	0.18652393693238	\\
0.43755768733828	&	0.50000000000491	&	0.81347606306762	\\
0.75983867330321	&	0.33021931166255	&	0.27367095309814	\\
0.24016132637704	&	0.33021931164076	&	0.72632904667574	\\
0.25215304685612	&	-0.00000000008615	&	0.73238871787919	\\
0.24784695314388	&	0.50000000008615	&	0.26761128212081	\\
0.74016132669679	&	0.16978068833745	&	0.72632904690186	\\
0.25983867362296	&	0.16978068835924	&	0.27367095332426	\\
0.25983867330321	&	0.83021931166255	&	0.27367095309814	\\
0.74016132637704	&	0.83021931164076	&	0.72632904667574	\\
0.75215304685612	&	0.49999999991385	&	0.73238871787919	\\
0.74784695314388	&	0.00000000008615	&	0.26761128212081	\\
0.24016132669679	&	0.66978068833745	&	0.72632904690186	\\
0.75983867362296	&	0.66978068835924	&	0.27367095332426	\\
\end{tabular}
\end{table}

\begin{table}[h!]
\centering
\caption{15 GPa POSCAR file}
\label{tab:15 GPa POSCAR file }
\begin{tabular}{lll}
150kBar/15GPa 	&		&		\\
1	&		&		\\
5.64370227203906	&	0.00000000116960	&	0.00018665744424	\\
0.00000000302346	&	9.79624677545720	&	-0.00000000439865	\\
-1.86196406469935	&	-0.00000000141210	&	5.70517450246358	\\
Zn	&	P	&	S	\\
4	&	4	&	12	\\
Direct	&		&		\\
0.49999999964609	&	0.16925190955916	&	-0.00000000105648	\\
0.00000000035391	&	0.33074809044084	&	0.00000000105648	\\
-0.00000000035391	&	0.66925190955916	&	-0.00000000105648	\\
0.50000000035391	&	0.83074809044084	&	0.00000000105648	\\
0.06366935615581	&	-0.00000000020133	&	0.18969858441911	\\
0.93633064384419	&	0.00000000020133	&	0.81030141558089	\\
0.56366935615581	&	0.49999999979867	&	0.18969858441911	\\
0.43633064384419	&	0.50000000020133	&	0.81030141558089	\\
0.76304704414533	&	0.32869643012194	&	0.27867177230403	\\
0.23695295511088	&	0.32869643056392	&	0.72132822766421	\\
0.25388431734843	&	-0.00000000016093	&	0.72760198473404	\\
0.74611568265157	&	0.00000000016093	&	0.27239801526596	\\
0.73695295585467	&	0.17130356987806	&	0.72132822769597	\\
0.26304704488912	&	0.17130356943608	&	0.27867177233579	\\
0.26304704414533	&	0.82869643012194	&	0.27867177230403	\\
0.73695295511088	&	0.82869643056392	&	0.72132822766421	\\
0.75388431734843	&	0.49999999983907	&	0.72760198473404	\\
0.24611568265157	&	0.50000000016093	&	0.27239801526596	\\
0.23695295585467	&	0.67130356987806	&	0.72132822769597	\\
0.76304704488912	&	0.67130356943608	&	0.27867177233579	\\
\end{tabular}
\end{table}

\begin{table}[h!]
\centering
\caption{20 GPa POSCAR file}
\label{tab:20 GPa POSCAR file }
\begin{tabular}{lll}
200kBar/20GPa	&		&		\\
1	&		&		\\
5.56928554528651	&	-0.00000000062449	&	0.00003524755760	\\
0.00000000018450	&	9.67195256165614	&	-0.00000000028183	\\
-1.82917913717499	&	0.00000000152903	&	5.60936466145858	\\
Zn	&	P	&	S	\\
4	&	4	&	12	\\
Direct	&		&		\\
0.49999999980189	&	0.16936548398874	&	-0.00000000048266	\\
0.00000000019811	&	0.33063451601126	&	0.00000000048266	\\
-0.00000000019811	&	0.66936548398874	&	-0.00000000048266	\\
0.50000000019811	&	0.83063451601126	&	0.00000000048266	\\
0.06452826034467	&	-0.00000000002177	&	0.19182625461856	\\
0.93547173965533	&	0.00000000002177	&	0.80817374538144	\\
0.56452826034467	&	0.49999999997823	&	0.19182625461856	\\
0.43547173965533	&	0.50000000002177	&	0.80817374538144	\\
0.76555982351116	&	0.32736742676409	&	0.28223382421792	\\
0.23444017630448	&	0.32736742687102	&	0.71776617577335	\\
0.75543118646772	&	0.49999999994005	&	0.72382459799926	\\
0.74456881353228	&	0.00000000005995	&	0.27617540200074	\\
0.73444017648884	&	0.17263257323591	&	0.71776617578208	\\
0.26555982369553	&	0.17263257312898	&	0.28223382422665	\\
0.26555982351116	&	0.82736742676409	&	0.28223382421792	\\
0.73444017630448	&	0.82736742687102	&	0.71776617577335	\\
0.25543118646772	&	-0.00000000005995	&	0.72382459799926	\\
0.24456881353228	&	0.50000000005995	&	0.27617540200074	\\
0.23444017648884	&	0.67263257323591	&	0.71776617578208	\\
0.76555982369553	&	0.67263257312898	&	0.28223382422665	\\
\end{tabular}
\end{table}

\begin{table}[h!]
\centering
\caption{25 GPa POSCAR file}
\label{tab:25 GPa POSCAR file}
\begin{tabular}{lll}
250kBar/25GPa	&		&		\\
1	&		&		\\
5.50374811342113	&	0.00000000935147	&	0.00292018292723	\\
-0.00000001004724	&	9.55979032318767	&	0.00000000540405	\\
-1.80360728851282	&	-0.00000000737491	&	5.53590197699186	\\
Zn	&	P	&	S	\\
4	&	4	&	12	\\
Direct	&		&		\\
0.49999999953076	&	0.16934222892342	&	0.99999999924109	\\
0.00000000046924	&	0.33065777107658	&	0.00000000075891	\\
0.99999999953076	&	0.66934222892342	&	0.99999999924109	\\
0.50000000046924	&	0.83065777107658	&	0.00000000075891	\\
0.06516296782288	&	-0.00000000000896	&	0.19330657979132	\\
0.93483703217712	&	0.00000000000896	&	0.80669342020868	\\
0.56516296782288	&	0.49999999999104	&	0.19330657979132	\\
0.43483703217712	&	0.50000000000896	&	0.80669342020868	\\
0.76764409310942	&	0.32617558050002	&	0.28493906486023	\\
0.23235590612214	&	0.32617558083506	&	0.71506093474169	\\
0.75676239719844	&	0.49999999974467	&	0.72063381808178	\\
0.74323760280156	&	0.00000000025533	&	0.27936618191822	\\
0.73235590689058	&	0.17382441949998	&	0.71506093513977	\\
0.26764409387786	&	0.17382441916494	&	0.28493906525831	\\
0.26764409310942	&	0.82617558050002	&	0.28493906486023	\\
0.73235590612214	&	0.82617558083506	&	0.71506093474169	\\
0.25676239719844	&	0.99999999974467	&	0.72063381808178	\\
0.24323760280156	&	0.50000000025533	&	0.27936618191822	\\
0.23235590689058	&	0.67382441949998	&	0.71506093513977	\\
0.76764409387786	&	0.67382441916494	&	0.28493906525831	\\
\end{tabular}
\end{table}

\begin{table}[h!]
\centering
\caption{100 GPa POSCAR file}
\label{tab:100 GPa POSCAR file}
\begin{tabular}{lll}
1000kBar/100GPa	&		&		\\
1	&		&		\\
4.87075238911411	&	0.00000000423863	&	0.02813037248142	\\
-0.00000000513346	&	8.48328930274853	&	-0.00000001516037	\\
-1.57491504013091	&	0.00000000912071	&	5.15984580354141	\\
Zn	&	P	&	S	\\
4	&	4	&	12	\\
Direct	&		&		\\
0.00000000026329	&	0.34014695575980	&	0.00000000040516	\\
-0.00000000026329	&	0.65985304427520	&	-0.00000000040516	\\
0.49999999974270	&	0.15985376556620	&	-0.00000000040516	\\
0.50000000026929	&	0.84014623445680	&	0.00000000040516	\\
0.55677446916586	&	0.50000000004304	&	0.19600273933515	\\
0.05677467640986	&	0.00000000003104	&	0.19600277772315	\\
0.94322532371214	&	-0.00000000003104	&	0.80399722227685	\\
0.44322553092914	&	0.49999999998096	&	0.80399726064485	\\
0.77689767488034	&	0.31336611576029	&	0.29986395937143	\\
0.77689767544678	&	0.68663388413106	&	0.29986395962840	\\
0.27689742776678	&	0.18663398199106	&	0.29986396290840	\\
0.27689742720034	&	0.81336601788828	&	0.29986396265143	\\
0.72575806918191	&	0.00000000017405	&	0.31958718891262	\\
0.22575834120391	&	0.50000000018605	&	0.31958780921762	\\
0.77424170461610	&	0.49999999983795	&	0.68041232918638	\\
0.27424197665309	&	-0.00000000017405	&	0.68041294947238	\\
0.72310257290167	&	0.18663398213472	&	0.70013603734757	\\
0.72310257233522	&	0.81336601803194	&	0.70013603709060	\\
0.22310232464922	&	0.31336611590394	&	0.70013604035160	\\
0.22310232521566	&	0.68663388427471	&	0.70013604060857	\\
\end{tabular}
\end{table}

\clearpage

\bibliographystyle{unsrt}
\bibliography{references}

\end{document}